\documentclass[a4paper,fleqn]{cas-sc}

\usepackage{graphicx}
\usepackage{amsmath}
\usepackage{float}
\usepackage{caption}
\usepackage{afterpage}
\usepackage{subcaption}
\usepackage{algorithmic}
\usepackage{todonotes}
\usepackage{booktabs}
\usepackage{siunitx}
\usepackage{tikz}
\usepackage{adjustbox}
\usepackage{soul}
\usepackage{amsmath}
\usepackage{caption}
\usepackage{subcaption}
\usepackage[T1]{fontenc}
\usepackage{makecell}
\usepackage{ulem}

\usepackage{draftwatermark} 

\SetWatermarkText{Preprint}          
\SetWatermarkFontSize{5cm}        
\SetWatermarkColor[gray]{0.9}     
\SetWatermarkAngle{45}            
\SetWatermarkScale{1}             

\def\tsc#1{\csdef{#1}{\textsc{\lowercase{#1}}\xspace}}
\tsc{WGM}
\tsc{QE}
\tsc{EP}
\tsc{PMS}
\tsc{BEC}
\tsc{DE}

\usepackage{xcolor}
\definecolor{darkblue}{rgb}{0.0, 0.0, 0.5} 

\usepackage{natbib}  
\usepackage{hyperref}
\renewcommand{\citep}[1]{\textup{(}\citeauthor{#1} \citeyear{#1}\textup{)}}
\renewcommand{\citet}[1]{\textup{\citeauthor{#1} \citeyear{#1}}}
\setcitestyle{semicolon}

\hypersetup{
    colorlinks=true,
    linkcolor=darkblue, 
    citecolor=darkblue, 
    urlcolor=darkblue   
}

\usepackage{todonotes} 

\usepackage{booktabs}
\usepackage{algorithm,algorithmic}

\begin{document}
\let\WriteBookmarks\relax
\def\floatpagepagefraction{1}
\def\textpagefraction{.001}

\shorttitle{TetraSub : Harmonic separation of directional focused wave group }

\shortauthors{Sithik Aliyar et~al.}

\title [mode = title]{Directional focused wave group response of a Floating Wind Turbine: Harmonic separation in experiment and CFD}                      



%
\author[1]{Sithik Aliyar}[type=editor,
                        auid=000,bioid=1,
                        orcid=0000-0003-0188-4613]



\ead{sitali@dtu.dk}


\credit{Conceptualization of this study, Methodology, Software, Writing}

\affiliation[1]{organization={Department of Wind and Energy systems, Technical University of Denmark},
    city={Kgs. Lyngby},
    postcode={2800}, 
    country={Denmark}}

\author[1]{ Henrik Bredmose}
\ead{hbre@dtu.dk}
\credit{Conceptualization of this study, Methodology, Writing}

\author[2] { Johan Roenby}
\ead{johan@stromning.com}
 \affiliation[2]{organization={Stromning Aps},
  addressline={Luftmarinegade 62}, 
   city={ København K},
    postcode={DK-1432}, 
     country={Denmark}}
     \credit{Methodology, Software}

\author[3] { Pietro Danilo Tomaselli}
 \ead{dto@dhigroup.com}
 \affiliation[3]{organization={DHI, Marine and Hydraulic Structures},
 addressline={Agern Alle 5}, 
    city={Hørsholm},
     postcode={2970}, 
   country={Denmark}}
    \credit{Software, Writing – Review $\&$ Editing}

\author[1]{ Hamid Sarlak}
\ead{hsar@dtu.dk}
\credit{Software, Writing – Review $\&$ Editing}




\begin{abstract}
The offshore wind sector relies on floating foundations for deeper waters. However, these face challenges from harsh conditions, nonlinear dynamics, and low-frequency resonant motions caused by second-order difference-frequency hydrodynamic loads. We analyze these dynamics and extract such higher harmonic motions for a semisubmersible floating foundation under extreme wave conditions using experimental and numerical approaches. Two distinct, focused wave groups, with and without wave spreading, are considered, and experimental data is obtained from scaled physical model tests using phase-shifted input signals to provide the harmonic decomposition of the floating foundation wave responses. The measured responses are reproduced numerically using a novel Computational Fluid Dynamics (CFD) based rigid body solver called FloatStepper, achieving generally good agreement. The study quantifies the effects of wave severity, spreading, and steepness on odd and even harmonics of the surge and pitch responses of the floating foundation and mooring line tensions. The focused wave group of a stronger sea state showed a notable increase in the amplitudes of odd harmonics for surge and pitch. In addition, the pitch subharmonic response, less noticeable in the milder sea states, became more apparent. Wave spreading primarily influenced the overall response of the spreading case, with a more pronounced effect observed on odd and even superharmonic responses. The results also reveal a front-back asymmetry in the tensions of the mooring lines, with the back lines experiencing greater tension than the front. Similarly to the effect of wave severity, a strict increase in wavegroup amplitude led to pronounced shifts in both subharmonic and superharmonic responses, transitioning from predominantly low-frequency surge-dominated behavior to a coupled surge-pitch interaction. The underlying cause of this pitch dominance is identified and discussed through CFD.
\end{abstract}


\maketitle

\section{Introduction}

Offshore winds are stronger and more consistent, and with the advancement of technology and the size of wind turbines, the Levelized Cost of Energy (LCOE) is expected to decrease \citep{IRENA2024}. Initially, offshore wind farms were designed as fixed structures (monopiles, tripods, jackets, gravity-based foundations) in shallow waters (less than $50$–$70$ m), with an average distance from the shore of approximately $20$ km \citep{diaz2020review}. As the industry evolves, floating structures (e.g., semi-submersibles, tension-leg platforms, spars) are being deployed in deeper waters (greater than $200$ m), complementing fixed structures where seabed-mounted foundations may be less feasible \citep{IRENA2024}. The development of floating wind turbines is expected to play an increasingly significant role in the future energy landscape, contributing to the larger goal of integration of renewable energy and supporting the Sustainable Development Goals (SDG) outlined in the UN Agenda for Sustainable Development $2030$ \citep{UN_SDGs_2015}.


 A key requirement of floating foundation design is the ability to withstand loads in highly non-linear wave conditions, particularly during extreme wave events. When floating wind turbines encounter such waves, the response becomes non-linear due to the interaction between the motion of the floating foundation and the wave forces \citep{pegalajar2019reproduction}. To mitigate direct wave excitation, floating wind turbines are engineered with long natural periods, typically exceeding $25$–$30$ s for heave, pitch, roll, and yaw motions, and even longer periods for surge and sway. This can be achieved by either reducing hydrostatic stiffness, lowering mooring stiffness, or employing both methods, which help position the system's natural frequencies away from the primary wave frequency range. However, their low hydrodynamic damping makes them highly resonant and vulnerable to excitation from second-order difference-frequency hydrodynamic loads. In recent years, considerable research interest has been in studying the effects of such nonlinear responses on floating wind turbines. When the turbine is not operating (with an idled rotor and feathered blades), difference-frequency wave forcing has been shown to govern global motion responses \citep{coulling2013importance}. As an illustration, in their research on semisubmersible floaters, \citet{coulling2013importance} used numerical simulations to replicate an experiment considering second-order wave loads. They analyzed the dynamics of stationary floaters under harsh sea conditions both with and without wind, demonstrating that nonlinear wave forces were the primary influence, whereas the impact of wind was minimal. \citet{roald2013effect} also suggested that including second-order difference-frequency wave forcing might not be important for operational turbine conditions but is important for parked turbines, based on a numerical study involving a spar floating foundation. In model-scale experiments by \citet{Goupee2014}, three floating foundations (tension leg platforms, spar, and semi-submersible) were tested, each supporting a scaled 5 MW turbine. The experiments revealed notable resonant subharmonic surge and pitch responses in the spar and semi-submersible floaters when subjected to the test of different sea states and wind conditions. \par
 \citet{orszaghova2021wave} investigated the impact of higher-order subharmonic responses on a Tetraspar soft-moored floating wind turbine under irregular wave conditions. They employed a phase-manipulated decomposition technique for harmonic separation, categorizing odd and even harmonics. Odd harmonics are odd multiples of the fundamental frequency (e.g., first, third, fifth), while even harmonics are even multiples (e.g., second, second difference, fourth, sixth). Their study revealed that second-order difference-frequency forces dominated the even subharmonic pitch response, while drag-induced forces drove the odd slow-drift resonant responses. This finding underscores the importance of considering unexpected odd resonant pitch motions in extreme sea states, as they may lead to floating wind turbine failure. All these studies focused on wave incidence aligned with the surge axis of the FOWT, whereas real-world ocean waves are multi-directional. While some studies have attempted to incorporate wave directionality into numerical models (\citet{alkarem2021complemental}, \citet{li2022short}), \citet{Aref2023} experimentally investigated the effect of multi-directional waves on a semi-submersible platform, highlighting the importance of wave spreading in floating foundation motion. However, the nonlinear effects of wave spreading on floating wind turbines remain largely unexplored. In the present work, we employ a harmonic separation methodology similar to \citet{orszaghova2021wave}, focusing on the effects of wave spreading on the floating foundation's nonlinear response under focused wave groups.

Numerical studies have highlighted challenges in accurately predicting floating wind turbines' subharmonic resonant loads and motions. State-of-the-art engineering tools for offshore wind systems have been verified but are still reported to underpredict nonlinear wave loads \citep{robertson2017oc5}. Initiatives such as the Offshore Code Comparison Collaboration (OC5, OC6 and OC7) and research by \citet{azcona2019low} and \citet{li2021experimental} have shown that models often underestimate resonant surge and pitch responses compared to wave-basin experiments. However, experimental and numerical data, as presented by \citet{wang2021oc6}, indicate that the previously understated nonlinear difference frequency wave loads of other engineering models can be accurately predicted using high-fidelity CFD tools. \citet{li2021experimentalb}  conducted a detailed study to examine the impact of nonlinear wave loads on the OC5-DeepCWind platform. Using OpenFOAM as the CFD tool and SIMA based on potential theory and Morison’s equation, their study showed that CFD provided more accurate predictions of higher harmonic wave load responses. However, modelling and reproducing irregular waves still poses challenges for CFD tools, especially under extreme conditions due to numerical damping of high-frequency wave components \citep{aliyar2024robust}. \citet{aliyar2022numerical} proposed a solution by coupling the OpenFOAM CFD tool with the potential flow solver HOS (Higher-Order Spectral method), and this approach was validated through experiment results for the OC3 spar floating foundation for different cases of regular and irregular waves. \citet{zhou2021assessing} developed an aero-hydro-mooring CFD model to investigate the impact of wave type and wave steepness on the hydro/aerodynamic response of floating wind turbines by a reconstructed focused wave group. They highlight that the dynamic motion of the floating platform, particularly its pitch motion, significantly affects the turbine aerodynamics by altering the swept area. They also indicate that the variation in incident waves has less impact on turbine performance than the floater's motion. \citet{luo2020experimental} investigates the impact of freak waves on a TLP model, revealing that freak waves, including both breaking and non-breaking types, cause severe platform motions, large tether forces, and significant structural damage, with implications for platform stability under extreme wave conditions. Although there is considerable research on focused wave group interactions with floating wind turbines (\citet{aliyar2024robust}, \citet{ransley2020blind}, \citet{bredmose2017triple}), studies focusing on higher-order harmonic loads and low-frequency resonance responses of semi-submersible floating wind turbine platforms remain limited, particularly using high-fidelity CFD modelling. A recent study by \citet{zeng2023investigation} examined the application of the focused wave group in estimating higher harmonic wave loads and low-frequency resonance on the DeepCwind platform using STAR-CCM+ and a phase decomposition technique \citep{fitzgerald2014phase}. Their numerical findings indicate a strong correlation between the motion of odd harmonics and the incident wave, attributed to a linear forcing and drag-induced response from the wave. As expected, even harmonics are primarily influenced by second-order difference-frequency resonance effects, oscillating at their natural periods. Our research follows a similar methodology, using CFD modelling to validate the harmonic separation analysis but based on experimental data while also investigating the effect of spreading. Furthermore, CFD was crucial in providing deeper insight into understanding wave-floating foundation interactions through detailed rendering and visualization.


The present paper aims to deliver an analysis of the dynamic motion of a novel semi-submersible floating wind configuration under nonlinear focused wave groups using experiment \citep{Aref2023} and a high-fidelity CFD tool, FloatStepper \citep{aliyar2024robust}. We employ harmonic separation, achieved through phase-manipulated realization in both experimental and numerical wave tanks, showing both even-harmonic and odd-harmonic components of the responses. We also investigated the effect of spreading in focused wave events and how it impacts the harmonic content of both the responses and the mooring line tensions. The structure of this paper is as follows. Section \ref{Section2} presents the characteristics of the floating foundation and details the experiment. Section \ref{Section3} describes the computational methods and formulation used in the analysis. Section \ref{Section4} provides an overview of the CFD numerical wave tank setup, including verification and validation procedures for the focused wave group, along with the model decay test. Sections \ref{Section5} and \ref{Section7} examine the harmonic analysis of focused wave groups and the floating foundation response characteristics. Sections \ref{Section8} through \ref{Section10} explore the effects of wave severity, mooring line configurations, and wave steepness on higher harmonics and wave spreading. Finally, Section \ref{Section11} presents a summary and conclusion. 

\section{Experimental setup}\label{Section2}

\begin{figure}[tb]
    \centering
    \begin{subfigure}[b]{0.64\textwidth}
        \centering
        \includegraphics[width=0.99\textwidth]{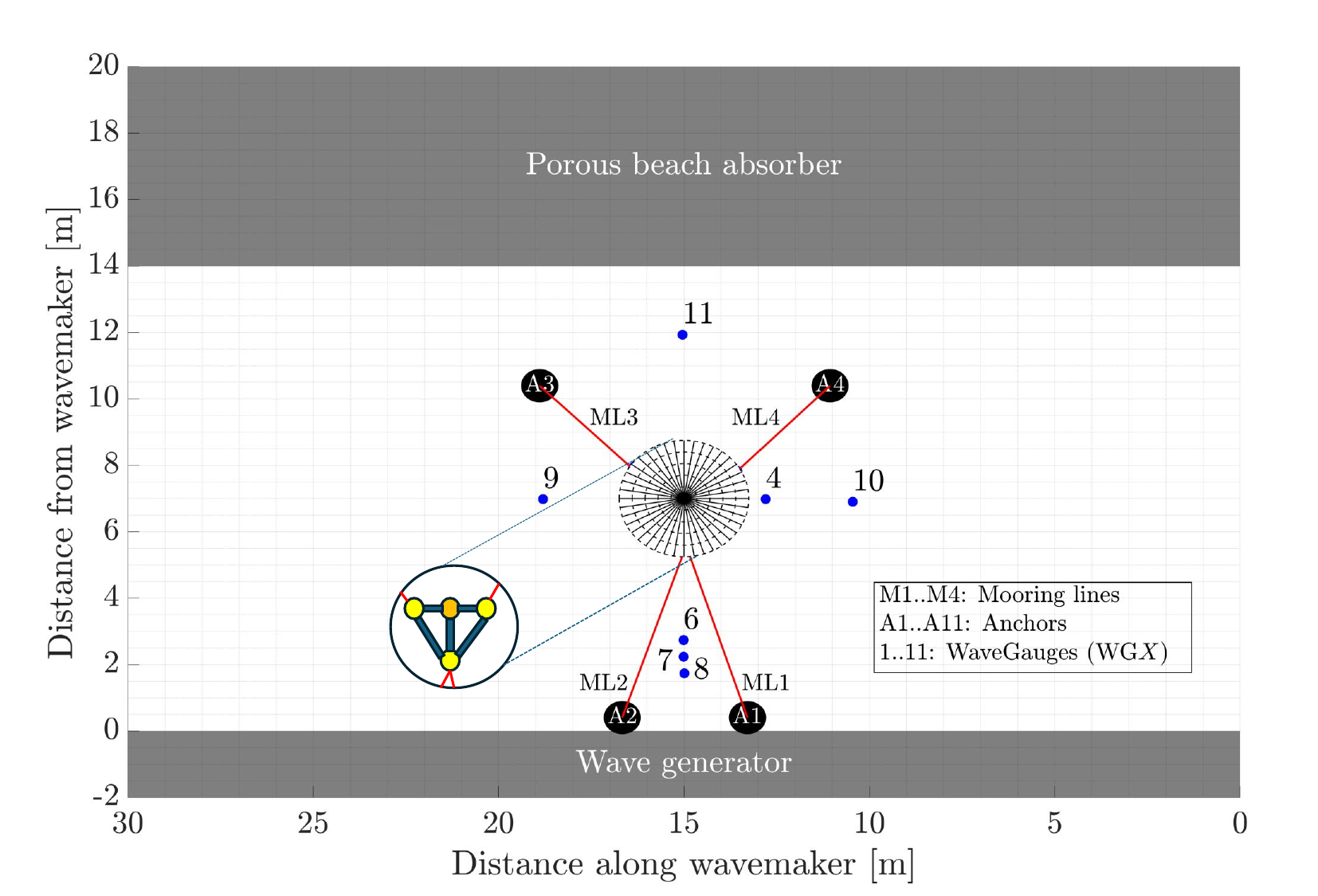}
    \end{subfigure}
    \hfill
    \begin{subfigure}[b]{0.34\textwidth}
        \centering
        \includegraphics[width=\textwidth]{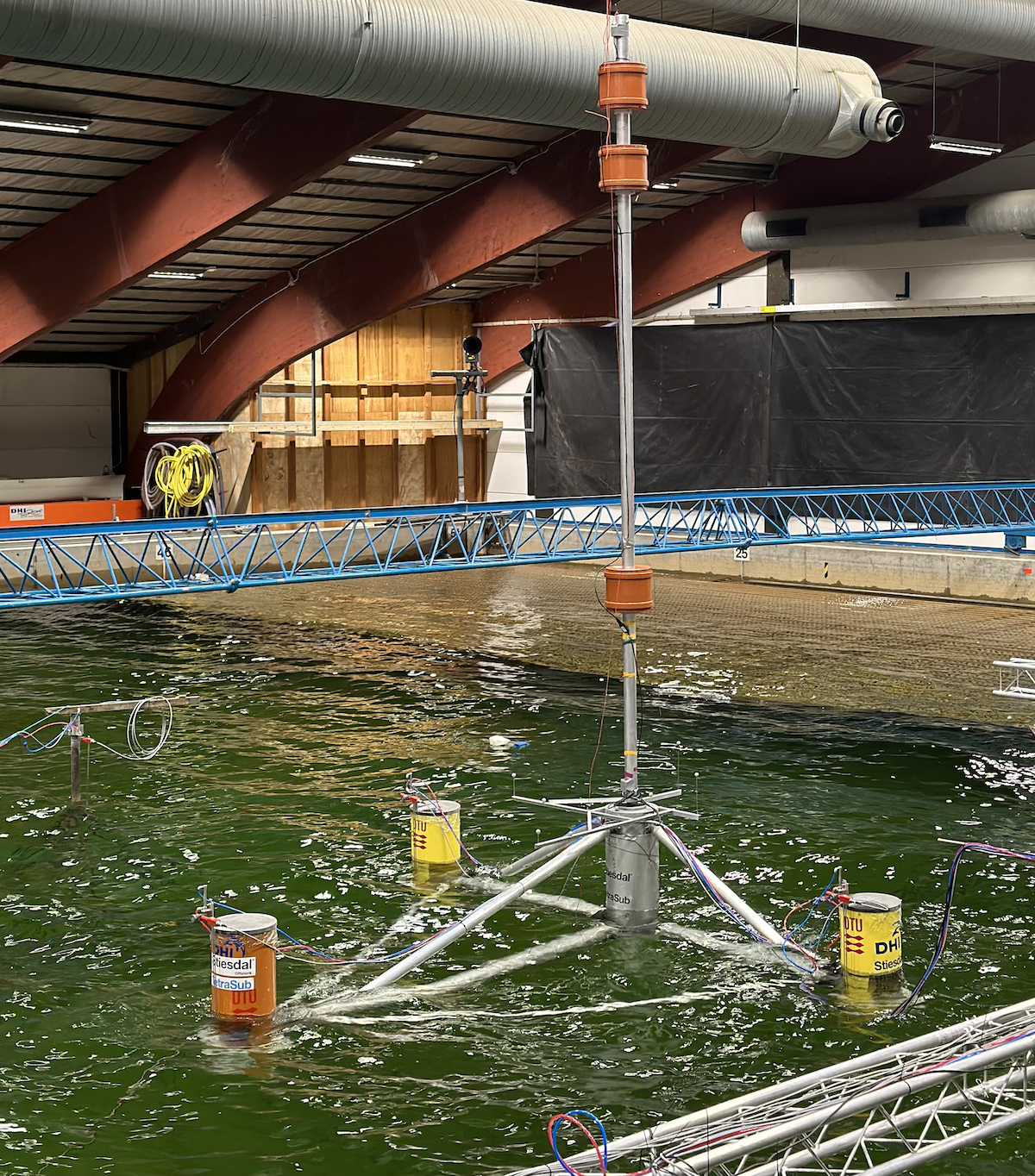}
    \end{subfigure}
    \caption{(a) Layout of the wave basin, with wave gauges (blue) and the model with the mooring lines (b) Photograph of the experimental set-up with floating foundation in the basin}
    \label{fig:ExperimentalSetup}
\end{figure}

The experimental campaign was carried out as part of the FloatLab project in the DHI\footnote{\label{footnote1}https://testfacilities.eu/listings/dhi-deep-water-ocean-basin/} Denmark deep water basin facility \citep{Aref2023}. The main goal was to evaluate the hydrodynamic performance of the $1$:$40$ model-scale floating wind turbine configuration across different wave conditions and investigate the effect of directional wave spreading. The basin measures $30$ m wide and $20$ m long, with a water depth of $3$ m. It is equipped with an articulated flap-type wavemaker, hinged $1.5$ m above the floor, and a parabolic-shaped perforated steel beach located in the last $6$ m of the basin to minimize reflections from the basin's wall. 

The floating foundation tested was a design variant of the Stiesdal Offshore TetraSub concept, constructed from aluminum, PVC, and 3D printed parts (Figure \ref{fig:ExperimentalSetup}b). It featured three vertical tanks connected by five braces to the central column and tanks, upon which the tower was mounted. Each vertical tank had a heave plate attached. The total length of the floating foundation tested was approximately $3$ m, with a total system mass of around $150$ kg. The height of the top of the tower relative to the bottom of the middle tank was about $4$ m. Calibrated weight plates were added to each tank to bring the floating foundation to its neutral equilibrium position, ensuring stability before the experiment. An inclined taut mooring system was used, with front mooring lines ML1 and ML2 connecting to the front tank and ML3 and ML4 attaching to the rear tanks. The mooring lines are symmetrically placed relative to the wave propagation direction but are asymmetric when viewed along the direction normal to wave propagation, as the front and back mooring lines differ (Figure \ref{fig:ExperimentalSetup}a). Due to the bracing layout, the floating foundation behaved as a rigid body and no attempt was made to quantify elastic deformation or local sectional loads. The weight of the rotor nacelle assembly, along with the tower, was replaced with a set of lumped masses calibrated to match the first natural frequency of the coupled floater-tower mode. The floating foundation was placed such that the central axis of the tower was $7.73$ m from the wavemaker and equidistant from the side walls of the basin (Figure \ref{fig:ExperimentalSetup} a). 

Eleven resistance-type wave probes were placed at different locations around the floating foundation, as shown in Figure \ref{fig:ExperimentalSetup}a. The coordinates of these probes are detailed in Table \ref{tab:wg_coordinates}. The main probe used is WG9, which was positioned laterally away from the model to minimize interference from scattered and radiated wave fields from the floating foundation. The tensions in the mooring lines were measured by $1$-DoF tension force gauges connected at the fairleads attached to tanks. The mooring lines used were made of $2$ mm diameter Dyneema rope, with a dry weight of $0.00625$ kg/m. The lengths of the back mooring lines were around $3.5$ m, while the front mooring lines were $4.5$ m. Four individually calibrated springs were attached at each fairlead to represent the scaled mooring stiffness. During the hydrostatic check, the floating foundation was strongly pretensioned by the mooring lines, ensuring stability and proper tension distribution. A non-intrusive motion tracking system by Qualisys recorded the movement of the floating foundation in $6$ DoF at the connection point between the tower and the floating foundation, with a sampling frequency of $200$ Hz. All motion data were filtered using a low-pass filter with a cut-off frequency of $20$ Hz for post-processing. For more details on the experimental setup, refer to \citet{Aref2023}, \citet{aliyar2024robust}.

\begin{figure}[tb]
    \centering
        \centering
        \includegraphics[width=0.9\textwidth]{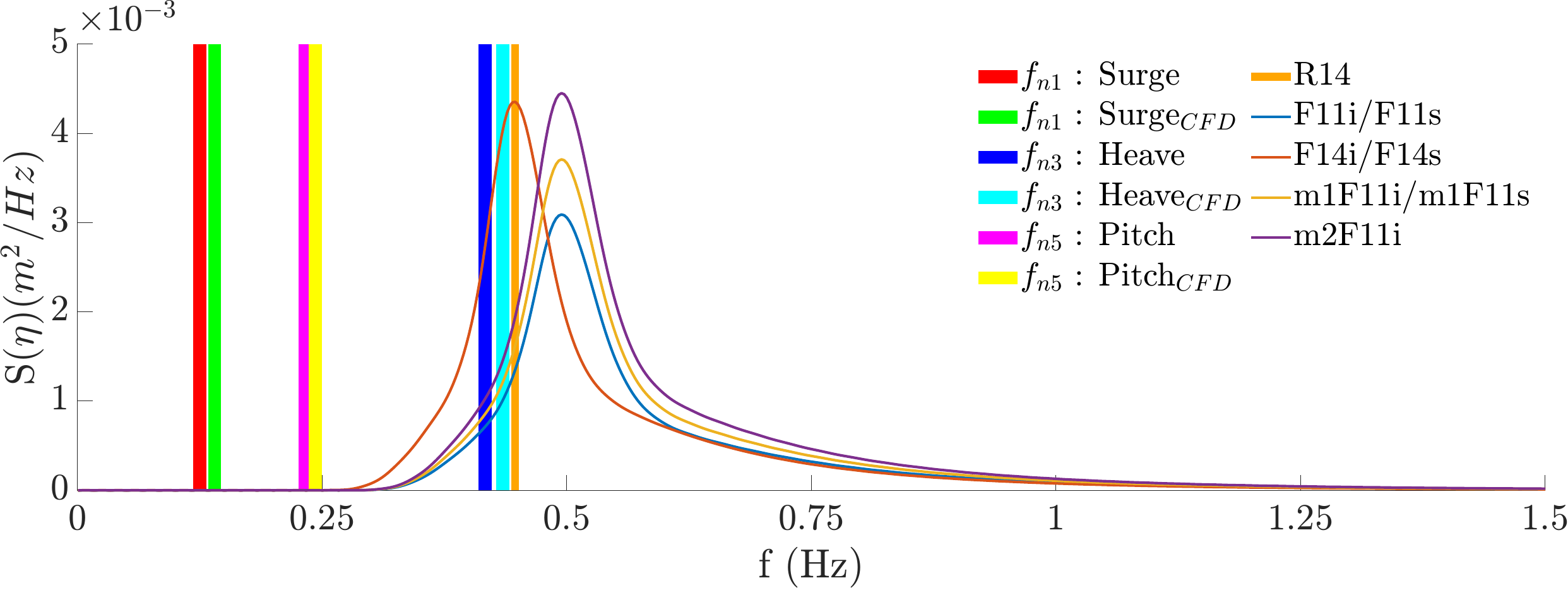}
        \caption{Target (solid)  free surface elevation spectra $S_{\eta}$), together with the experiment and CFD floating-system rigid-body natural frequencies}
        \label{fig:TargetSpectrum}

\end{figure}

The waves tested during the campaign included regular waves, focused wave groups, irregular long-duration sea states, and swell. This paper addresses only focused wave groups in sea states {F11} and {F14} (Table \ref{tab:TestMatrix}). Both focused wave groups follow the JONSWAP spectrum (Figure \ref{fig:TargetSpectrum}) and were tested under both non-spread and $20^\circ$ spread conditions. The terms {m1F11} and {m2F11}(Table \ref{tab:TestMatrix}) refer to a modified version of the wave cases of F11, where the normal crest height was increased by $20\%$ and $44\%$. Although this modified scenario was not tested experimentally, it was included in the CFD simulations to evaluate the effects of the steepness of the waves. Each focusing event lasted for $1$ minute and was repeated $7$ times, with a $5$-minute interval between the repetitions. Minimal variability was observed across the repetitions, and therefore only the first group was used for further analysis. To facilitate the separation of individual harmonics, each test was conducted through two realizations: the first with a fixed phase corresponding to focused wave groups and the second with an inverted paddle signal.  Due to the proprietary nature of the floating foundation, all translational motion signals have been normalized with a fixed number, $\xi^*_1$, and all rotations by another fixed number, $\xi^*_5$ in this paper. Comparative analysis between focused wave groups is possible since $\xi^*_1$ and $\xi^*_5$ are kept constant for all focused wave groups. Furthermore, the mooring line tensions were also normalized by $T^*$. 

\begin{table}[ht]
\centering
\begin{tabular}{@{}l*{11}{S[table-format=2.1]}@{}}
\toprule
\textbf{WG} & {1} & {2} & {3} & {4} & {5} & {6} & {7} & {8} & {9} & {10} & {11} \\
\midrule
\textbf{Distance from wavemaker} (\si{\meter}) & 5.58 & 7.83 & 7.83 & 6.98 & $NA$ & 2.74 & 2.24 & 1.74 & 6.98 & 6.9 & 11.93 \\
\textbf{Distance along wavemaker} (\si{\meter}) & 14.94 & 13.56 & 16.32 & 12.8 & $NA$ & 15.01 & 15.01 & 14.99 & 18.8 & 10.45 & 15.04 \\
\bottomrule
\end{tabular}
\caption{Coordinates of Wave Gauges (WG) in the wave basin}
\label{tab:wg_coordinates}
\end{table}



\begin{table}[ht]
\centering
\begin{tabular}{@{}clccccccc@{}} 
\toprule
\textbf{Wavetype} & \textbf{Label} & H$_s$ ($m$) & $\eta_{\text{peak}}$ (m) & T$_p$ ($s$) & $k_p \eta_{\text{peak}}$ & $k_p h$ & Phase & $\sigma_{\theta}(^\circ)$ \\ 
\midrule
\multirow{4}{*}{\textbf{F11}} & \textbf{F11} & $0.165$ & $0.196$ & $2.024$ & $0.192$ & $2.946$ & $0$ & $0$ \\
                              & \textbf{F11i} & $0.165$ & $0.196$ & $2.024$ & $0.192$ & $2.946$ & $180$ & $0$ \\
                              & \textbf{F11s} & $0.165$ & $0.196$ & $2.024$ & $0.192$ & $2.946$ & $0$ & $20$ \\
                              & \textbf{F11si} & $0.165$ & $0.196$ & $2.024$ & $0.192$ & $2.946$ & $180$ & $20$ \\
\cmidrule(lr){1-9}
\multirow{4}{*}{\textbf{m1F11}} & \textbf{m1F11} & $0.198$ & $0.235$ & $2.024$ & $0.231$ & $2.946$ & $0$ & $0$ \\
                               & \textbf{m1F11i} & $0.198$ & $0.235$ & $2.024$ & $0.231$ & $2.946$ & $180$ & $0$ \\
                               & \textbf{m1F11s} & $0.198$ & $0.235$ & $2.024$ & $0.231$ & $2.946$ & $0$ & $20$ \\
                               & \textbf{m1F11si} & $0.198$ & $0.235$ & $2.024$ & $0.231$ & $2.946$ & $180$ & $20$ \\
\cmidrule(lr){1-9}
\multirow{2}{*}{\textbf{m2F11}} & \textbf{m2F11} & $0.238$ & $0.282$ & $2.024$ & $0.277$ & $2.946$ & $0$ & $0$ \\
                               & \textbf{m2F11i} & $0.238$ & $0.282$ & $2.024$ & $0.277$ & $2.946$ & $180$ & $0$ \\
\cmidrule(lr){1-9}
\textbf{R14}                  & \textbf{R14} & $0.218$ & $0.11$ & $2.245$ & $0.0889$ & $2.397$ & $0$ & $0$ \\
\cmidrule(lr){1-9}
\multirow{4}{*}{\textbf{F14}} & \textbf{F14} & $0.218$ & $0.245$ & $2.245$ & $0.196$ & $2.397$ & $0$ & $0$ \\
                              & \textbf{F14i} & $0.218$ & $0.245$ & $2.245$ & $0.196$ & $2.397$ & $180$ & $0$ \\
                              & \textbf{F14s} & $0.218$ & $0.245$ & $2.245$ & $0.196$ & $2.397$ & $0$ & $20$ \\
                              & \textbf{F14si} & $0.218$ & $0.245$ & $2.245$ & $0.196$ & $2.397$ & $180$ & $20$ \\
\bottomrule
\end{tabular}
\caption{Environmental conditions used are described in terms of significant wave height (H$_s$), peak amplitude ($\eta_{\text{peak}}$), peak period ($T_p$), $k_p \eta_{\text{peak}}$, and $k_ph$. The notation used included '$F$' for focused wave group,'$\eta_{\text{peak}}$' for peak crest amplitude, '$s$' for spread waves,'$i$' for nominal phase inverted, and '$m$' for the modified version.}
\label{tab:TestMatrix}
\end{table}

\section{Computational Methods}\label{Section3}
This section briefly summarizes the computational methods used. First, the governing equations of the two-phase Navier-Stokes solver are outlined, followed by an explanation of the techniques used for wave generation and absorption. Next, the FloatStepper algorithm is introduced, covering the rigid body dynamic solver and its coupling with the mooring system.

\subsection{Two-phase Navier Stokes solver}\label{sec:fluidEOMs}

The two-phase incompressible Navier-Stokes equations are solved using OpenFOAM-based interIsoFoam solver, which employs a geometric Volume of Fluid (VoF) method \citep{roenby2016computational} to track the air-water interface. The governing equations consist of the continuity and the momentum conservation equation, expressed as

\begin{subequations}\label{eq:fluidEOMS}
\begin{gather}
	\nabla \cdot \mathbf{u} = 0, \label{eq:continuity} \\
	\frac{\partial (\rho \mathbf{u})}{\partial t} + \nabla \cdot (\rho \mathbf{u} \mathbf{u}) = -\nabla p + \nabla \cdot (\mu \nabla \mathbf{u}) + \rho \mathbf{g} + \mathbf f, \label{eq:NS}
\end{gather}
\end{subequations}
where $\mathbf{u}$ is the velocity field, $p$ is the pressure, $\rho$ is the density, $\mu$ is the dynamic viscosity, $\mathbf{g}$ is the gravitational acceleration and $\mathbf f$ represents other forces, which in this work are modelled as an additional resistance for the porous medium (refer to Eqn. \ref{eqn:darcy1}). The transport equation governs the advection of the VoF field,

\begin{equation}\label{eq:alpha_transport}
    \frac{\partial \alpha}{\partial t} + \nabla \cdot (\alpha \mathbf{u}) = 0,
\end{equation}

Here, $\alpha$ represents the volume fraction of water, with $\alpha = 1$ indicating water and $\alpha = 0$ representing air. To solve Eqn. \eqref{eq:alpha_transport}, the \textit{isoAdvector} method is applied. Unlike conventional approaches, which rely on flux-limiting schemes, \textit{isoAdvector} introduces a geometric approach in the advection step to accurately reconstruct the shape of the interface.  This method preserves the sharpness of the interface and conserves the volume of each phase, even on unstructured meshes, as detailed in \citet{roenby2016computational}. All results presented in this paper were computed using the laminar model in OpenFOAM. In some cases, the $k\text{-}\omega$ SST turbulence model was attempted, but had minimal impact on the results and was thus omitted.

\subsection{Numerical wave generation and absorption}
Accurate numerical wave modelling requires the precise generation of waves at the inlet and effective absorption at the outlet. We are primarily looking at focused wave groups. The generation of focused wave groups is based on the superposition principle, where multiple linear wave components with specific amplitudes, frequencies, and phases combine. The free surface elevation, \(\eta(\mathbf x,t)\), at any point in space and time is expressed as

\begin{equation}
    \mathbf{\eta}(\mathbf x, t) = \sum_{j=-N}^{N} {A}_j e^{i(\omega_j t - \mathbf{k}_j \cdot \mathbf{x} +\phi_j)},
\end{equation}
where \(A_j\) is the amplitude, \(\mathbf{\omega}_j\) the angular frequency and \(\phi_j\) the phase of the \(j\)-th wave component. To focus the wave energy at a particular point \(\mathbf x_0\) and time \(t_0\), the phase condition is set as \(\phi_j = -\mathbf k_j \cdot \mathbf x_0 + \omega_j t_0\), ensuring that all components interfere constructively at the desired location and time. In the case of directional waves, wave number \(\mathbf{k}_j\) represents the directional components of the wave number as follows,

\begin{equation}
    \mathbf{k}_j = k_j \begin{pmatrix} \cos\theta \\ \sin\theta \end{pmatrix},
\end{equation}
The angle \(\theta\) characterizes the direction of wave propagation relative to a reference axis. A small value of \(\theta\) indicates that the wave predominantly propagates in the reference direction, while larger values suggest a more substantial component in the perpendicular direction.

This superposition technique is used to generate a directional focused wave group in the experiment, and the resulting physical wavemaker displacement is then used in CFD simulations by dynamically moving the inlet boundary to replicate the behavior of a physical flap wavemaker (similar to the method implemented in the OlaFlow solver \citep{higuera2017olaflow}. Wave absorption at the outlet is handled using a shallow-water equation-based active absorption technique\citep{higuera2017olaflow}, assuming a uniform velocity profile over depth. Secondly, to prevent residual reflections from directional waves, a numerical dissipative porous beach was implemented with dimensions similar to the physical wave absorber in the wave tank. More details on the porous absorber are presented in Section \ref{section:waveabsorption}.

\subsection{Rigid body solver and the FloatStepper Algorithm}\label{sec:FloatStepper}


The FloatStepper \citep{roenby2023robust} is a recently developed enhanced non-iterative 'added mass' prediction method for coupling the dynamics of a rigid body with incompressible fluid flows in CFD simulations. Its primary aim is to eliminate the added mass instability that can occur when a lightweight floating structure interacts with a denser fluid. When the added mass exceeds the actual mass, the solver that governs the motion equations may struggle to accurately compute the body acceleration, as the effects of added mass are not directly included in the mass matrix. FloatStepper works by preceding every computational time step by a series of virtual time steps with predefined body motions. The fluid response to these motions reveals the instantaneous added mass matrix and the force and torque components arising from non-added mass effects. This removes the added mass instability in the subsequent calculation and integration of the body acceleration to obtain the updated body velocity and position. An overview of the algorithm is presented in \textbf{Algorithm} \ref{alg:FloatStepper1D}, where the virtual time steps are done in Steps 3-5. In Step 3, all forces and torques except the added mass contributions are calculated. In particular, this includes forces from the mooring lines. In Step 8, the body is moved, the body is moved, and the mesh of the surrounding fluid domain is changed accordingly using OpenFOAM's built-in mesh morphing technique. In this step, mesh points that are initially between a user-specified inner and outer distance from the surface of the floater are moved in each time step to accommodate the body motion. This is done using spherical linear interpolation (SLERP) between the rigid body motion of the mesh points inside the inner distance (following the floater motion) and the zero displacement of the mesh points outside the outer distance. Step 8 also updates the fluid boundary conditions on the floating body for the subsequent flow calculation. Here, a customized velocity boundary condition called \textit{floater\-Velocity} is used for the velocity field to ensure consistency with the instantaneous rigid body velocity. The boundary condition comes in a slip and a no-slip variant; here, we use the no-slip variant. For pressure, the built-in boundary condition called \texttt{fixed\-Flux\-Extrapolated\-Pressure} ensures proper accounting for the instantaneous acceleration of the body in pressure treatment. A complete description of the FloatStepper implementation is beyond the scope of this paper, and the reader is referred to \citet{roenby2023robust}.


\begin{algorithm}[h!]
\caption{The FloatStepper algorithm.\label{alg:FloatStepper1D}}
\begin{algorithmic}[1]
\STATE Initialize the body and fluid state.
\STATE Increment time by one-time step. \label{stp:FStimeInc}
\STATE Take a probe time step with zero acceleration and measure the resulting force and torque. \label{stp:zeroAccSteo}
\STATE Rewind the body, mesh, and fluid states to the state just before Step \ref{stp:zeroAccSteo}.
\STATE Calculate the updated added mass matrix.
\STATE Calculate the body acceleration with the updated added mass.
\STATE Time-integrate the acceleration to get the updated body motion and position.
\STATE Move the body (and mesh) and update fluid boundary conditions on its surface accordingly.
\STATE Calculate the new fluid state as if the found body motion and displacement were prescribed.
\STATE If the end time is reached, stop; otherwise, go to Step \ref{stp:FStimeInc}.
\end{algorithmic}
\end{algorithm}




\subsection{FloatStepper and mooring}\label{sec:EqMooring}
FloatStepper is coupled with the dynamic mooring line tool MoorDyn \citep{hall2015moordyn}, which is an open-source library designed to be coupled with rigid body solvers. The fundamental routines of MoorDyn are built into a shared library \citep{hall2015moordyn} to allow dynamic loading of functions within FloatStepper. The coupling between FloatStepper and MoorDyn follows a methodology consistent with that used in \citet{aliyar2022numerical} and \citet{roenby2023robust}. The FloatStepper algorithm is integrated with the MoorDyn solver using a weak coupling approach, enabling the exchange of information between the two solvers at each time step. The FloatStepper module shares the floating foundation's position and velocity with MoorDyn. MoorDyn, in turn, computes the fairlead kinematics and updates the states of the mooring system, including position, velocity of the mooring line node, and segment tension. Subsequently, MoorDyn returns the mooring restraining forces and moments resulting from the sum of all the fairlead tensions to the body motion solver to update the body’s acceleration.

\section{Re-modelling in OpenFOAM}\label{Section4}
 Here, we present the numerical setup with wave reproduction and the absorption techniques implemented. The results of the numerical decay tests are then presented and compared with the experimental results. Furthermore, model validation with regular wave-floating foundation interactions is presented.

\subsection{Numerical setup}

The dimensions of the computational domain replicate those of the physical wave basin (Figure \ref{fig:ExperimentalSetup}). An unstructured nonconformal grid was produced and refined with OpenFOAM's snappyHexMesh (sHM) tool. Mesh refinement was concentrated around the floating foundation to balance computational cost and accuracy. The floating body dynamic motion was handled through mesh deformation without changing its topology. In line with the OC6 Phase Ia meshing guidelines \citep{wang2022oc6}, the grid was refined around both the floating foundation and the heave plate.

\begin{table}[ht]
\centering
\caption{Boundary conditions used in simulations}
\begin{tabular}{cccc} 
\toprule
\textbf{Boundary}  & \textbf{Velocity}                   & \textbf{Pressure}         & \textbf{Alpha}         \\ 
\midrule
Inlet      & movingWallVelocity                & fixedFluxExtrapolatedPressure & zeroGradient     \\
Outlet     & waveAbsorption3DVelocity          & fixedFluxExtrapolatedPressure & zeroGradient     \\
Side       & noslip                            & fixedFluxExtrapolatedPressure & zeroGradient     \\
Atmosphere & pressureInletOutletVelocity       & totalPressure               & inletOutlet       \\
Floater    & floaterVelocity (no-slip)                  & fixedFluxExtrapolatedPressure & zeroGradient     \\
Bottom     & noslip                            & fixedFluxExtrapolatedPressure & zeroGradient     \\
\bottomrule
\end{tabular}
\label{Boundarycondition}
\end{table}

\begin{table}[ht]
\centering
\caption{Numerical schemes used in simulations}
\begin{tabular}{ccc} 
\hline
\textbf{Term}        & \textbf{Scheme}                 & \textbf{Order}   \\ 
\hline
\textbf{Gradient}    & {Gauss linear}           & {Second}  \\

\textbf{Divergence}~ & {Gauss vanLeer}          & {Second}  \\

\textbf{Laplacian}   & {Gauss linear corrected} & {Second}  \\

\textbf{Time}        & {Euler}                  & {First}   \\
\hline
\end{tabular}
\label{OFSchemes}
\end{table}

To ensure the uniqueness of the solution to the governing equations, appropriate boundary conditions must be prescribed at the domain boundaries. Table \ref{Boundarycondition} lists the boundary conditions used. Most of them are standard OpenFOAM boundary conditions. The \textit{movingWallVelocity} specifies a patch with a predefined velocity profile, allowing it to move with a specified velocity, which is useful for simulating cases where the boundary itself is in motion. In our case, it is a wavemaker. The \textit{waveAbsorption3DVelocity} absorbs the waves by adjusting the velocity field. The \textit{pressureInletOutletVelocity} is a blend of the \textit{pressureInletVelocity} and \textit{inletOutlet} boundary conditions. In \textit{pressureInletVelocity}, the velocity is calculated from the difference between the total and static pressure, where the direction is normal to the faces of the patch. The \textit{floaterVelocity} boundary condition is an in-house development that incorporates both \textit{movingWallVelocity} and \textit{no-slip} features, tailored to simulate the translation and rotational motions of the body. This boundary condition corrects the flux due to the mesh motion, so the total flux through the moving wall is always zero. The \textit{fixedFluxExtrapolatedPressure} extrapolates the pressure field at the boundary based on the flux values, ensuring that the pressure is adjusted to maintain a consistent flux across the boundary, and the \textit{totalPressure} boundary condition sets the pressure at zero at the atmospheric boundary. A more detailed description of the boundary conditions can be found in \citet{OFmanual}. One outer (SIMPLE) pressure-momentum correction loop and three inner (PISO) pressure correction loops were adopted. The adopted residuals limit was $10^{-6}$ for all simulations. Table \ref{OFSchemes} summarizes the methods of the discretization schemes in space and time. All simulations employ variable time-stepping, constrained by the Courant–Friedrichs–Lewy (CFL) condition to maintain numerical stability. In adherence to the OC6 guidelines \citep{wang2022oc6}, the maximum CFL number is kept below 0.5 and 0.25 near the free surface. Only the Euler scheme has been implemented and tested in FloatStepper, as the inclusion of higher-order temporal schemes is planned for future work.

\subsection{Focused wave group in CFD} \label{section:waveconvergence}
 Accurate wave field reproduction is essential for any wave-structure interaction simulation. The modelling of an incident-focused wave group relies on wavemaker motion for wave generation and very fine meshes for wave propagation. Two cases are investigated here: F14 and F14s (see Table \ref{tab:TestMatrix}). By selecting the unidirectional F14 case (Table \ref{tab:TestMatrix}) and using its peak period to determine a wavelength, a mesh convergence study was carried out. This study ranged from $37$ cells per wavelength to $75$, $150$, $300$ and $450$ cells per wavelength, corresponding to $1$, $2$, $4$, $8$, and $12$ cells per significant wave height (H$_s$), respectively. To name the investigated cases, the format MXHY is adopted here, where M represents the mesh, X indicates the number of cells per wavelength in the wave propagation direction, H represents the wave height, and Y denotes the number of cells per wave height. The investigation focused on comparing the data from the probes at the focusing location. For the F14s spreaded case, the setup closely resembles the previous case, with the distinction that the mesh is now three-dimensional. The M150H4 and M300H8 mesh setups in the unidirectional case are extended perpendicularly to the wave propagation direction to investigate the aspect ratios of $1$, $2$, and $4$ in the spread case.

\begin{figure}[tb]
    \centering
    \includegraphics[width=\textwidth]{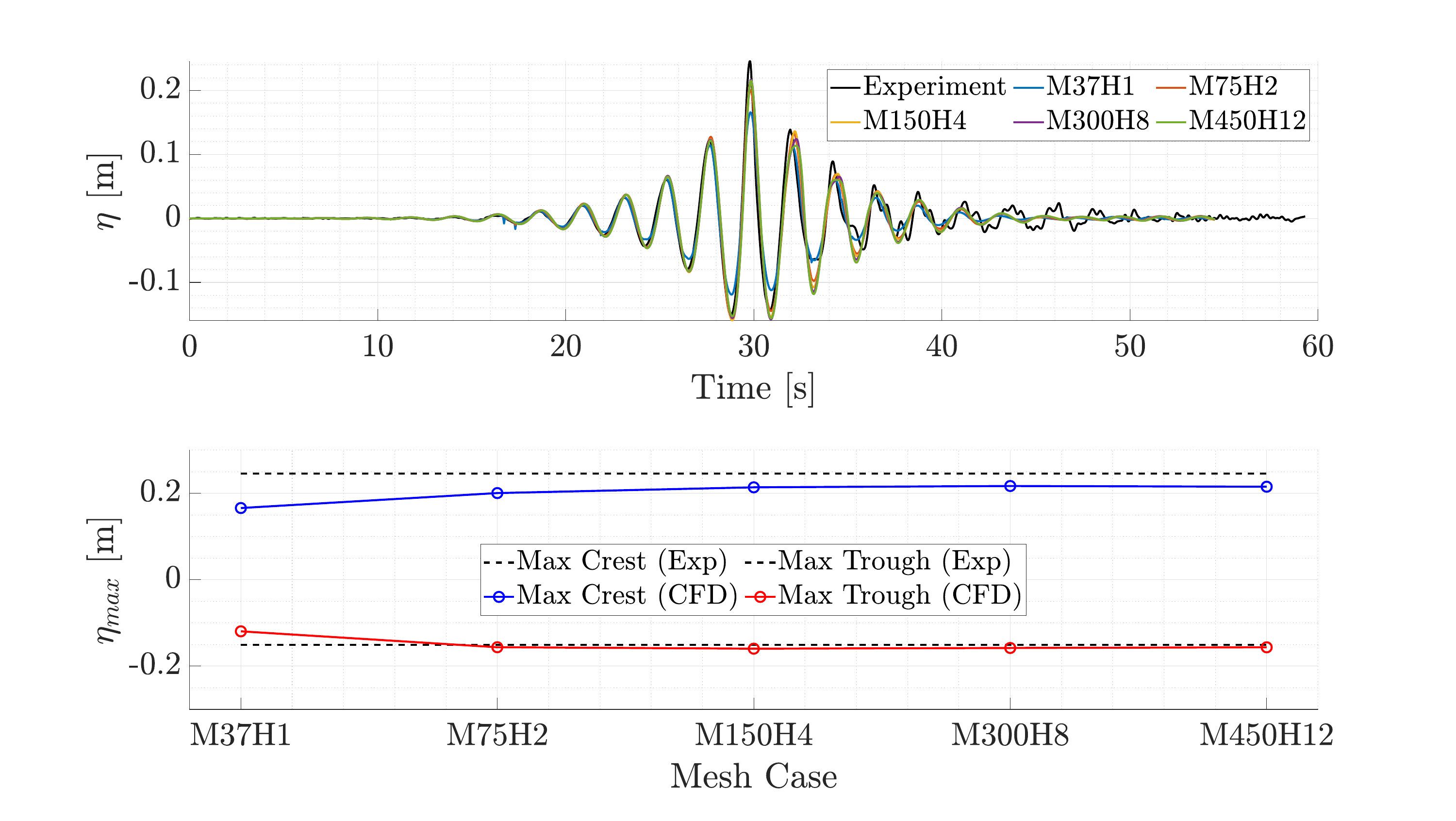}

    \caption{2D Mesh Convergence study to understand the required mesh size for accurate wavefield reproduction. The upper figure shows the time series comparison at the focusing point, while the lower figure depicts the convergence based on the maximum crest and trough comparison with the experiment. \label{fig:MeshConvergenceF14}}
\end{figure}

Unfortunately, a dedicated wave-only experiment was not conducted for the focused wave group. As a result, wave validation was performed using the probe located farthest from the floating foundation in the lateral direction, yet closest to the desired focusing position (designated WG9), which was selected for analysis. Figure \ref{fig:MeshConvergenceF14} compares the experiment and numerical waves of different mesh configurations. The upper figure shows the time series comparison at the focusing point, while the lower figure depicts the convergence based on the maximum crest and trough comparison. Convergence was observed from the M150H4 case, after which the remaining tested cases achieved crest values that closely approached the experimental maximum, with only minor difference from the experiment(lower figure in Figure \ref{fig:MeshConvergenceF14}). However, the maximum trough is well captured even with the M75H2 mesh. Hence, for the remainder of the paper, the M150H4 mesh will be used. We also tested the spread effects of the directional focused wave group by varying the aspect ratios of the cells parallel to the wavemaker. For brevity, this parametric study is not presented here. An aspect ratio of four with respect to the $x$ and $z$ directions proved to be efficient, showing minimal to no differences in the waves generated compared to the standard aspect ratio of one and two. Therefore, for further 3D investigations, the M150H4 configuration with an aspect ratio of four in the $y$  direction was used.

\subsection{Wave absorption} \label{section:waveabsorption}
Wave absorption is modelled both as a boundary condition and also through a porous absorber following the Darcy-Forchheimer equation for flow through a porous medium. This amounts to adding the following force to Eqn. \eqref{eq:NS} within the fluid region of the selected zone in the NWT.
\begin{equation}\label{eqn:darcy1}
\mathbf f = -\nabla p_{\text{Darcy}} = -\left( \frac{\mu}{d} \right) \mathbf{u} - \left( \frac{\rho f}{\sqrt{d}} \right) |\mathbf{u}| \mathbf{u}
\end{equation}
Here, $\mathbf f$ represents the other forces from Eqn. \eqref{eq:NS}, \(\nabla p_{\text{Darcy}}\) is the pressure gradient specific to the Darcy-Forchheimer model, \(\mu\) is the dynamic viscosity, \(d\) is the darcy permeability parameter of the porous medium, \(f\) is the Forchheimer coefficient. \({\mu}/{d}\) represents the viscous drag term (Darcy term) and \({\rho f}/{\sqrt{d}}\) represents the inertial drag term (Forchheimer term).



\begin{figure}[tb]
    \centering
    \begin{subfigure}[b]{0.34\textwidth}
    	\centering
	    \includegraphics[width=\textwidth]{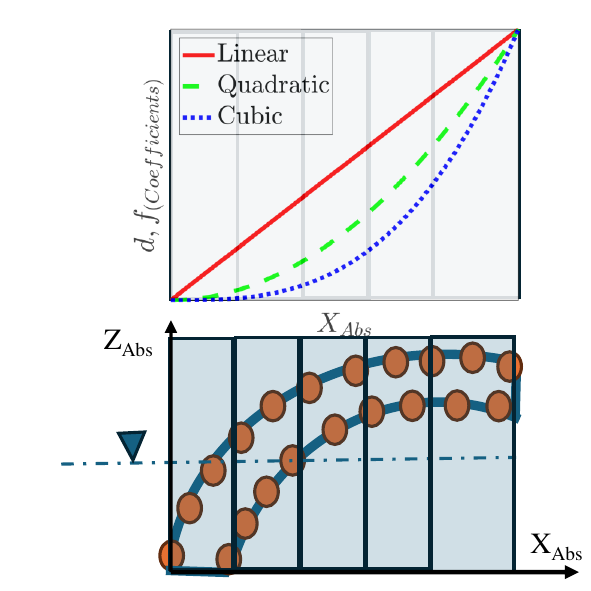}
	    \caption{}
	\end{subfigure}
    \begin{subfigure}[b]{0.64\textwidth}
    	\centering
	    \includegraphics[width=\textwidth]{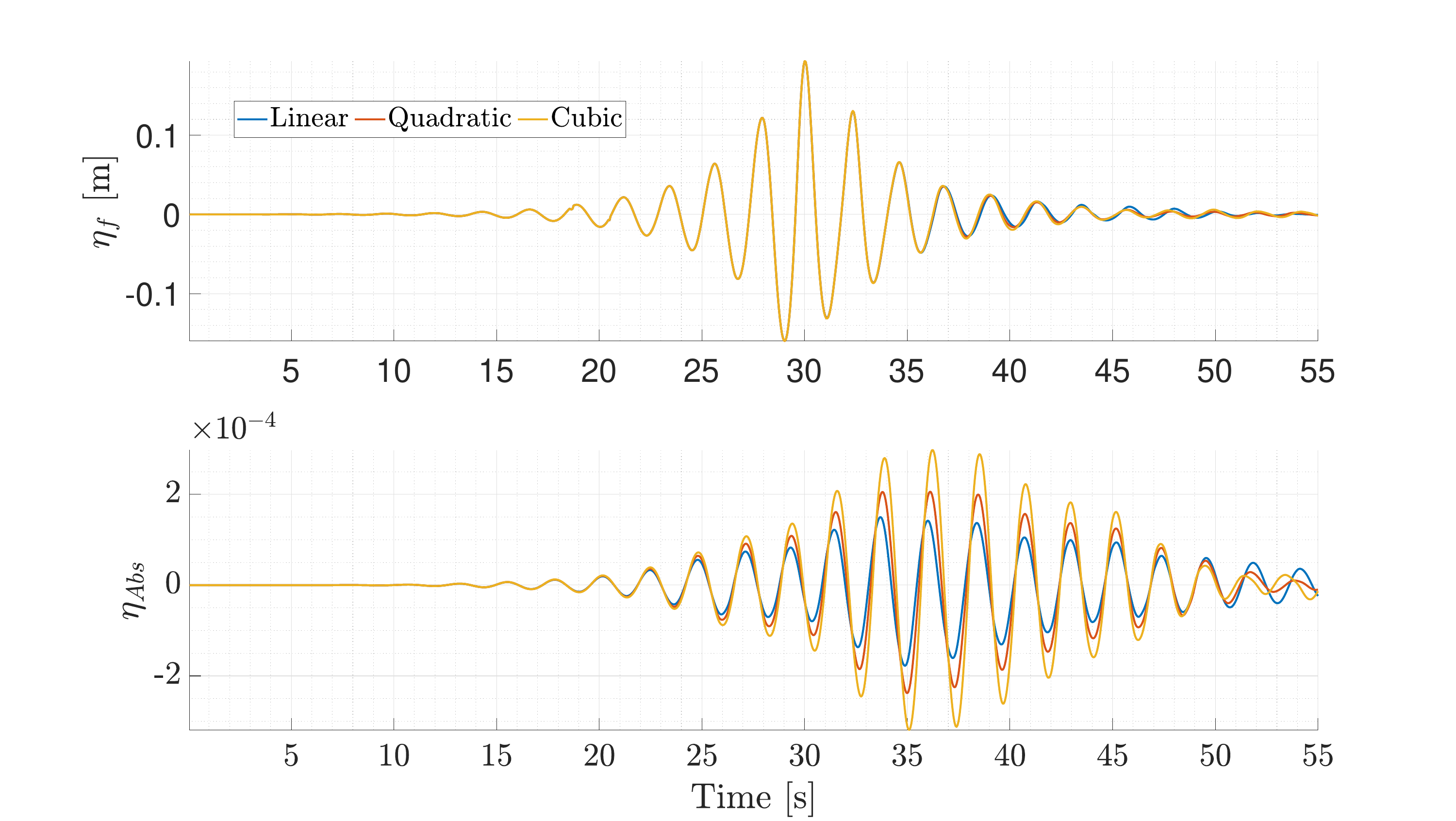}
	    \caption{}
	\end{subfigure}
    \caption{Left: Side cross-section of porous wave absorber model and coefficient variation. Right: Comparison of free surface elevation at focusing location and a probe positioned in the middle of the absorber}\label{fig:AbsorberFigs}
\end{figure}

The size and shape of the absorber are selected on the basis of the experimental wave basin absorber. A parametric study was conducted on the F11 wave by dividing the absorber into five zones and varying the parameters \( d \) (from 1000 to 10000) and \( f \) (from 10 to 250). These parameters were adjusted according to linear, quadratic, and cubic variations along the length of the absorber to identify the optimal coefficients for effective wave absorption.
The focused wave group, traversing the numerical wave tank (NWT), was monitored by probes placed every 0.25 meters across the domain, with specific probes positioned at the focusing location and the end of the domain. The left side of Figure \ref{fig:AbsorberFigs} shows a typical representation of the absorber, while the right side presents the focused wave group at the focusing location and the probe at the middle of the absorber. The study indicates that reducing \( d \) and increasing \( f \) effectively controls wave absorption, with linear variation showing to be the most efficient in minimizing reflections and maximizing absorption. Information on the application of porous media in OpenFOAM and the coefficients used is available in \citet{jensen2014investigations}.

\subsection{Decay Tests}\label{section:decay}



\begin{figure}[tb]
    \centering
    \begin{subfigure}[b]{0.48\textwidth}
    	\centering
	    \includegraphics[width=\textwidth]{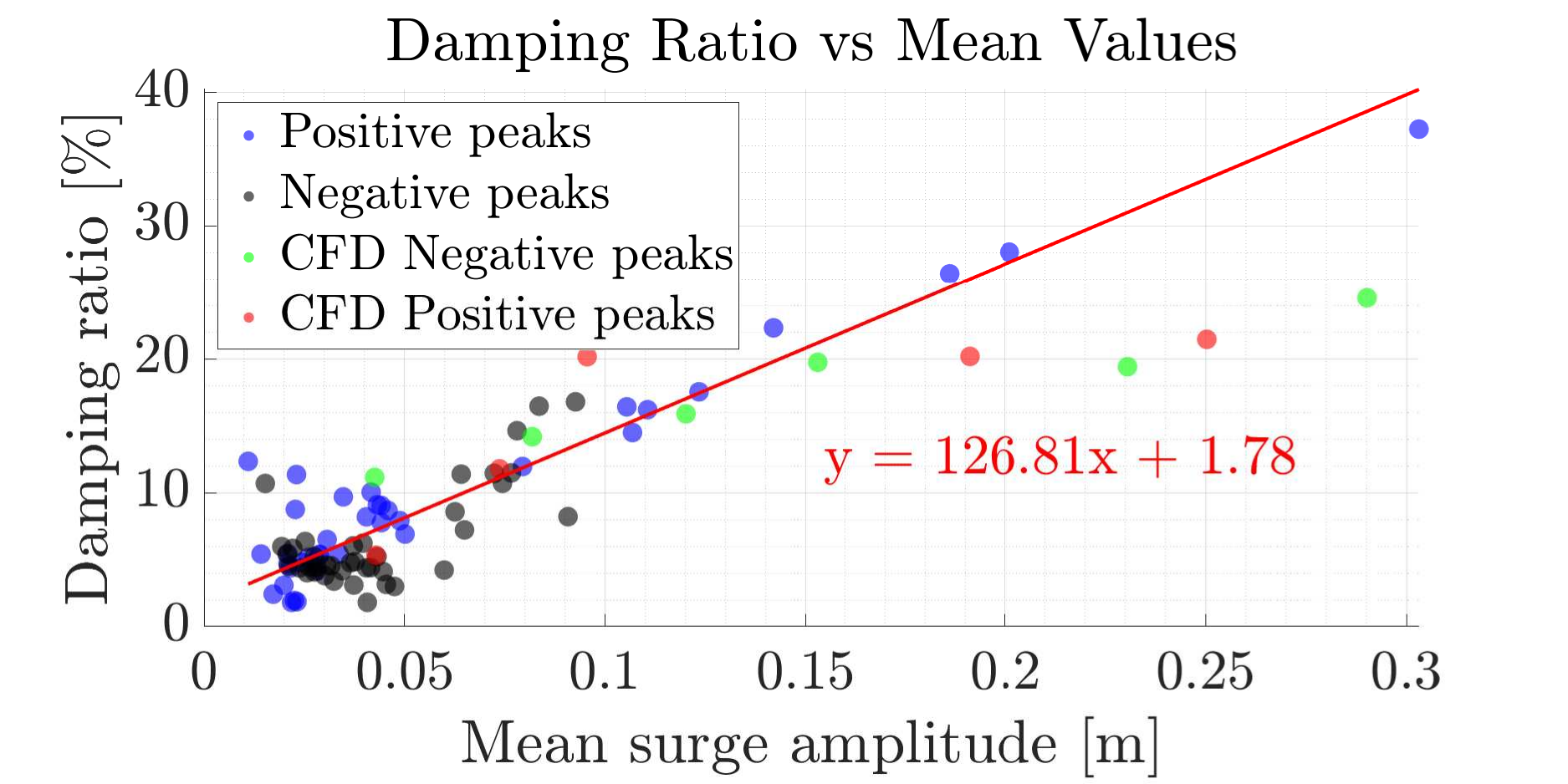}
	    \caption{\label{fig:SurgeDecay}}
	\end{subfigure}
    \begin{subfigure}[b]{0.48\textwidth}
    	\centering
	    \includegraphics[width=\textwidth]{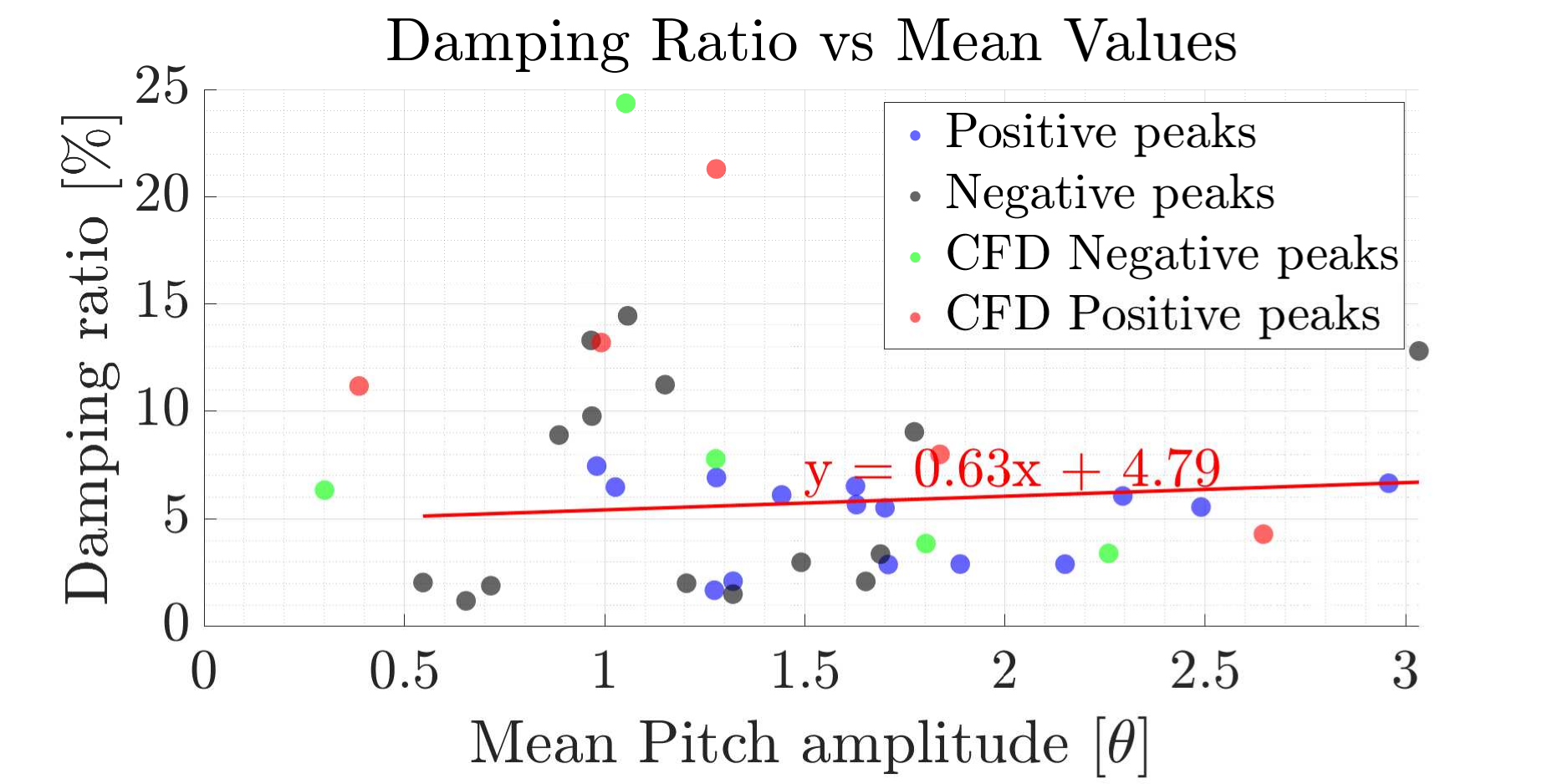}
	    \caption{\label{fig:PitchDecay}}
	\end{subfigure}
    \caption{Hydrodynamic damping observed in both the experiment and CFD for the design variant of Stiesdal Offshore’s TetraSub floating foundation, compared to the mean amplitudes\label{fig:DecayTestplots}}
\end{figure}

In the experiment, the natural frequencies of the rigid body were determined through moored decay tests conducted in still water. For each DoF, tests were repeated six times for different amplitudes. These tests induced initial displacements by pulling lines attached to different locations in the physical model. A bandpass filter was applied to each decay test to mitigate distortions from coupling with other DOFs. Given the focus of this study on surge and pitch, only these modes are discussed. The CFD natural frequencies (presented in Figure \ref{fig:TargetSpectrum}) were in good agreement with the natural frequencies measured for the heave and pitch DOFs. In particular, higher discrepancies were observed in the surge DOF, attributed to uncertainties about the mooring system, particularly in the anchor point's locations on the basin floor. Figure \ref{fig:DecayTestplots} presents the damping ratio for each half-decay cycle against the mean motion amplitude of the corresponding peaks. The experimental surge decay analysis follows a linear trend, and the damping ratios are stronger for positive peaks than for negative ones at smaller amplitudes. The CFD results followed a similar linear trend but showed consistently lower damping for positive peaks at smaller amplitudes. To further improve the results, a smaller y-plus value and more refined mesh regions around the heave plate could improve the accuracy of the damping ratio calculation. However, this refinement would significantly increase the cell count, making the CFD methodology adopted computationally prohibitive. We proceeded with the understanding that CFD typically predicts slightly lower damping than observed in experiments. The chosen filtering technique significantly influences the observed trend and scatter points in the pitch decay results, resulting from the strong coupling of pitch with heave and surge. This coupling often leads to different scatter patterns in the analysis. However, after applying a careful filtering process to isolate the pitch response, it was determined that pitch damping can best be represented by a constant damping ratio of approximately $5\%$ to $6\%$ for most amplitudes, both in experimental data and CFD simulations, although still with appreciable scatter. 

\begin{figure}[tb]
    \centering
    \begin{subfigure}[b]{0.35\textwidth}
    	\centering
	    \includegraphics[width=\textwidth]{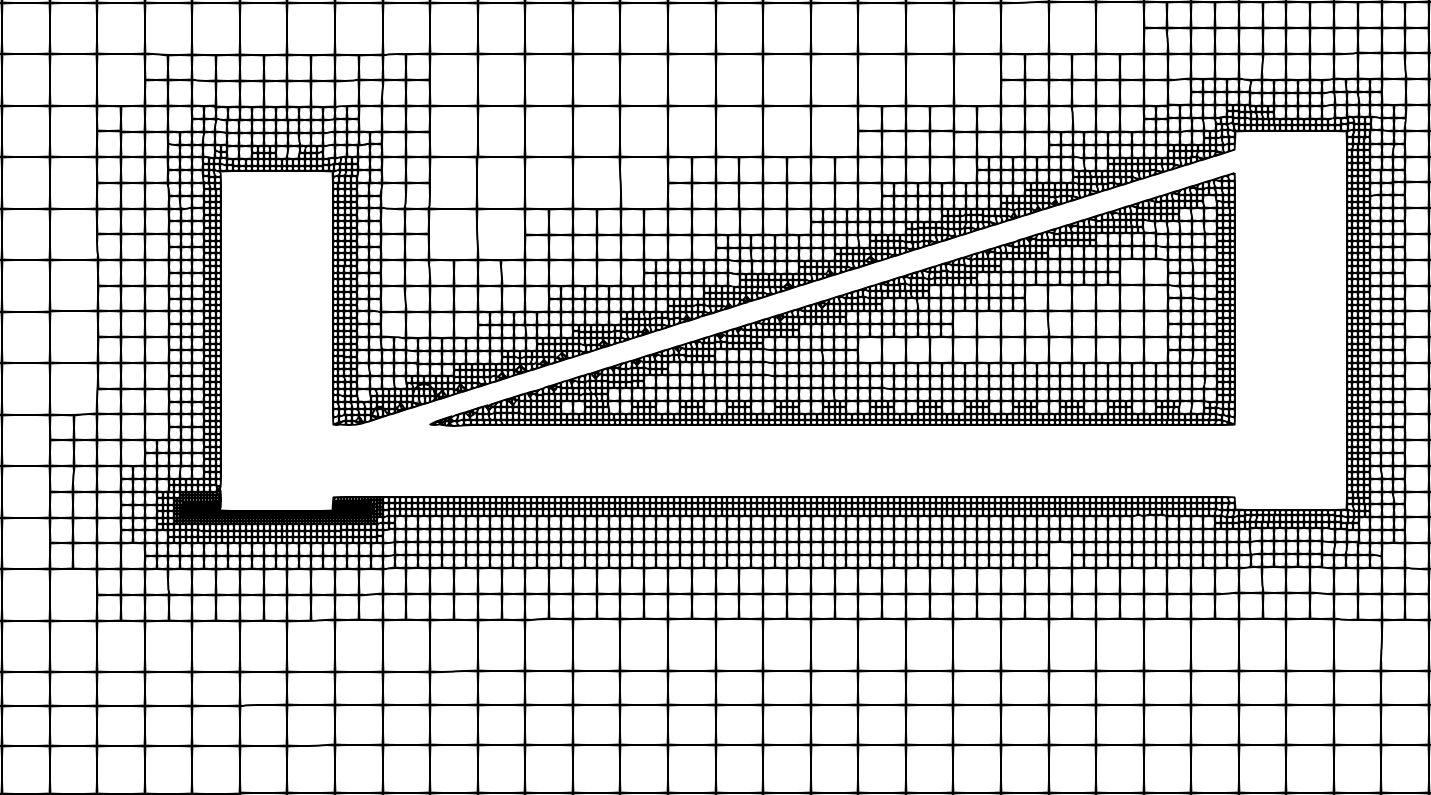}
	    \caption{}
	\end{subfigure}
    \begin{subfigure}[b]{0.49\textwidth}
    	\centering
	    \includegraphics[width=\textwidth]{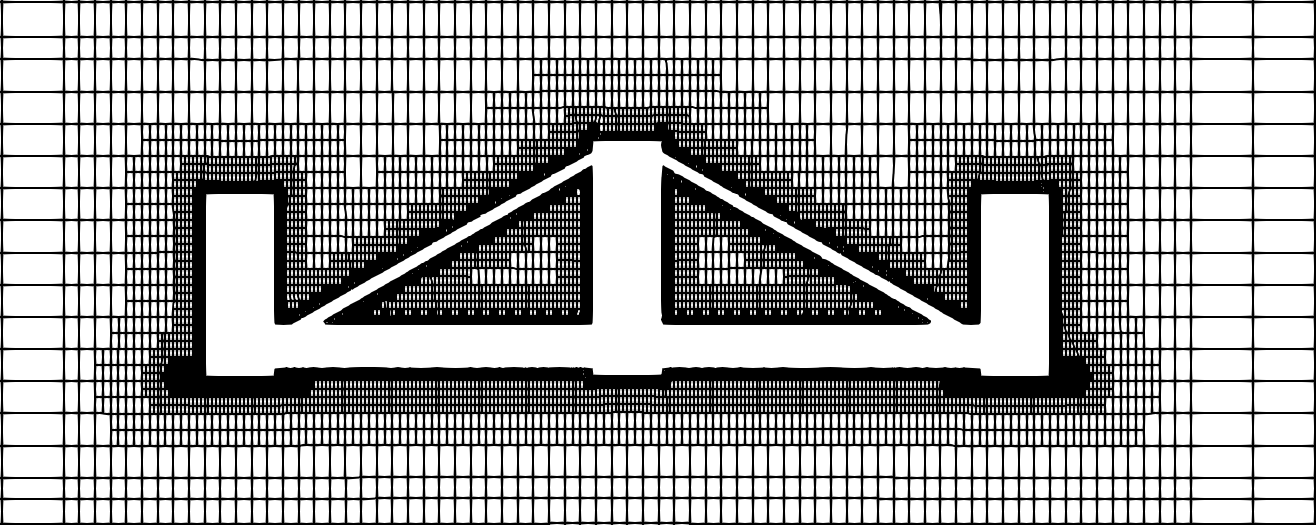}
	    \caption{}
	\end{subfigure}
    \caption{Front and side cross-sections of the mesh around the floating foundation and heave plate, detailing the spatial resolution and distribution of the mesh elements}\label{fig:CFDMesh}
\end{figure}

\begin{figure}
    \centering
        \includegraphics[width=\textwidth]{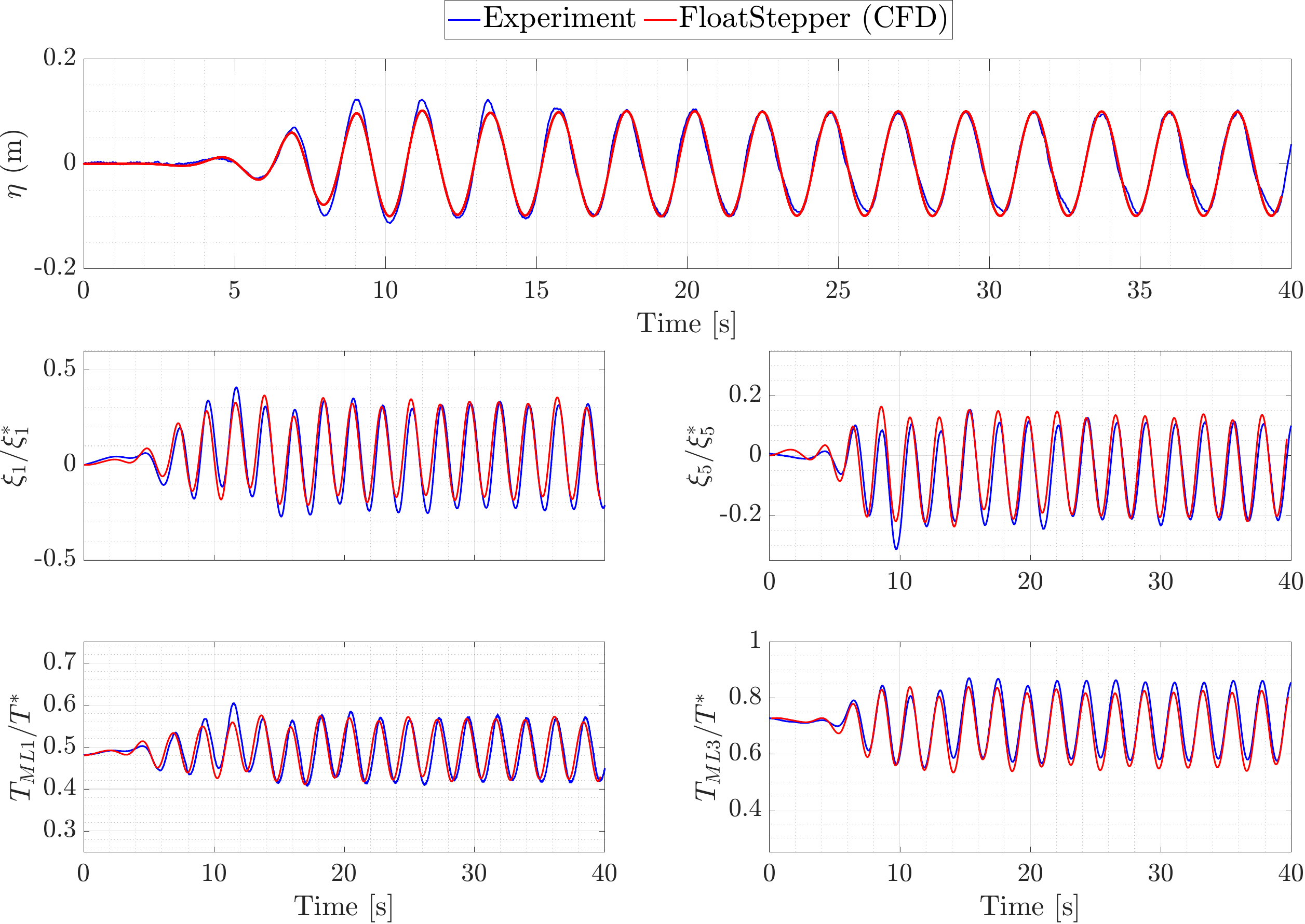}
    \caption{Regular wave (R14) floating foundation interaction comparison for experiment and CFD. \newline Non-dimensionalized by $\xi^{*}_{1}$, $\xi^{*}_{5}$ and $T^*$ for surge, pitch, front, and back mooring line tensions, except $\eta$ (free surface elevation) respectively.}
    \label{fig:R11}
\end{figure}

\subsection{Validation with regular wave interaction}

Although the decay tests showed close agreement in their frequencies, regular wave validation (R14 - refer to Table \ref{tab:TestMatrix} or Figure \ref{fig:TargetSpectrum}) was conducted to assess the performance of the setup within the wave frequency regime and to verify its ability to replicate the expected motions accurately. The selected mesh resolution for our investigation is made up of 9 million cells, as detailed in the refinement study presented in our conference paper, \citet{aliyar2024robust}. Furthermore, the mesh resolution adheres to the recommendations of our wave convergence study (Section \ref{section:waveconvergence}), which specified a resolution of 150 cells per wavelength and 4 cells per wave height, with an aspect ratio of 4 in the direction parallel to the wave maker. The x- and y-plane cross-sections of the chosen mesh are shown in Figure \ref{fig:CFDMesh}, while the results of the wave floating foundation interaction are presented in Figure \ref{fig:R11}. The comparison between the experiment and the CFD model shows good agreement in free surface elevation and motion responses. The surge and pitch responses accurately capture both transient and steady-state conditions. The pitch response also aligns well in both amplitude and phase. Furthermore, the mooring line tensions of the front and back mooring lines (Figure \ref{fig:ExperimentalSetup}) demonstrate minimal differences, which confirms the validity of the prepared numerical setup. Given the similar characteristics of the waves used (Table \ref{tab:TestMatrix}) in this investigation, the same mesh configuration was applied throughout. 



\section{Harmonic decomposition for incident wave field}\label{Section5}

When analyzing the hydrodynamic loads and responses of floating structures, understanding the role of harmonic components is important, since natural frequencies can resonate with wave-induced excitations in the nonlinear frequency regime. Under the assumption that the response can be modelled using Stokes perturbation theory, where higher harmonic contributions are captured through a Taylor series expansion, the phase separation method (\citet{fitzgerald2014phase}, \citet{walker2004shape}, \citet{jonathan1997irregular}) provides a robust approach to isolate and identify subharmonic and superharmonic loads and responses, particularly in extreme wave conditions. A brief overview of the approach is given here. 

Let \( X_0(t) \) and \( X_{180}(t) \) represent two-time histories for a given hydrodynamic quantity from two experimental measurements involving incident waves generated by paddle signals, where \( X_0(t) \) is the original paddle signal and \( X_{180}(t) \) is the inverted signal that is $180^\circ$ out of phase. These time signals can be expressed as \citep{orszaghova2021wave}:

\begin{equation}
\begin{split}
X_{0/180}(t) = & \Re \left\{ \pm \sum_{n} A_n T_n^{(1)} e^{i 2\pi f_n t} + ... \right. \\
& \quad + \sum_{n}\sum_{m} A_n A^{*}_{m} T_{n,m}^{(2\pm)} e^{i 2\pi (f_n + f_m)t} \pm \sum_{n}\sum_{m}\sum_{p} A_n A^{*}_{m} A^{*}_{p} T_{n,m,p}^{(3\pm)} e^{i 2\pi (f_n \pm f_m \pm f_p)t} + O(A^4)\Bigg\}
\end{split}
\end{equation}
where \( A_n = |A_n| e^{i \phi_n} \) represents the complex amplitudes of the paddle movement, \( ^* \) denotes the complex conjugation, and \( T_n^{(1)} \) is the complex linear transfer function for the frequency \( f_n \), with \( \phi_n \) representing the phase. Double and triple summations cover both superharmonics and subharmonics, with \( T_{n,m}^{(2)} \) and \( T_{n,m,p}^{(3)} \) indicating the second and third-harmonic transfer functions.

Using the two phase-shifted realizations, odd and even harmonics can be separated by subtracting and adding the two signals \citep{fitzgerald2014phase} : 

\begin{align}
\text{Odd:} \quad &\frac{1}{2}\left(X_0 - X_{180}\right) = \Re\left\{\sum_{n} A_n T_n^{(1)} e^{i 2\pi f_n t} + \sum_{n}\sum_{m}\sum_{p} A_n A^{*}_{m} A^{*}_{p} T_{n,m,p}^{(3\pm)} e^{i 2\pi (f_n \pm f_m \pm f_p)t} + O(A^5)\right\} \\
\text{Even:} \quad &\frac{1}{2}\left(X_0 + X_{180}\right) = \Re\left\{\sum_{n}\sum_{m} A_n A^{*}_{m} T_{n,m}^{(2\pm)} e^{i 2\pi (f_n \pm f_m)t} + O(A^4)\right\}
\end{align}

The odd component captures linear, third-harmonic, and higher odd-harmonic interactions, while the even component contains second-harmonic, fourth-harmonic, and higher even-harmonic interactions. Also, subharmonic difference frequency forcing occurs in even harmonics, while drag is part of odd harmonics. Further details on harmonic phase decomposition can be found in \citet{fitzgerald2014phase}, \citet{orszaghova2021wave} and \citet{hansen2024resonant}, where the latter discusses the recent implementation of a four-phase evaluation for a flexible floating structure. We will apply this harmonic separation approach to analyze the focused wave group here and, in the following sections, the responses of the floating foundation.

\begin{figure}[ht]
    \centering
        \includegraphics[width=\textwidth]{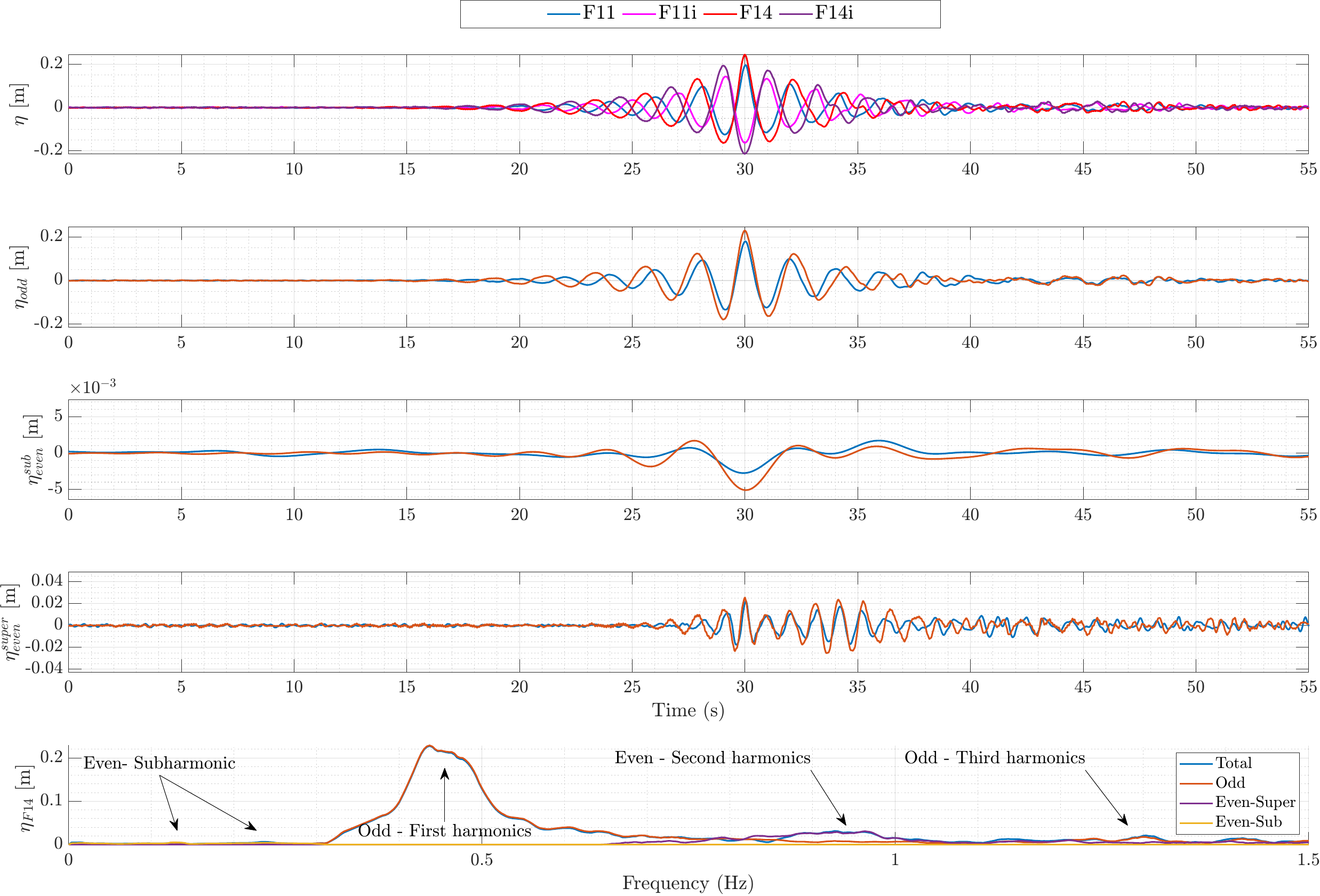}  
    \caption{Free surface elevation for two wave groups (F11 and F14) at the focusing location. The first row is the original waves, and the following rows are its harmonic decompositions: odd, even sub, and even superharmonics. The last row displays the spectral form of each time series}
    \label{fig:WaveonlyExptF11F14}
\end{figure}

 The two-phase harmonic decomposition requires crest- and trough-focused wave groups to effectively separate the signal into odd and even harmonics. The trough-focused signal is obtained by introducing a phase shift of  $180^\circ$ to the paddle signal of the crest-focused group, which is designed to have zero phase at the focused location. Consequently, the F11 and F14 waves (Table \ref{tab:TestMatrix}) are generated by introducing $0^\circ$ and $180^\circ$ phase shifts, and the results are shown in Figure \ref{fig:WaveonlyExptF11F14}. Both waves align well at the focusing time ($30$ s) as their phases are chosen to achieve peak alignment at that moment. Applying harmonic separation to these waves yields the odd (\(\eta_{\text{odd}}\)) and the even harmonics (\(\eta_{\text{even}}\)). In addition, the even signals have been spectrally separated into subharmonics (\(\eta^{\text{sub}}_{\text{even}}\)) and superharmonics (\(\eta^{\text{super}}_{\text{even}}\)), which are presented in the subsequent rows of the same figure.

  
It is observed that reproducing the odd harmonics seems to be essential to represent the wave group's overall evolution accurately. This is because linear harmonics, together with other odd harmonics, account for $92\%$ of the measured free surface elevation and influence the progression of bound harmonics in the present case. Even subharmonic waves correspond to bound wave components with longer wavelengths. These waves result in a set-down or depression in the water level below the primary wave group, creating a local trough regardless of the increase in the wave severity. Although the amplitude of the even subharmonic wave is relatively small (about $2\%$  of the maximum amplitude of odd harmonics), F14 exhibits twice the setdown compared to F11. The superharmonics, with a maximum amplitude of about $10\%$ of the maximum amplitude of odd harmonics, show similar results in both F11 and F14. Thus, at least in the present case, superharmonics are less sensitive to variations in wave severity than even subharmonics. In the final row of the figure, a spectral analysis of subharmonic and superharmonic waves for F14 is presented. The frequency is plotted along the horizontal axis, whereas the amplitude of free surface elevation is represented on the vertical axis. As noted previously, the first odd harmonics dominate the frequency spectrum, while the second and third harmonics are also clearly distinguishable. In contrast, the even subharmonic is barely noticeable because of its small energy. Additionally, another observation from the superharmonics time series is the presence of spurious higher-harmonic waves arising from the linear wave generation theory (\citet{barthel1983group}, \citet{hughes1993physical}, \citet{ghadirian2019investigation}) used in the wavemaker. Due to the lack of paddle motion for the bound superharmonics, a field of superharmonic free waves is released from the paddle to suppress the superharmonic wave motion locally.  Based on the deep water dispersion relation, the second harmonic free waves travel at half the group speed of the first harmonic waves. Hence, the peaks of the spurious free waves are expected to reach the focusing point at approximately $34$ s to $35$ s (refer to calculations in Appendix \ref{AnnexA}). We can observe that the arrival time is precisely as expected, and the magnitude of the waves are nearly identical to the bound super-harmonics of the focused wave group.

\begin{figure}[p]
    \centering
        \begin{tikzpicture}
        \node at (0,0) {\includegraphics[scale=0.2]{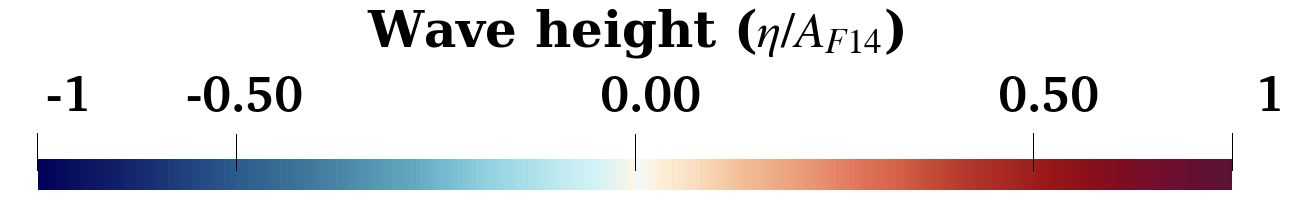}};
    \end{tikzpicture}

    \begin{subfigure}[b]{0.49\textwidth}
        \centering
        \includegraphics[width=\textwidth,height=0.2\textheight]{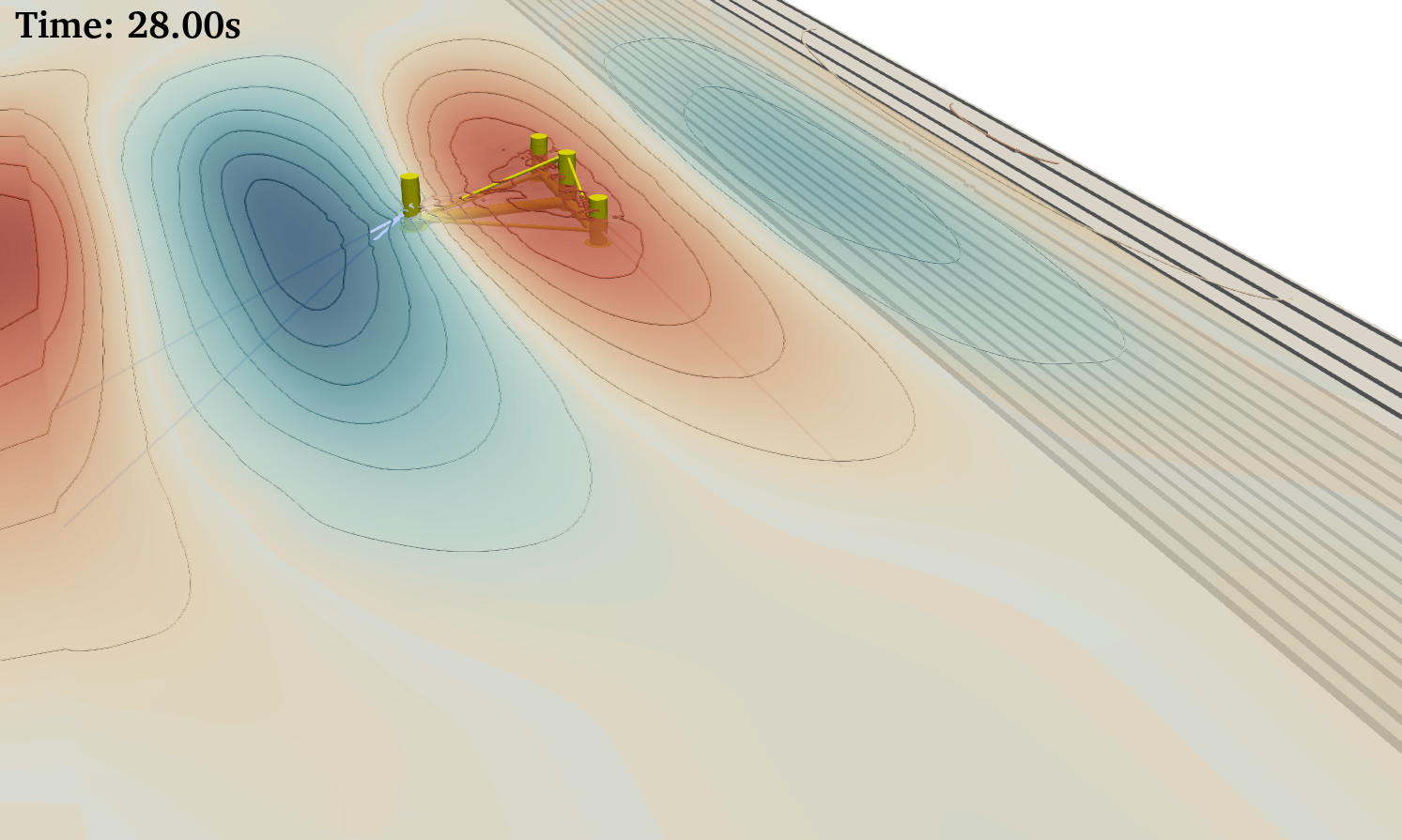}
    \end{subfigure}
    \hfill
    \begin{subfigure}[b]{0.49\textwidth}
        \centering
        \includegraphics[width=\textwidth,height=0.2\textheight]{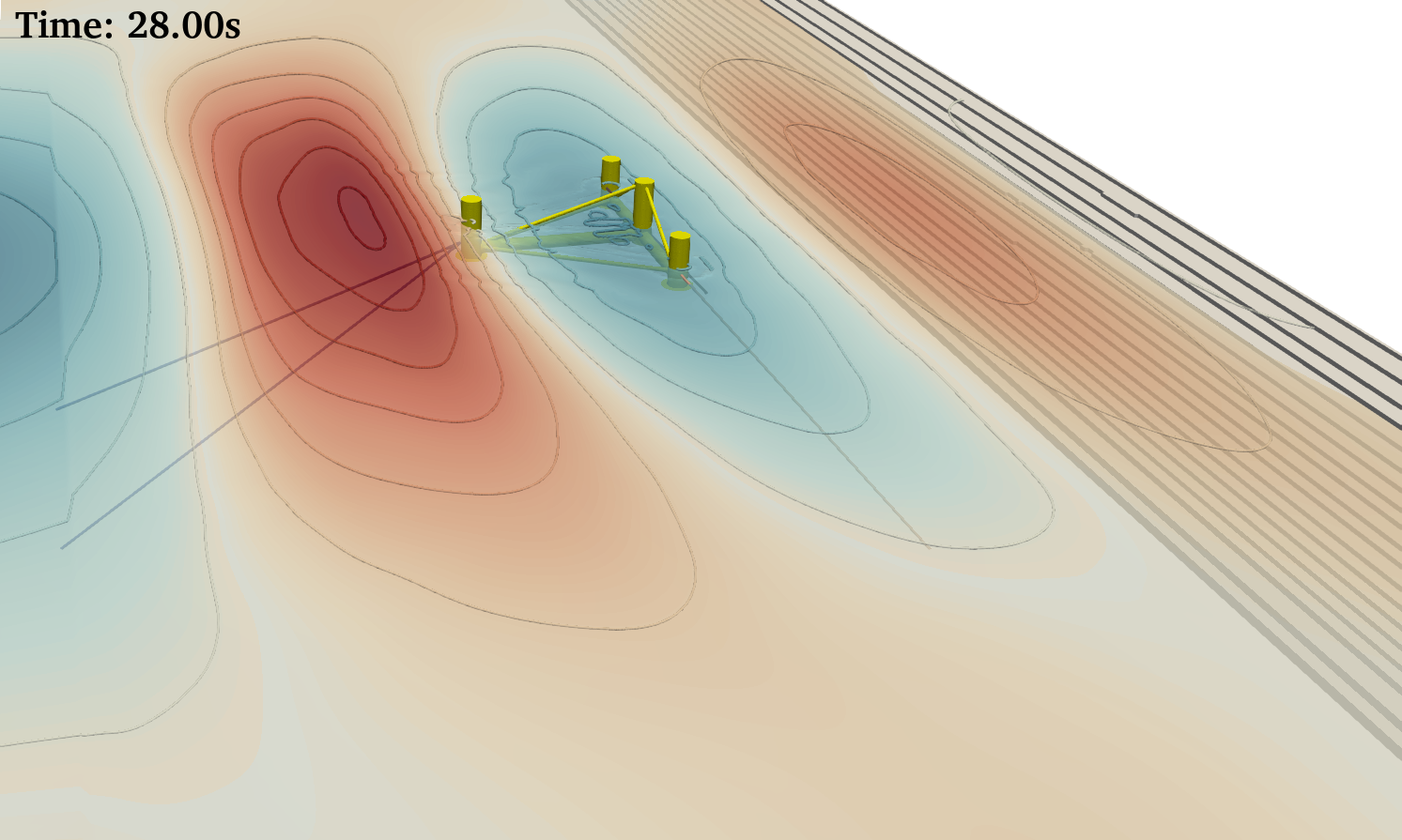}
    \end{subfigure}
    
    \vspace{0.1em}

        \begin{subfigure}[b]{0.49\textwidth}
        \centering
        \includegraphics[width=\textwidth,height=0.2\textheight]{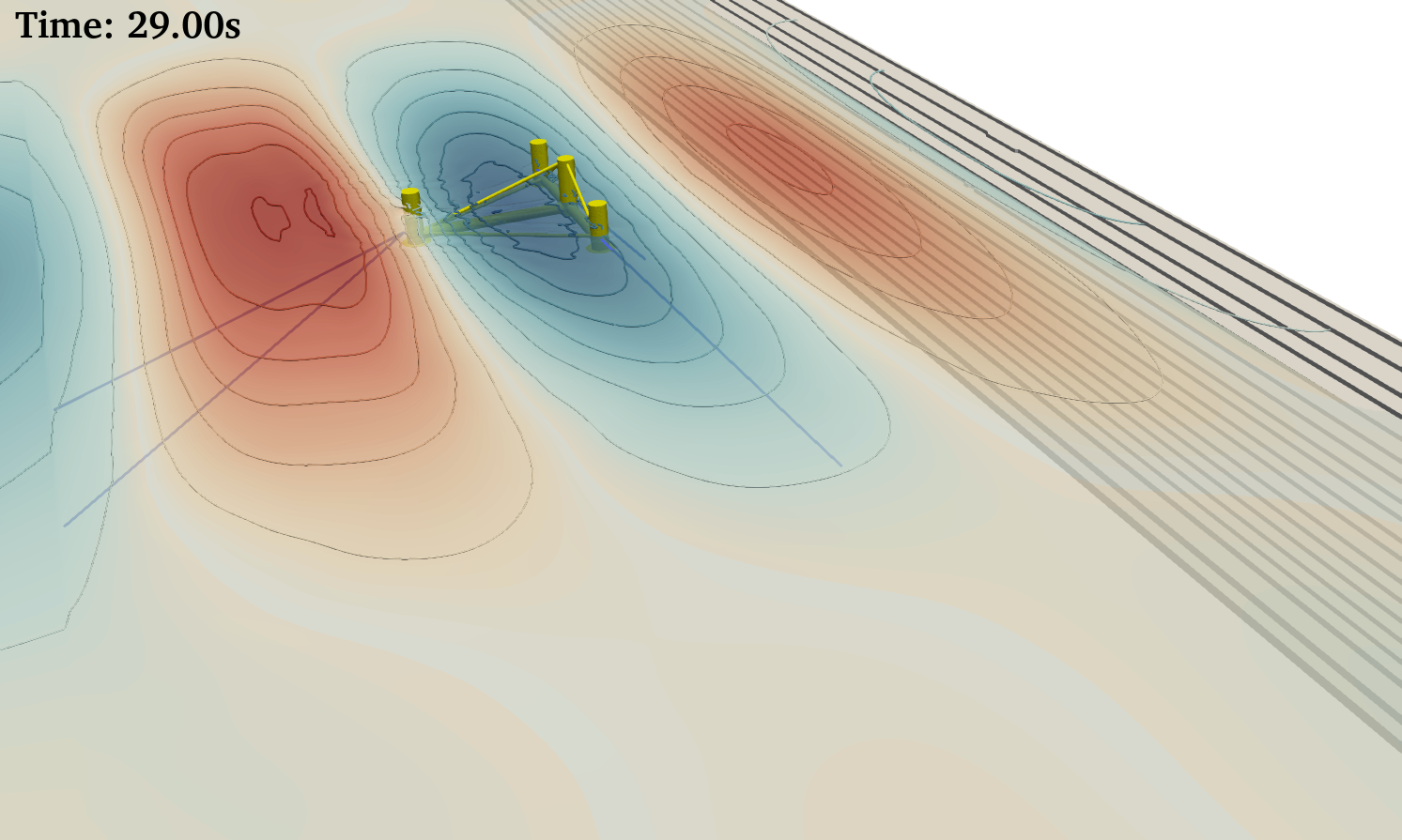}
    \end{subfigure}
    \hfill
    \begin{subfigure}[b]{0.49\textwidth}
        \centering
        \includegraphics[width=\textwidth,height=0.2\textheight]{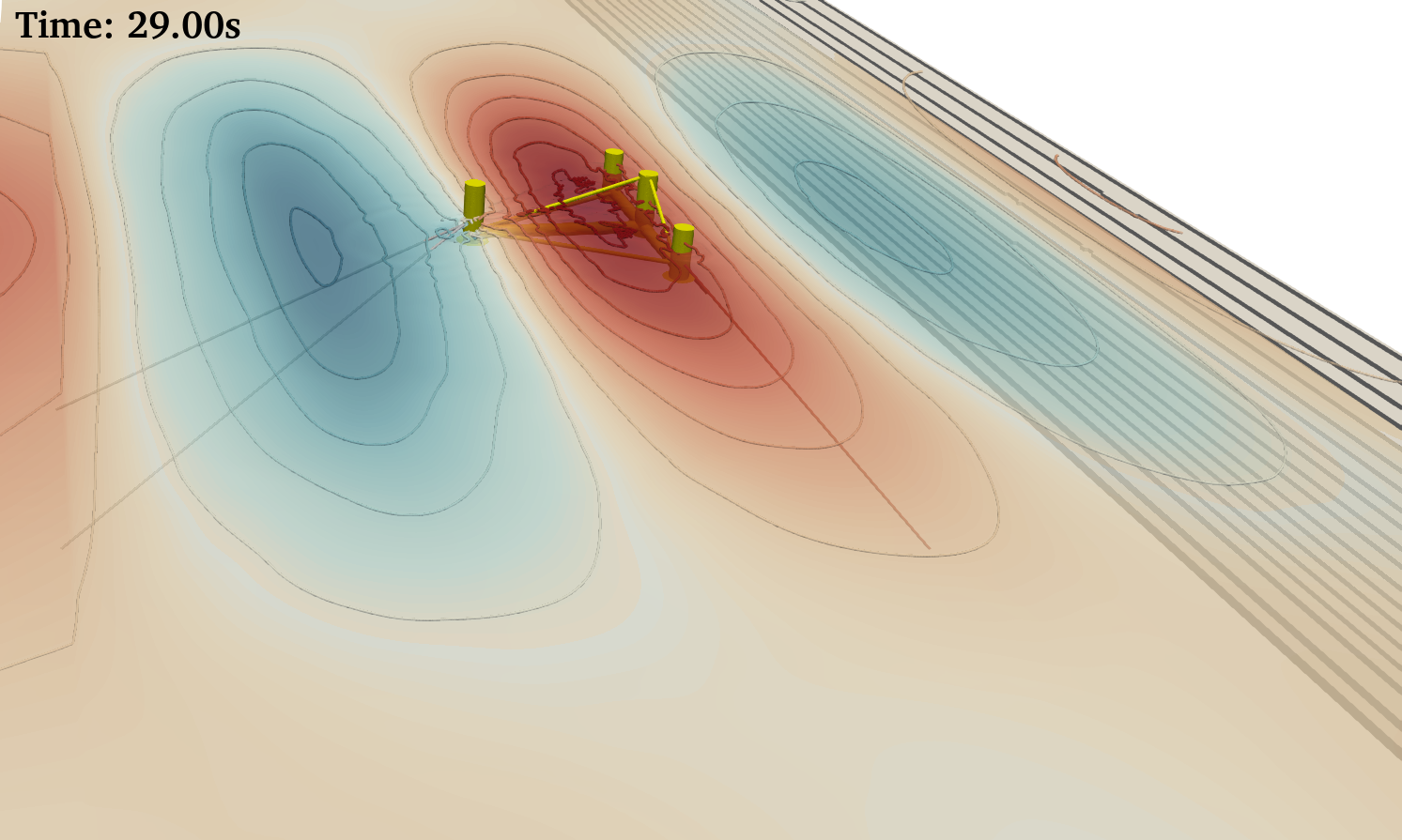}
    \end{subfigure}
    
    
    \begin{subfigure}[b]{0.49\textwidth}
        \centering
        \includegraphics[width=\textwidth,height=0.2\textheight]{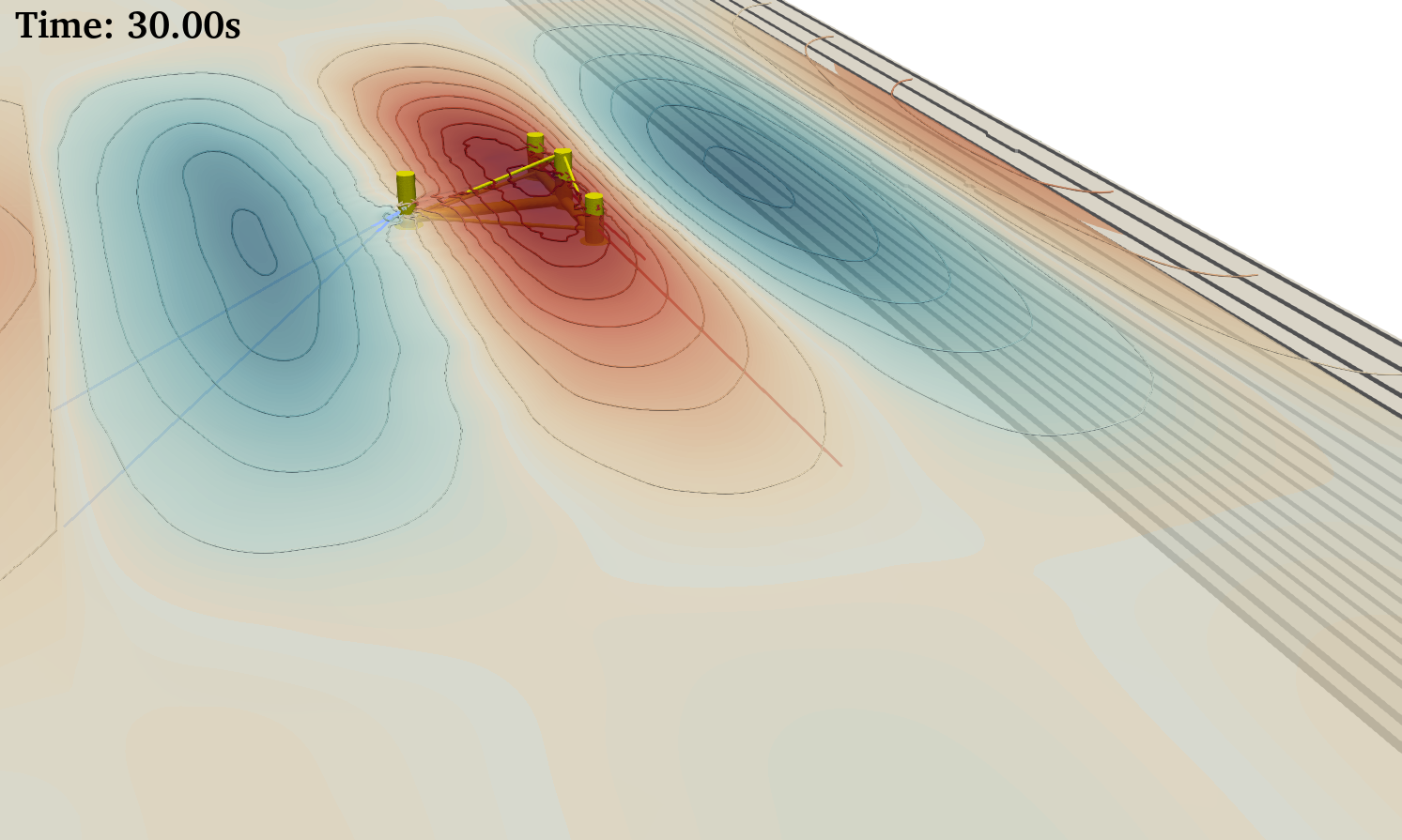}
    \end{subfigure}
    \hfill
    \begin{subfigure}[b]{0.49\textwidth}
        \centering
        \includegraphics[width=\textwidth,height=0.2\textheight]{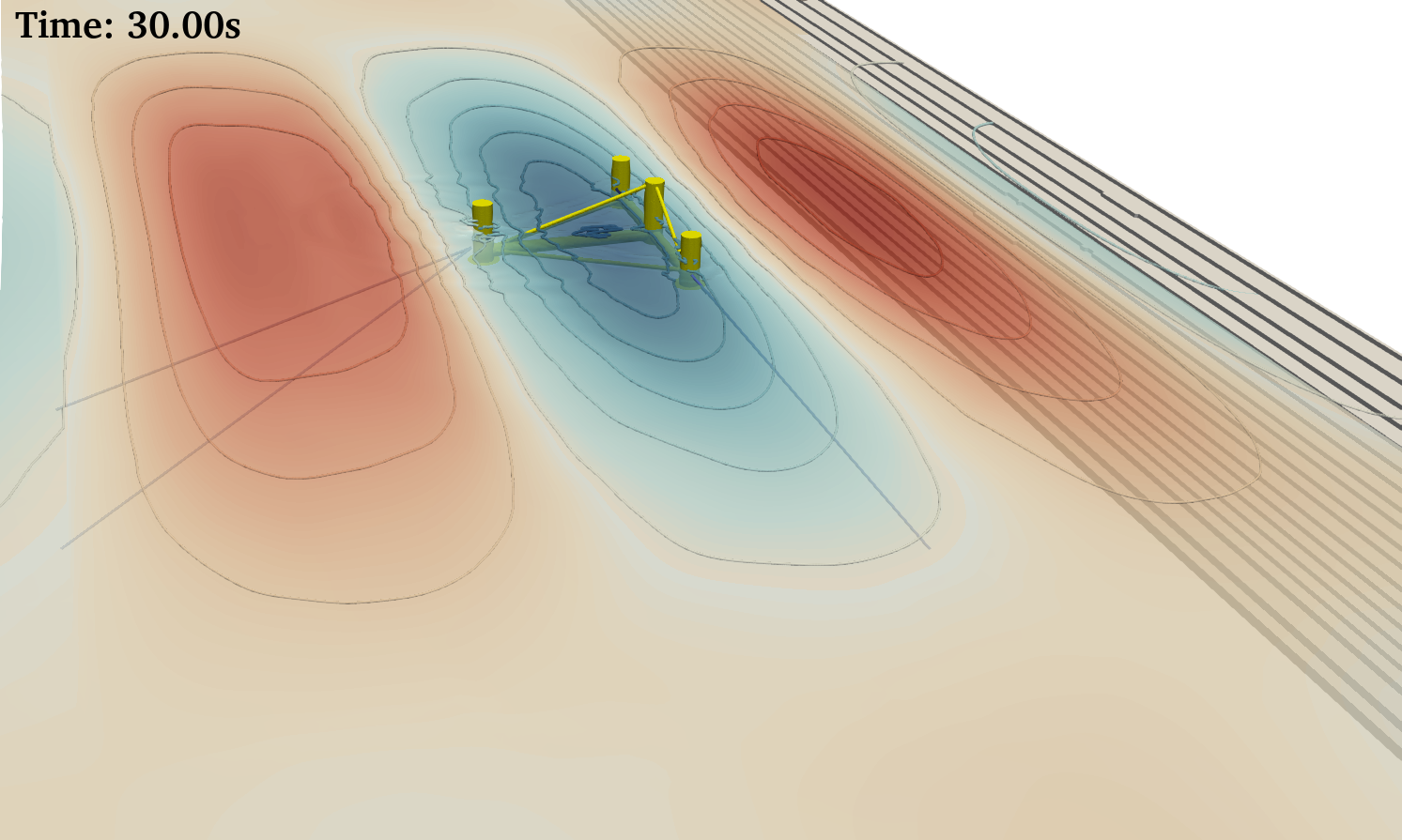}
    \end{subfigure}
    

        \begin{subfigure}[b]{0.49\textwidth}
        \centering
        \includegraphics[width=\textwidth,height=0.2\textheight]{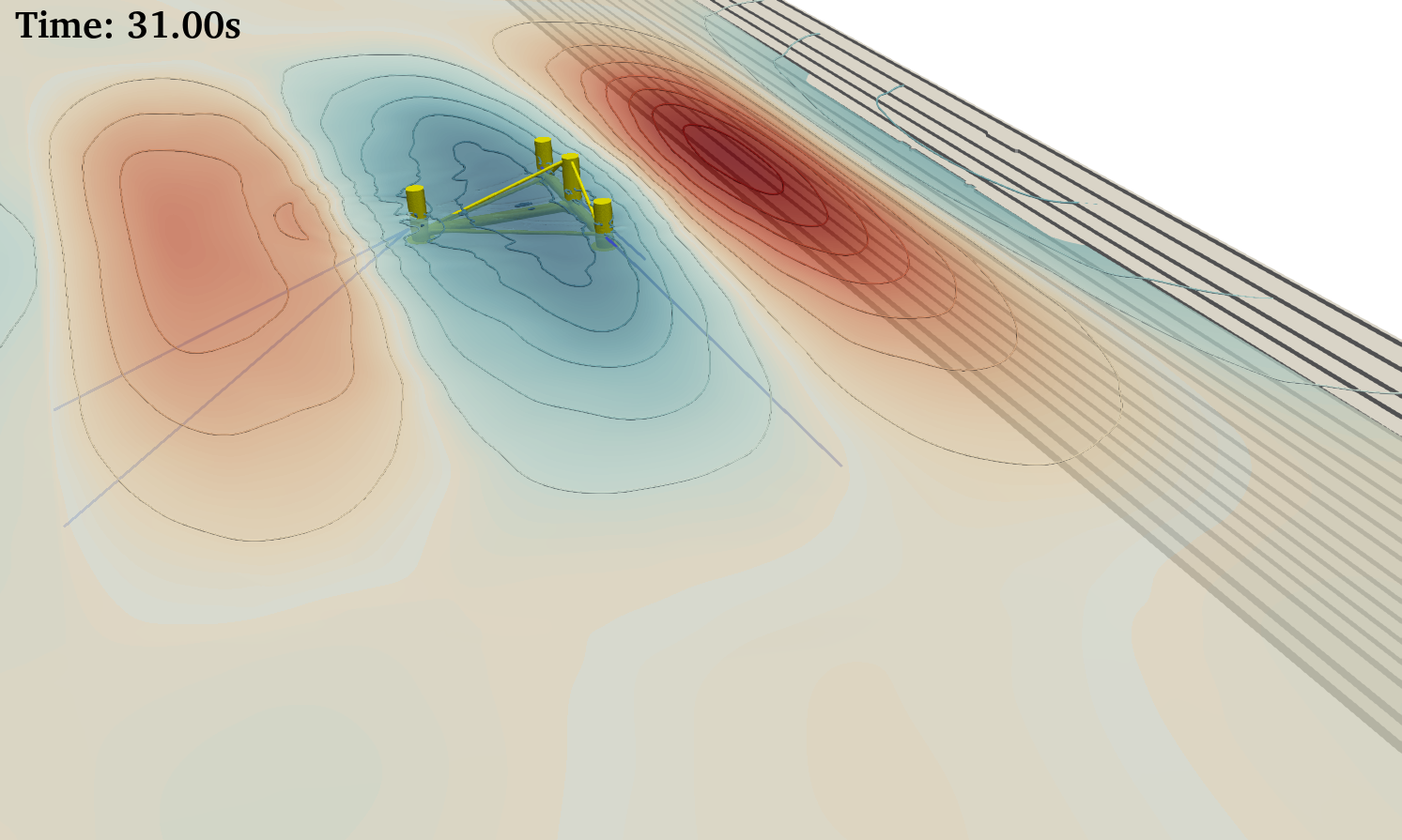}
    \end{subfigure}
    \hfill
    \begin{subfigure}[b]{0.49\textwidth}
        \centering
        \includegraphics[width=\textwidth,height=0.2\textheight]{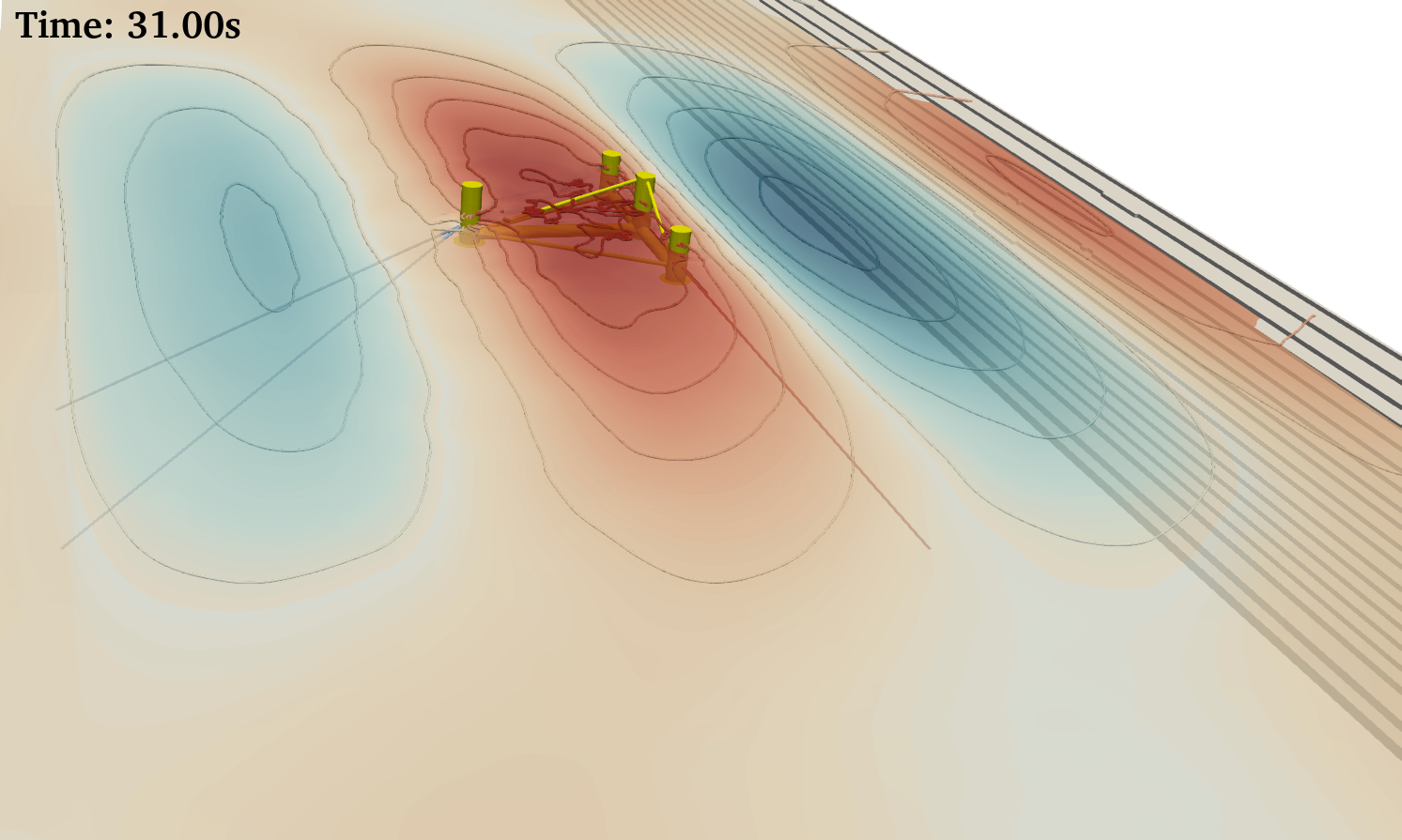}
    \end{subfigure}
        
    
    
    \caption{Timesnaps of focused wave group passing over the floating foundation in direct and phase-shifted cases, shown in the left and right column. The focusing time is set at $30$ s, with a spreading angle of $20^\circ$, where the red color represents the crest, blue represents the trough, and the contours depict the normalized free surface elevation}
    \label{fig:overall}
\end{figure}

\begin{figure}[p]
    \centering
        \begin{tikzpicture}
        \node at (0,0) {\includegraphics[scale=0.25]{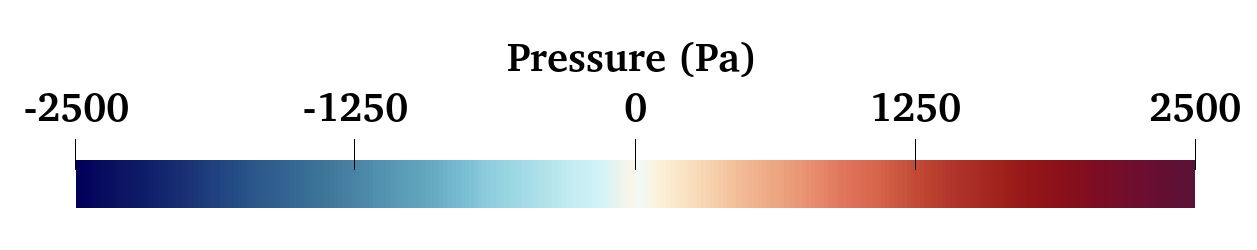}};
    \end{tikzpicture}

    \begin{subfigure}[b]{0.49\textwidth}
        \centering
        \includegraphics[width=\textwidth,height=0.2\textheight]{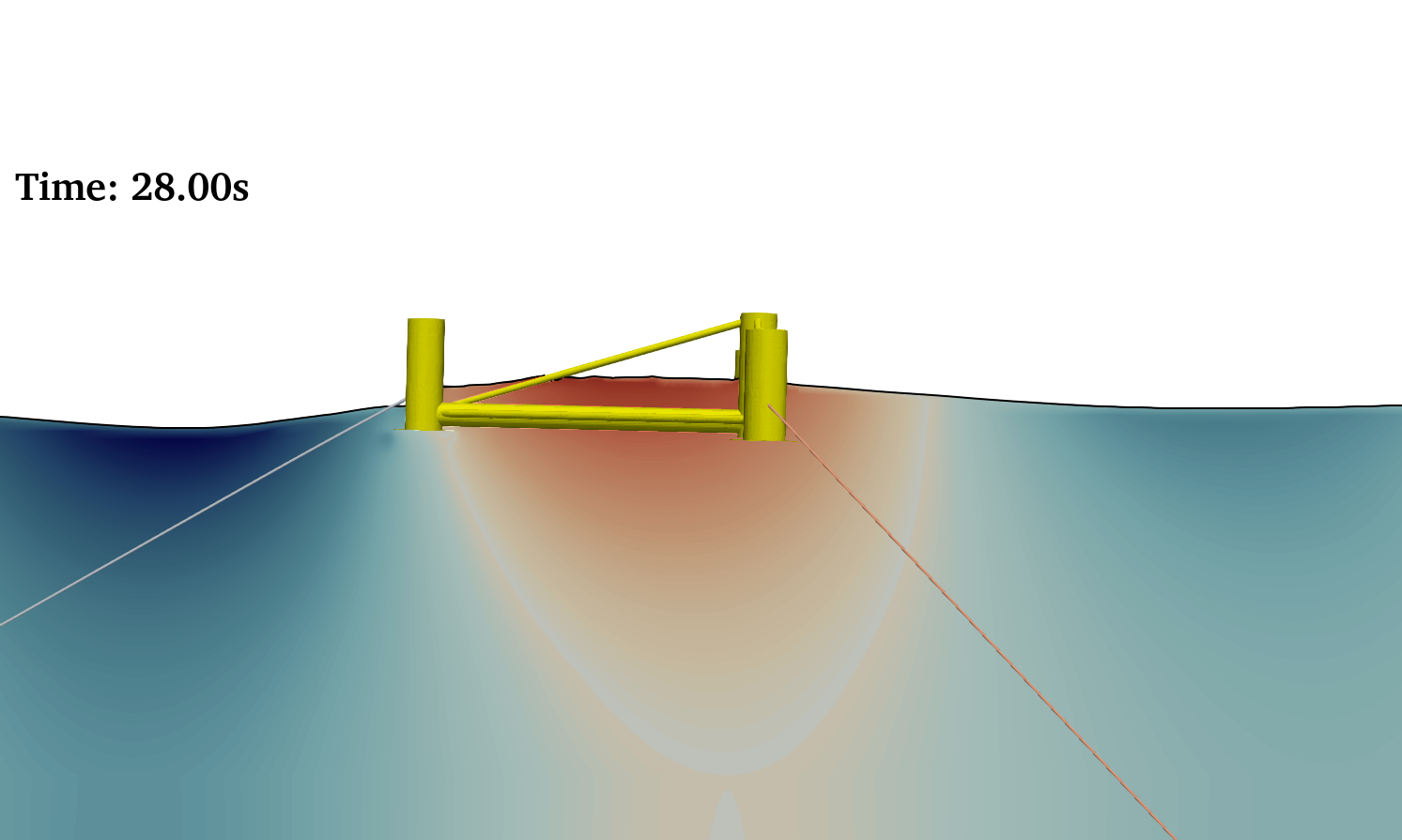}
    \end{subfigure}
    \hfill
    \begin{subfigure}[b]{0.49\textwidth}
        \centering
        \includegraphics[width=\textwidth,height=0.2\textheight]{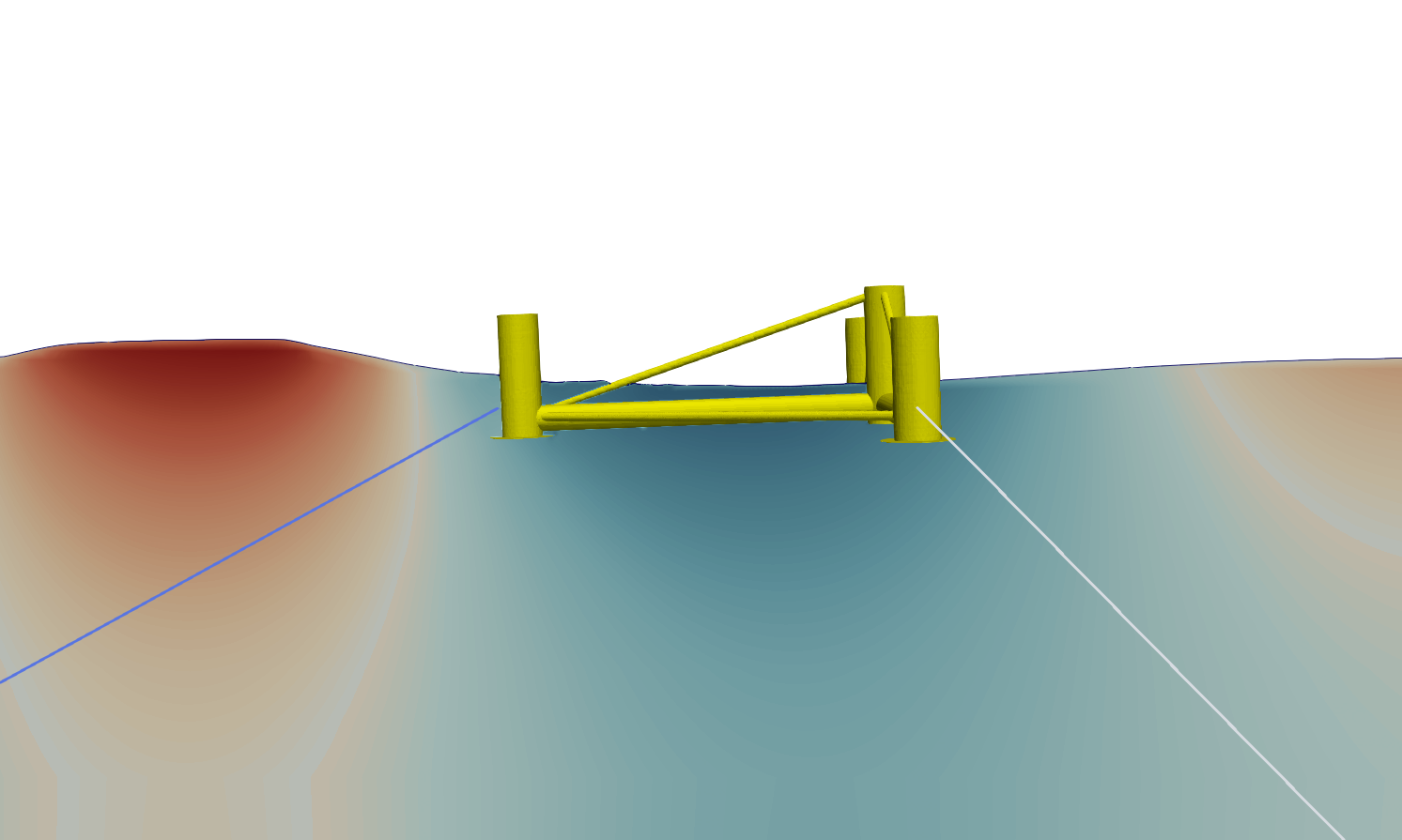}
    \end{subfigure}
    
    \vspace{0.1em}

        \begin{subfigure}[b]{0.49\textwidth}
        \centering
        \includegraphics[width=\textwidth,height=0.2\textheight]{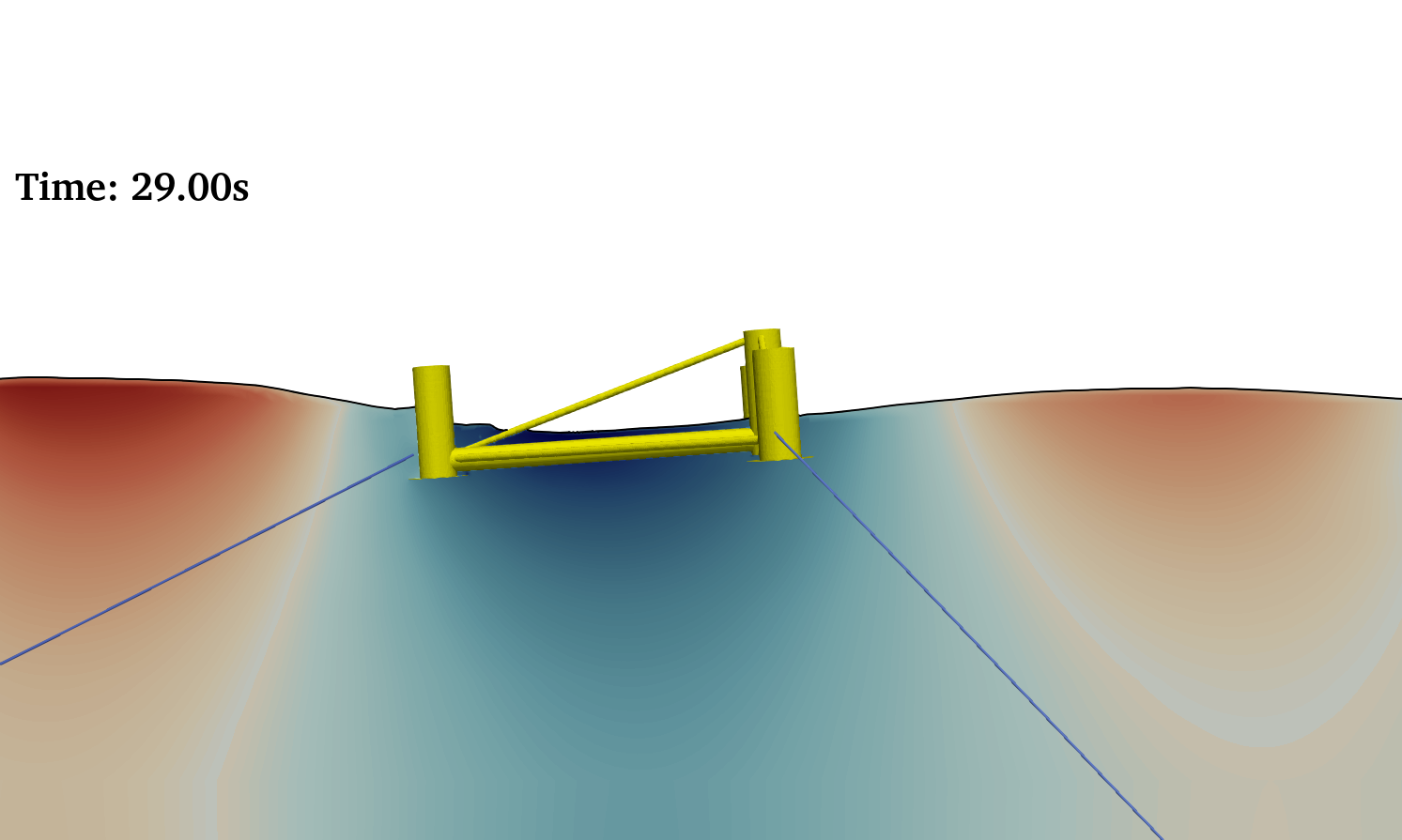}
    \end{subfigure}
    \hfill
    \begin{subfigure}[b]{0.49\textwidth}
        \centering
        \includegraphics[width=\textwidth,height=0.2\textheight]{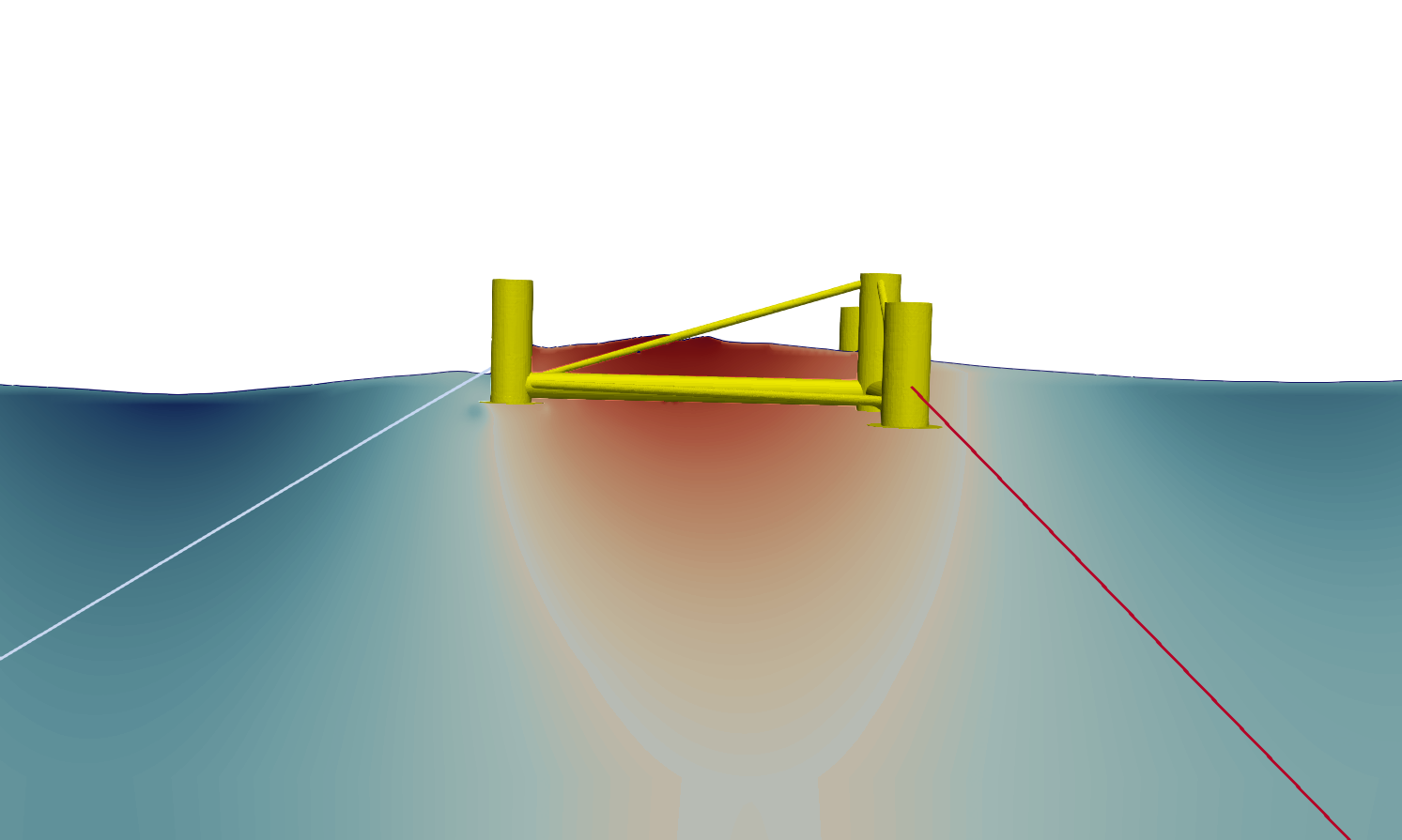}
    \end{subfigure}
    
    
    \begin{subfigure}[b]{0.49\textwidth}
        \centering
        \includegraphics[width=\textwidth,height=0.2\textheight]{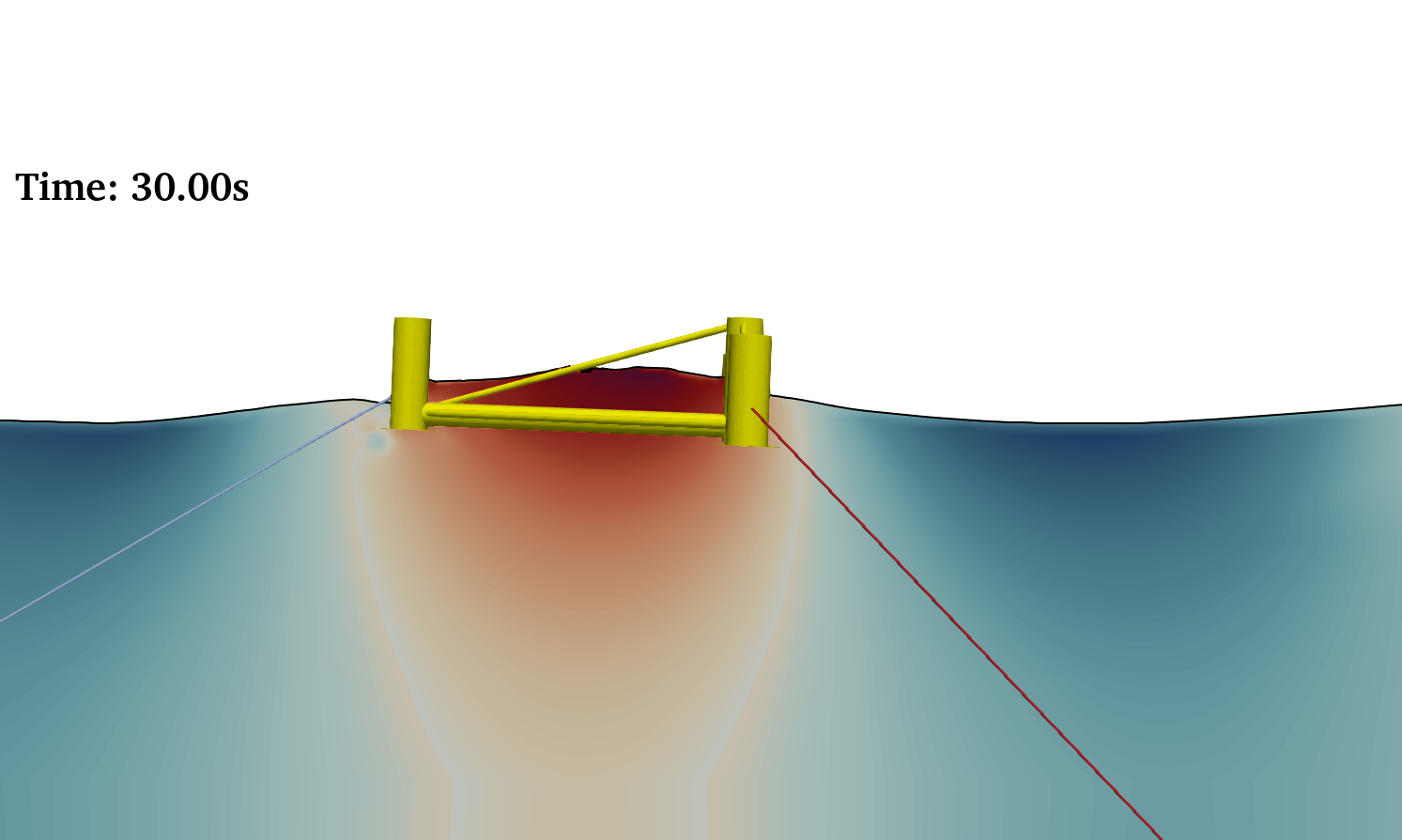}
    \end{subfigure}
    \hfill
    \begin{subfigure}[b]{0.49\textwidth}
        \centering
        \includegraphics[width=\textwidth,height=0.2\textheight]{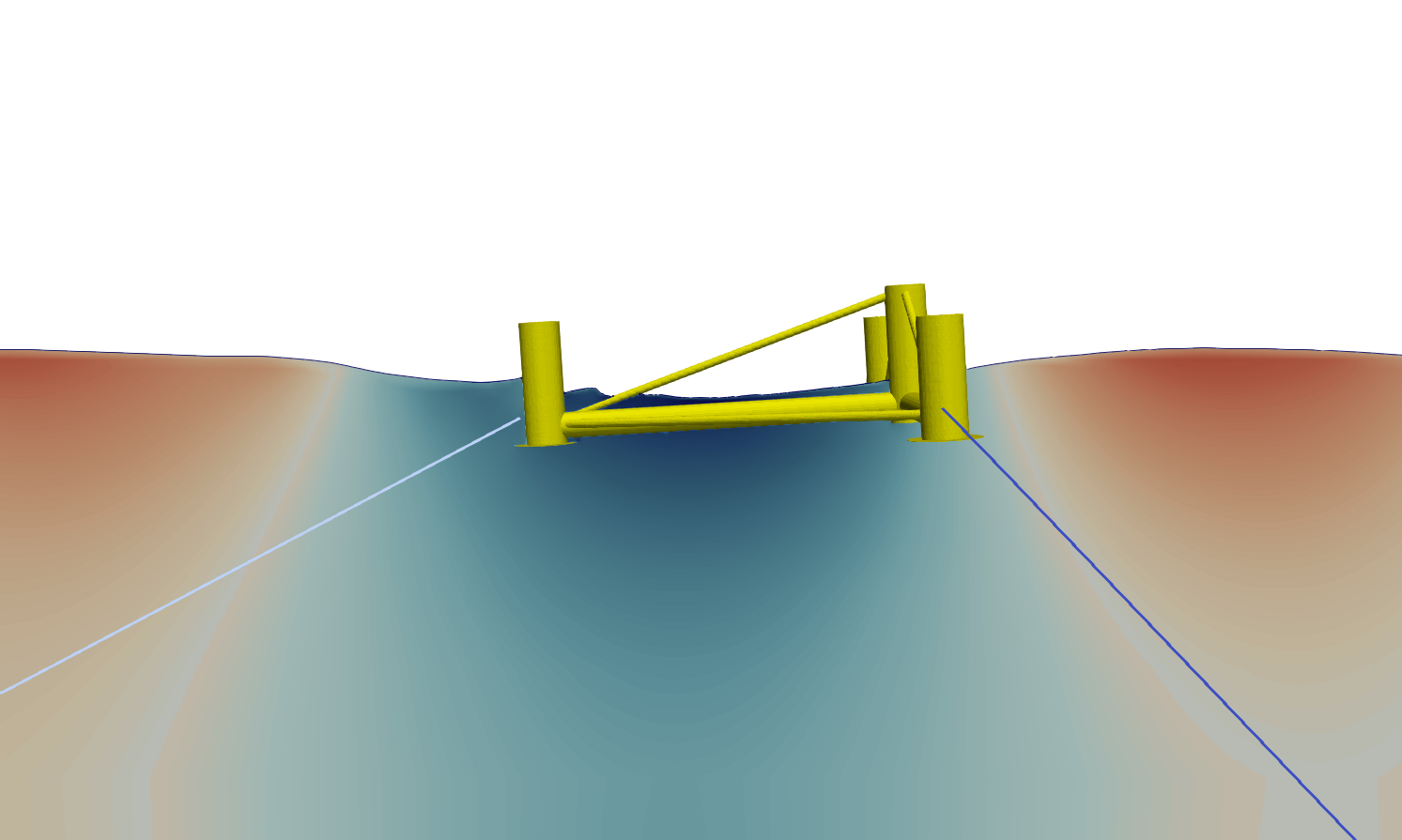}
    \end{subfigure}
    

        \begin{subfigure}[b]{0.49\textwidth}
        \centering
        \includegraphics[width=\textwidth,height=0.2\textheight]{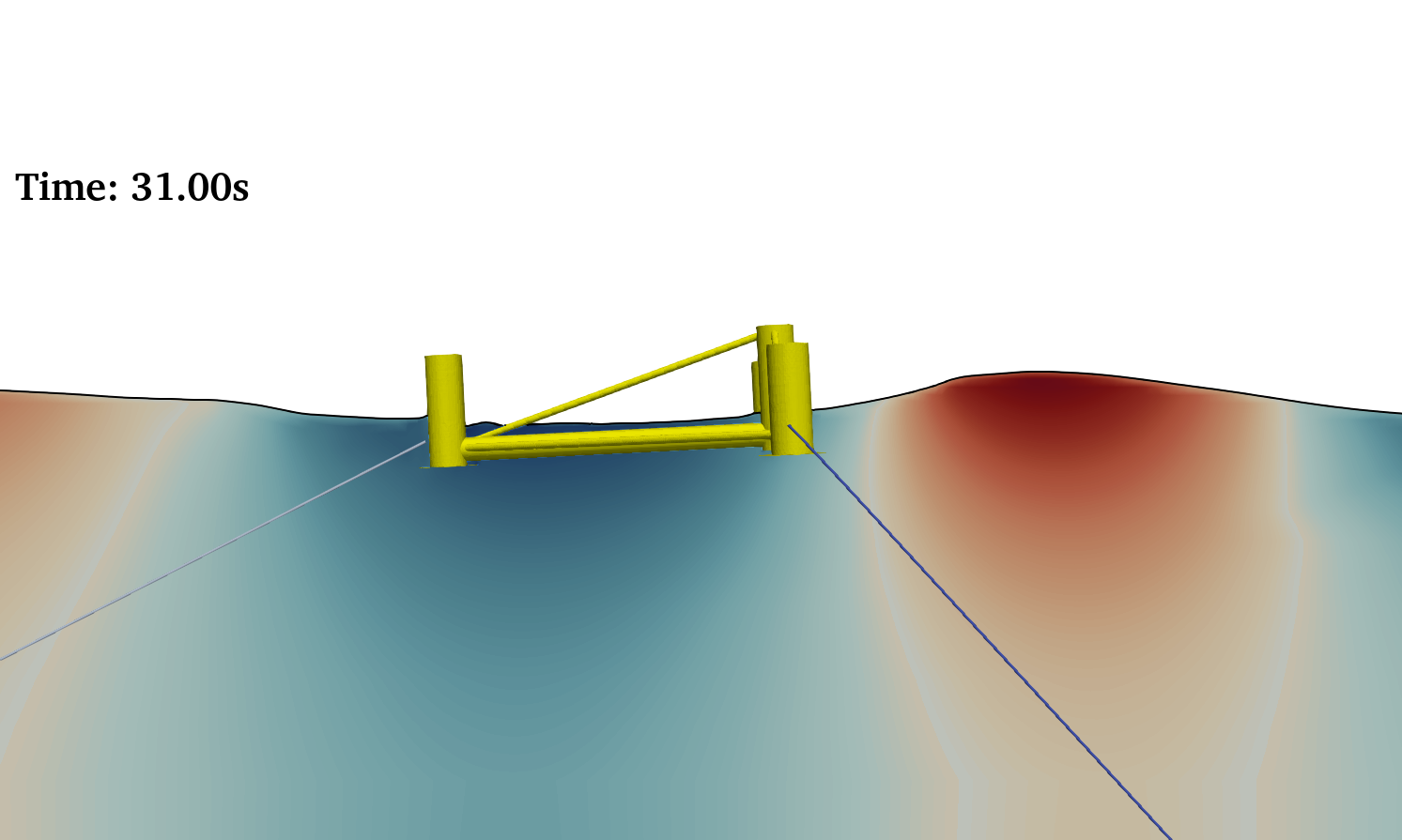}
    \end{subfigure}
    \hfill
    \begin{subfigure}[b]{0.49\textwidth}
        \centering
        \includegraphics[width=\textwidth,height=0.2\textheight]{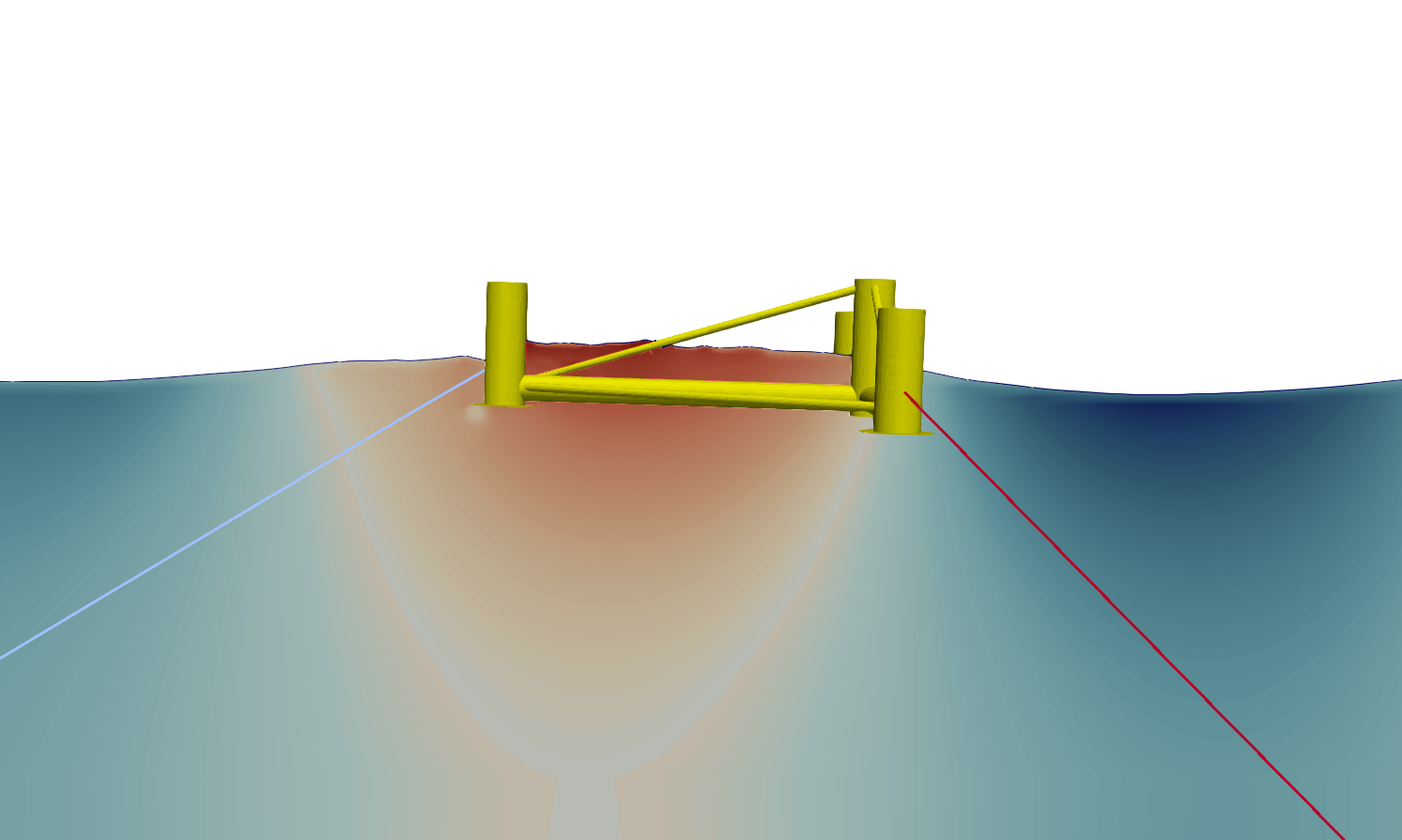}
    \end{subfigure}
        
    
    
    \caption{Sectional view of time snaps of focused wave group passing over the floating foundation in direct and phase-shifted cases, shown at the left and right of column. The focusing time is set at $30$ s, with a spreading angle of $20^\circ$, where the red color represents the crest, blue represents the trough.}
    \label{fig:overall2}
\end{figure}

\begin{figure}[p]
    \centering
    \begin{subfigure}[b]{0.9\textwidth}
        \centering
        \includegraphics[width=\textwidth]{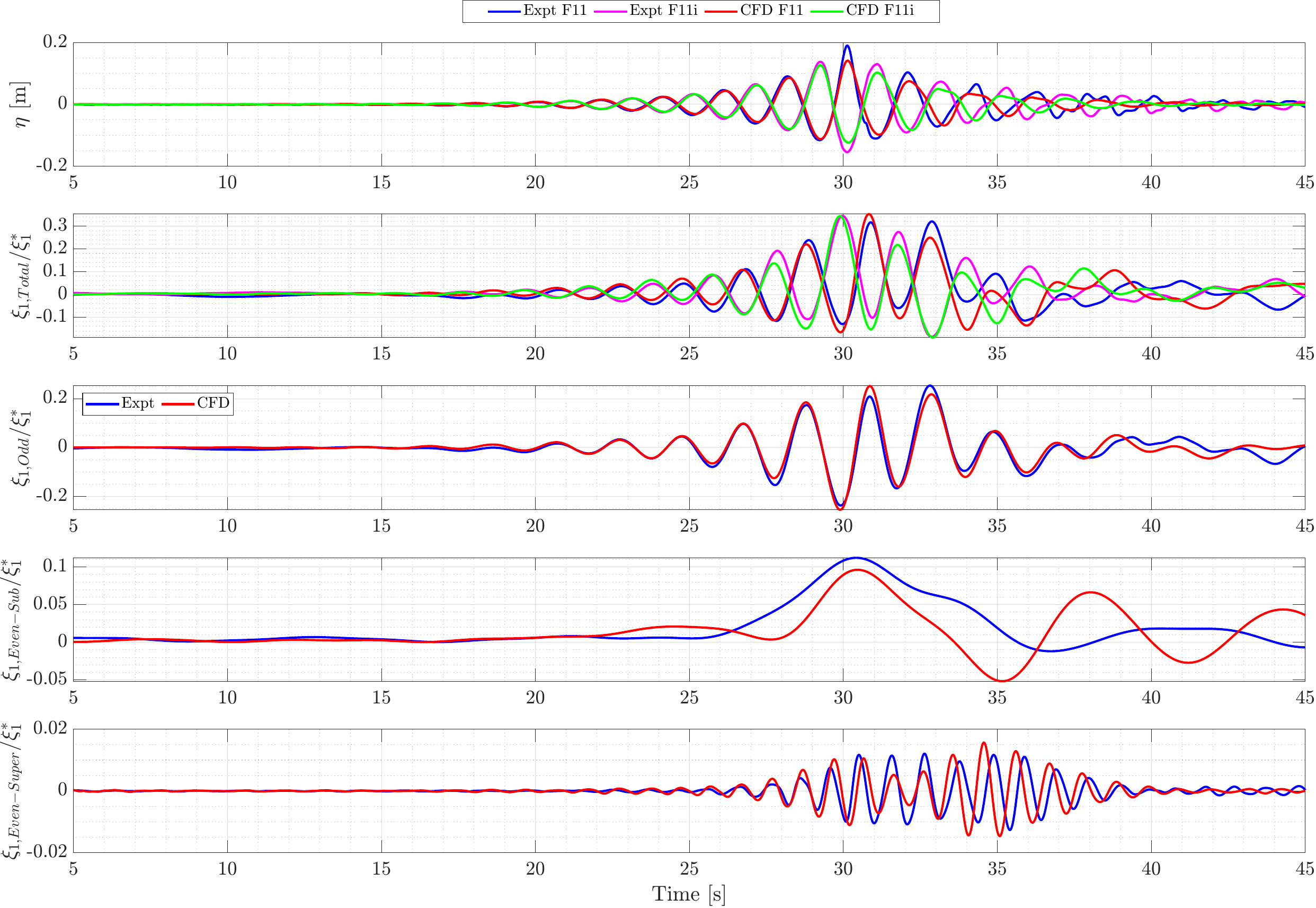}
        \label{fig:F110degSurge}
    \end{subfigure}
    
    
    \begin{subfigure}[b]{0.9\textwidth}
        \centering
        \includegraphics[width=\textwidth]{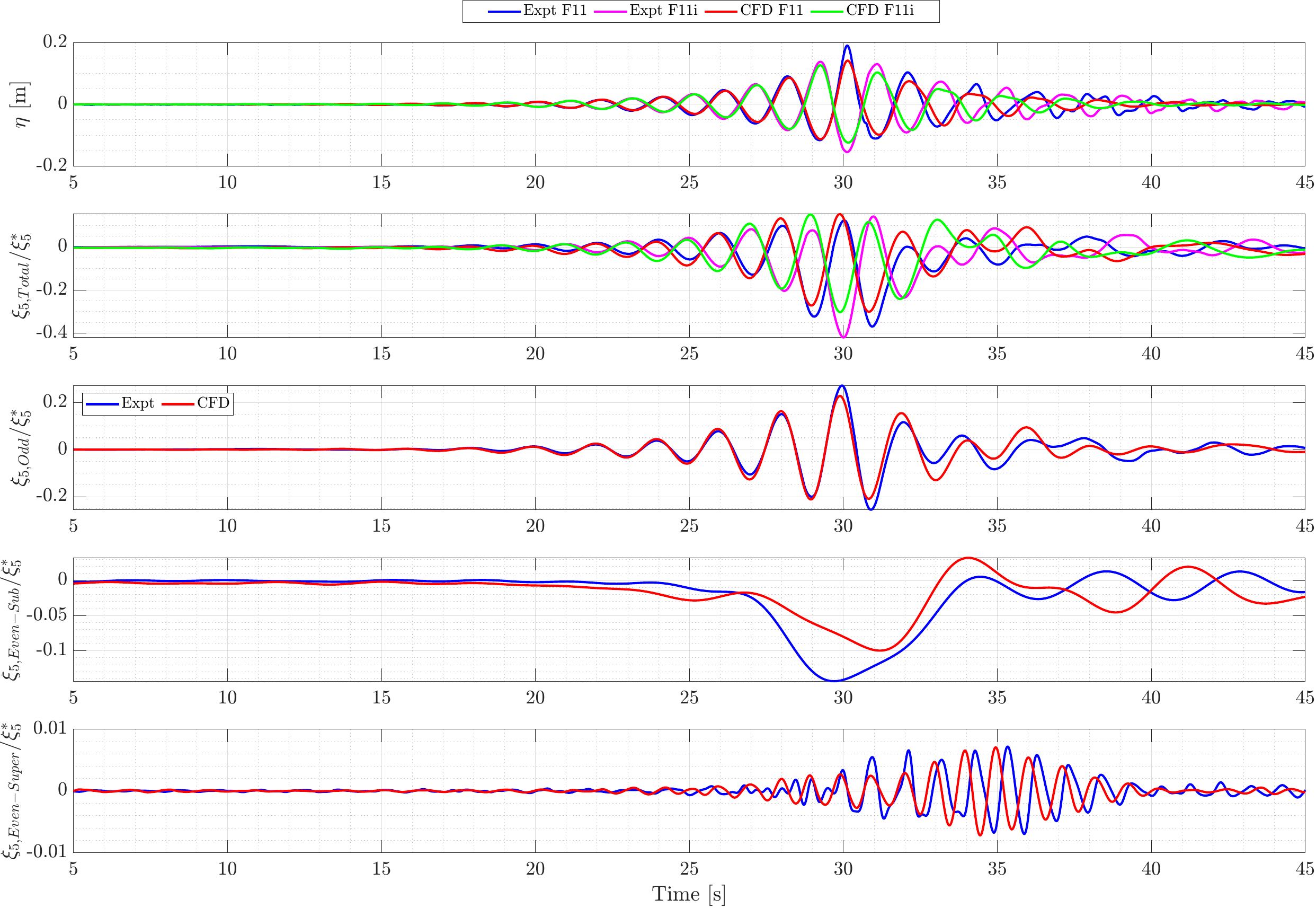}
        \label{fig:F110degPitch}
    \end{subfigure}
    \caption{Surge(top) and pitch(bottom) responses for group F11(nondimensionalized by $\xi^{*}_{1}$ and $\xi^{*}_{5}$ respectively). Row 1 shows the free surface elevation of the incident and phase-shifted waves; Row 2 displays the total response and rows 3,4 and 5 present odd, even subharmonic and superharmonic responses, respectively}

    \label{fig:F110degCombined}
\end{figure}

\section{Typical interaction Between Focused Waves and the Floating Foundation}

 An overview of the typical interaction between the directional focused wave (F14s) and the floating foundation is provided by the snapshots in Figure \ref{fig:overall} and Figure \ref{fig:overall2}.  A crest-focused wave is shown on the left, and a trough-focused wave is shown on the right. The contours in Figure \ref{fig:overall}, normalized by wave amplitude, depict the evolution of free surface elevation as it passes the floating foundation, highlighting the dynamics during the focusing event. In contrast, Figure \ref{fig:overall2} shows the pressure variation in the domain. The wave, focused directionally with a $20^{o}$ spreading, reaches its maximum crest or trough at the midpoint between the rear tanks.  A comparison of the left and right columns shows the difference in response due to group inversion. Specifically, the crest approaches the floating foundation at $29$ and $30$ s, while a deep trough impacts the floating foundation simultaneously in the inverted case. This also results in significant wave run-up at the front tank and a pronounced wave run-down for the phase-inverted waves. In particular, the focusing location is centered between the rear tanks, which means that the direct and phase-inverted waves initially interact with the front tank. This interaction leads to shadowing effects, resulting in the rear tank experiencing a slightly attenuated wave due to the presence of the front tank. This wave floating foundation interaction process outlines two distinct stages of the dynamic response. In the first stage, the focused wave group passes over the floating foundation, exerting wave loads that induce responses in the floating foundation (Figure \ref{fig:overall} and Figure \ref{fig:overall2}). The second stage, the decay response, follows after the passing of the wave group. During this stage, the floating foundation oscillates at its natural frequencies, with the amplitude of these oscillations gradually reducing due to damping effects. In the following section, we will apply harmonic separation to analyze the floating foundation's surge and pitch responses, allowing us to identify and differentiate between these two stages.

\section{Effect of wave severity on the sub and superharmonic response}\label{Section7}
\begin{figure}[p]
    \centering
    \begin{subfigure}[b]{0.9\textwidth}
        \centering
        \includegraphics[width=\textwidth]{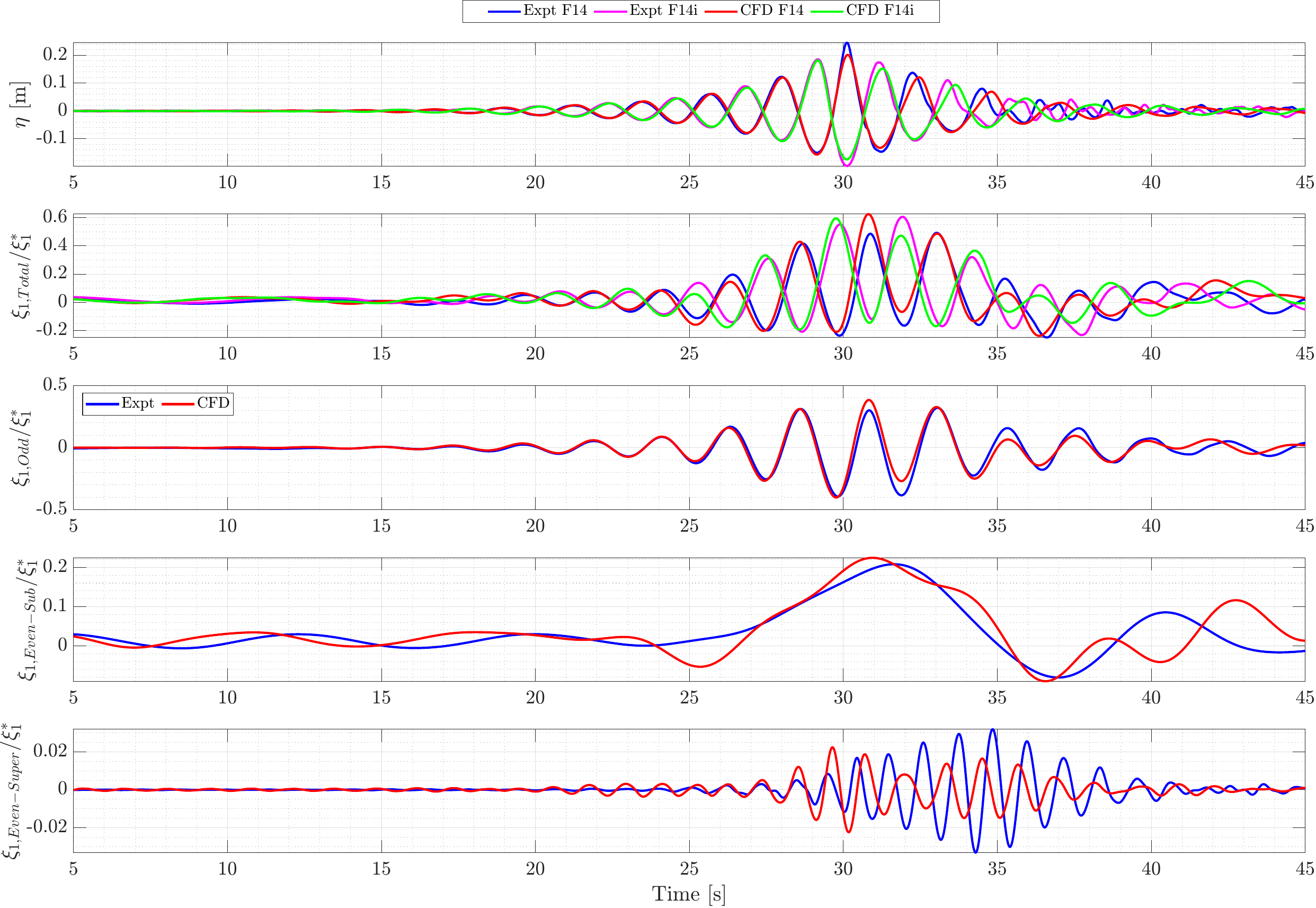}
        \label{fig:F140degSurge}
    \end{subfigure}
    
    
    \begin{subfigure}[b]{0.9\textwidth}
        \centering
        \includegraphics[width=\textwidth]{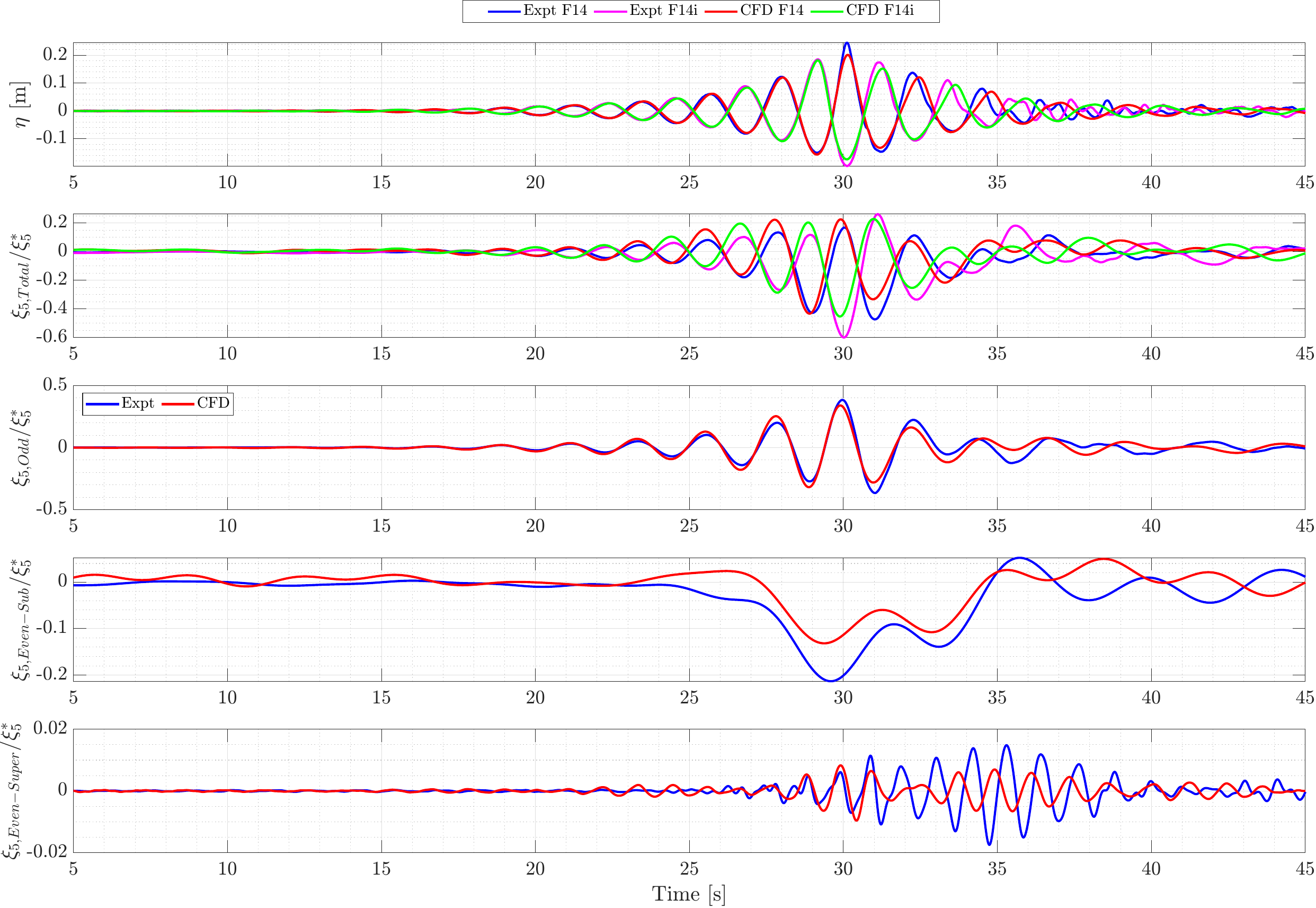}
        \label{fig:F140degPitch}
    \end{subfigure}
    \caption{F14: Surge and Pitch responses (nondimensionalized by $\xi^{*}_{1}$ and $\xi^{*}_{5}$ for surge and pitch). Row 1 shows the free surface elevation of incident and phase-shifted waves; Row 2 displays the total response; Row 3,4 and 5 presents odd, even sub and superharmonic responses respectively. }
    \label{fig:F140degCombined}
\end{figure}

Surge and pitch responses following the interaction of the focused wave group (Table \ref{tab:TestMatrix}) with the floating foundation are harmonically decomposed to distinguish between odd and even subharmonics and superharmonics. These decomposed results are presented in figures  \ref{fig:F110degCombined} and  \ref{fig:F140degCombined}, focusing on the non-spread F11, followed by the F14 case. Similar observations were noted for the spread focused wave group tests, so they are omitted for brevity. However, the comparison and effect of spreading are discussed in detail in Sections \ref{Section8} and \ref{Section9}. In Figure \ref{fig:F110degCombined} it is observed that the odd and even surge responses peak at $31$ s, whereas the pitch responses reach their maximum at $30$ s, coinciding with the wave group focusing time. This delay in the surge response can be explained by inertia effects associated with the relatively large surge mass, causing the maximum displacement to occur slightly after the crest has passed. In contrast, the pitch response responds more to instantaneous wave steepness and wave-induced moments, leading to a more immediate response. This behavior is also consistent in the case of an increased wave severity (F14), as shown in Figure \ref{fig:F140degCombined}. 

The wave groups drive a significant subharmonic response, which is well separated by the harmonic decomposition.  For the F11 case, the subharmonic surge responses are nearly $40\%$ of the maximum amplitude of the odd harmonics. This ratio also applies to the subharmonic pitch response, although this motion is in the opposite direction to the surge due to the mooring induced surge-pitch coupling. The superharmonic surge response is seen to follow the behaviour from the wave-only analysis with a bound response that coincides with the main wave group, partially overlapping with a free group, arriving around 4~s later. The amplitude of the resulting long response sequence is $5\%$ of the odd response level. In contrast, the bound superharmonic pitch response is only around $0.5\%$ of the odd harmonic amplitude, whereas the subsequent free superharmonic response is stronger, with an amplitude of up to $2.5\%$ of the odd amplitude. 

 The CFD accurately reproduces the free surface elevation and the odd harmonic responses for both surge and pitch. However, there is a slight reduction in the peak amplitude of the subharmonic surge responses and a lower damping compared to the experiments. This discrepancy will be further discussed in relation to the F14 case.  Also, for pitch, the subharmonic trough amplitudes are slightly lower than in the experiment. However, the superharmonic responses of surge and pitch align well in amplitude and trend, although the CFD shows spurious free wave responses slightly earlier than observed in the experiment.

The floating foundation response to the increase in wave severity in the F14 case is presented in Figure \ref{fig:F140degCombined}. Odd harmonics show a $20\%$ increase in peak amplitudes for both surge and pitch compared to their counterpart in the F11 responses. The pitch subharmonics demonstrate a significant change in focusing time. For F14, the increase in wave severity induces a pronounced motion at the natural pitch frequency, while the less severe case F11 is mainly influenced by low-frequency surge responses. The subharmonic response in the surge is observed to be $75\%$ of the maximum amplitude of odd harmonics, while the pitch peak-trough is around $45\%$. These numbers can be compared with the ratio of 40\% observed for the F11 case. Superharmonic responses in bound waves show a trend similar to that of F11, with maximum amplitudes comparable to odd harmonics and no significant increase. Linked to the nonlinear nature of the superharmonic reponses, spurious free wave responses reach up to $12\%$ of the maximum amplitude of odd harmonics in both surge and pitch, double the magnitude observed for F11. 

Regarding CFD reproduction, the free surface elevation of F14 waves is well captured. When decomposing the total responses, the odd harmonics of the surge show a good trend match, although the CFD overpredicts the peak response by $8\%$. For pitch, odd harmonics trend and amplitude are accurately reflected. The subharmonic even responses in both surge and pitch trends are well captured, although the pitch peak trough is slightly underpredicted compared to the experimental results. This is also the case for the superharmonic responses which are well captured in trend but are underpredicted in both surge and pitch. A potential explanation is the earlier arrival of spurious free waves in the CFD for the F14 case, causing a cancelation effect between the bound and free high-frequency motion. 

Consistent with the observations from the decay tests in Section \ref{section:decay}, the experimental damping for the subharmonic motion is stronger for F14 which has a stronger amplitude. This behaviour is also reproduced in the CFD results, although the damping level is smaller than in the experiment, as was also found in the reproduction of the decay tests. 

\begin{figure}[htb]
    \centering
	    \includegraphics[width=\textwidth]{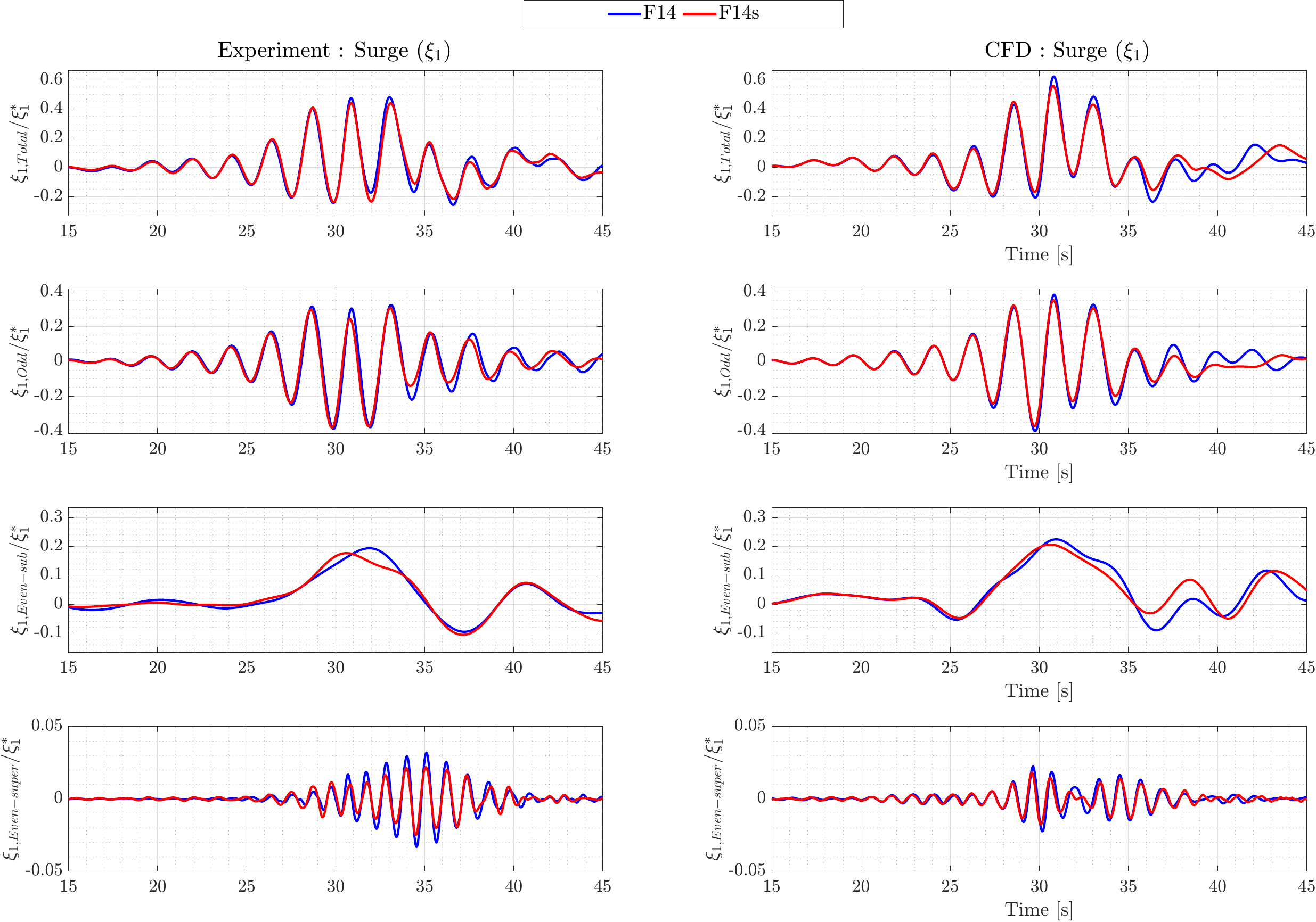}
    \caption{Comparison of total, odd, and even surge responses($\xi_{1}$) for experimental (Column 1) and CFD (Column 2) results at F14 and F14s. The nondimensionalized parameter is represented by $\xi^{*}_{1}$. \label{fig:Surge0and20}}
\end{figure}

\begin{figure}[tb]
    \centering
	    \includegraphics[width=\textwidth]{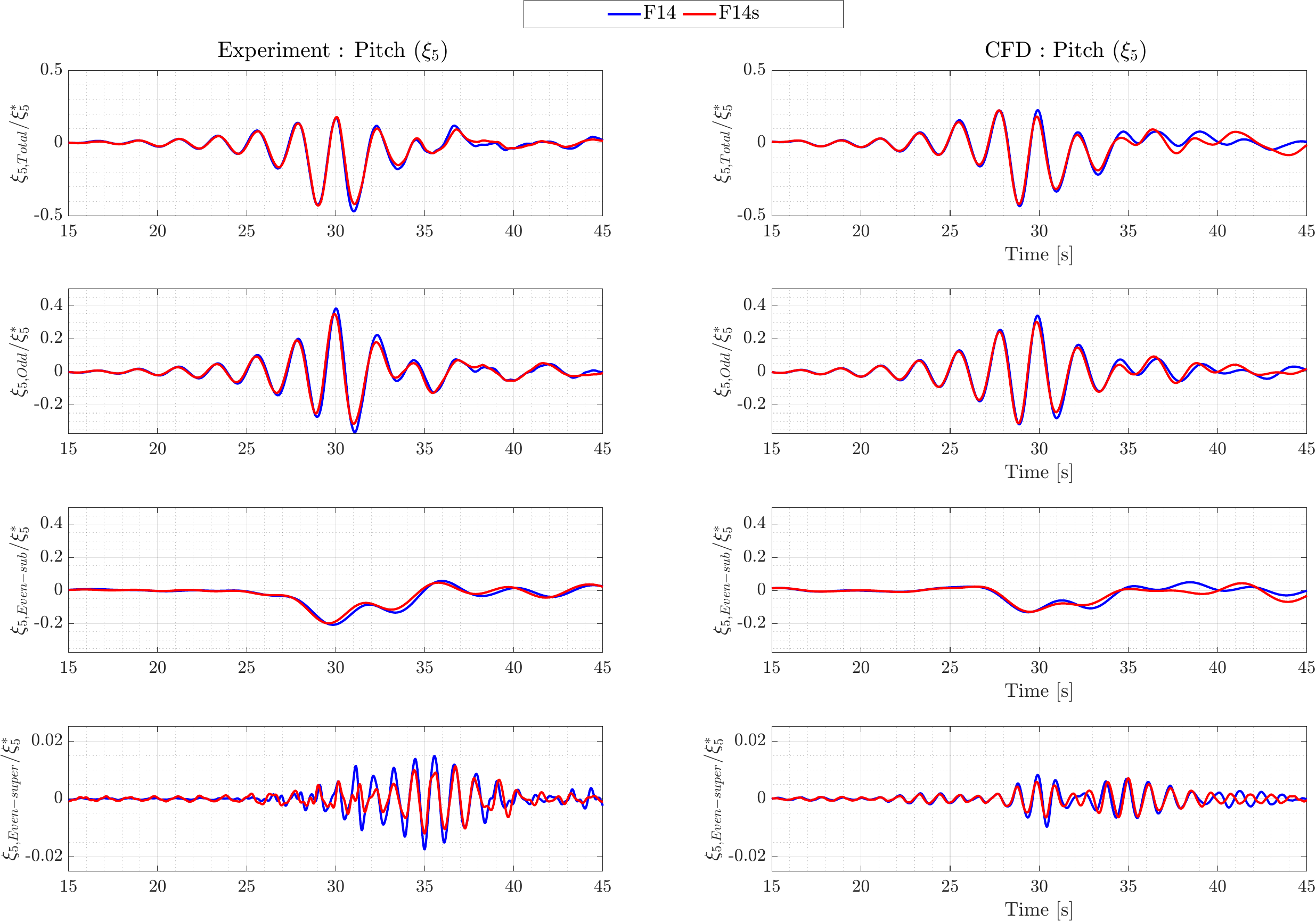}
    \caption{Comparison of total, odd, and even pitch responses($\xi_{5}$) for experimental (Column 1) and CFD (Column 2) results at F14 and F14s. The nondimensionalized parameter is represented by $\xi^{*}_{5}$. \label{fig:Pitch0and20}}
\end{figure}

\section{Effect of spreading on the sub and superharmonic response}\label{Section8}
 As outlined in Table \ref{tab:TestMatrix}, two types of wave conditions were tested: F11 without spreading, F11 with $20^\circ$ spreading (F11s), and similarly for F14. Both the spread and non-spread responses are presented for discussion using data from experiments and CFD simulations. Since the results for F11 and F14 exhibited similar characteristics, only the results for F14 are discussed here. 

\subsection{Surge}

The effect of wave spreading on the surge response for the F14 case is shown in Figure \ref{fig:Surge0and20}, with experimental results presented in the first column and CFD results in the second. The total surge response in the experiment shows that the spread case exhibits slightly smaller peaks than the non-spread case. This difference is primarily reflected in the odd harmonics, which aligns with the expectation from linear wave theory. The reduction in amplitude from spreading is also evident for superharmonic responses, where the spread focused wave group shows significantly smaller peaks, especially during spurious-free wave content near 35~s.

Differences between the unidirectional and spread cases are also seen for the subharmonic surge response. Here, the introduction of spreading in the experiment, however, manifests as a dip at the focussing time that resembles the pitch frequency. Therefore, a quantification of the increase or reduction in amplitude is not possible. 

The CFD results for total and odd surge harmonics are consistent with the experimental data, showing similar amplitude reductions for the spread case.  
Also, the amplitude reduction on the superharmonic response is reproduced, although the free spurious free waves arrive slightly earlier, causing a cancelation effect between the main wave response and the spurious free wave response. Furthermore, for the subharmonic response, the pitch-related dip for the spread case is not replicated numerically, and an overall amplitude reduction during the main wave passage is evident. The decay trend (after $35$ 
 s) does not align completely with the experimental results due to differences in damping behavior (see Section \ref{section:decay}).

\subsection{Pitch}

 The effect of wave spread on pitch response for the F14 case is presented in Figure \ref{fig:Pitch0and20}. Although the time histories of the experiment and CFD are not entirely identical, the effect of spreading is remarkably well reproduced.
 
 From the experimental results, the total pitch response shows minimal difference between spread and nonspread cases until the focusing time, after which a slight reduction is observed for the spread case. The same behavior is observed in the odd harmonic response, where the reduction in amplitude is explicitly visible. The even subharmonic pitch response also shows a slight amplitude reduction from spreading. Similarly to the surge case, the strongest amplitude reduction is seen for the superharmonic responses. As was also seen for surge,  the spurious-free wave response amplitudes dominate over the main wave response. The CFD results reflect a behavior similar to that of the experiment with a slight reduction in total and odd harmonics for the spread focused wave group. For even subharmonics, the CFD also shows a slight reduction in the amplitudes during the focusing time, but the difference is marginal after that. Additionally, the CFD superharmonic response shows an amplitude reduction for the spread focused wave group during the main wave group passage, with observable cancelation effects similar to those seen in surge results. 

\begin{figure}[tb]
    \centering
	    \includegraphics[width=0.825\textwidth]{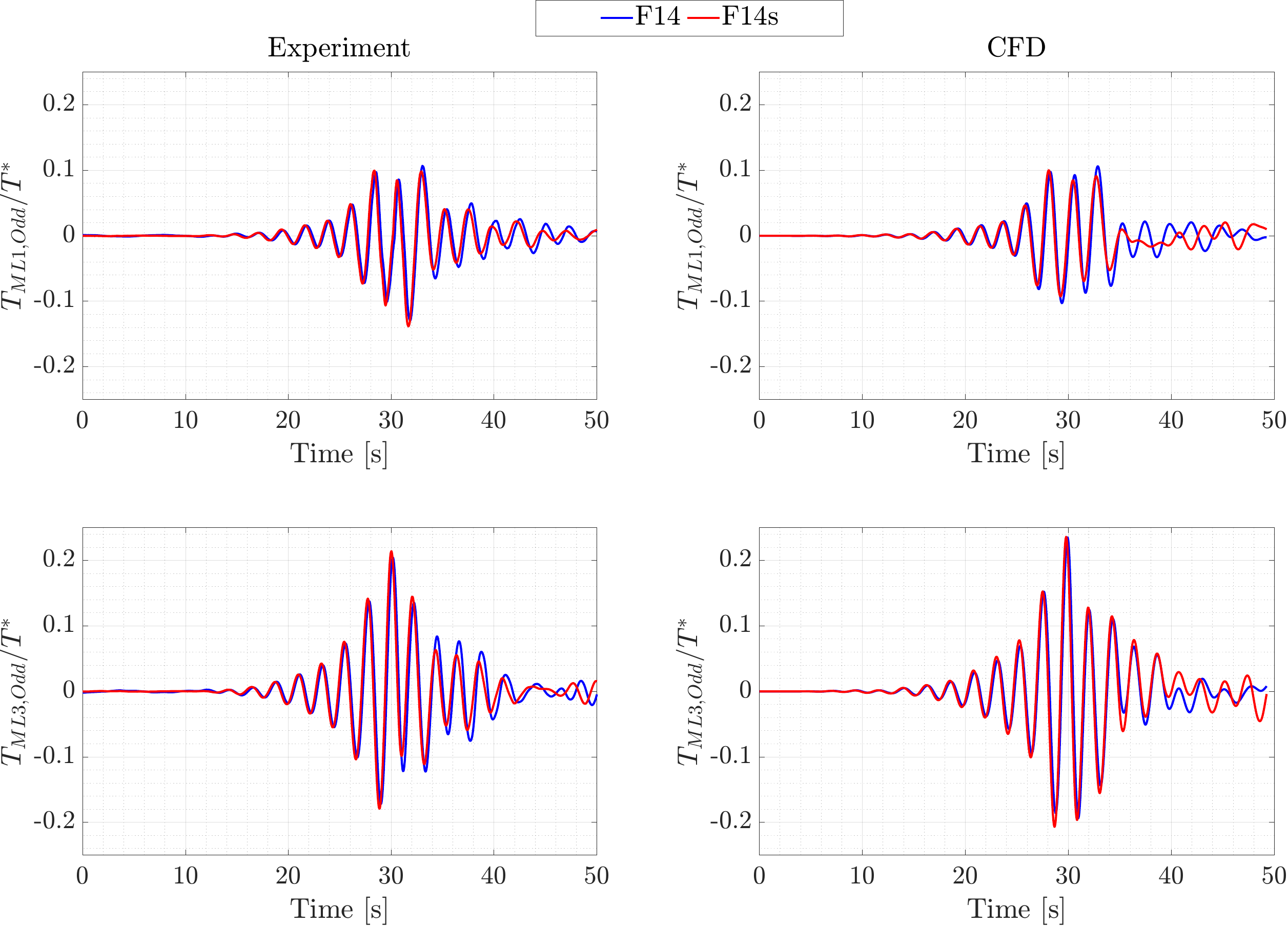}
    \caption{F14 and F14s: Row 1 displays experimental and CFD results for the front mooring line tensions due to odd harmonics, followed by the rear mooring line tensions in Row 2. The comparison includes cases with and without wave spreading.\label{fig:ML13Odd0and20}}
\end{figure}

\begin{figure}[p]
    \centering
	    \includegraphics[width=0.825\textwidth]{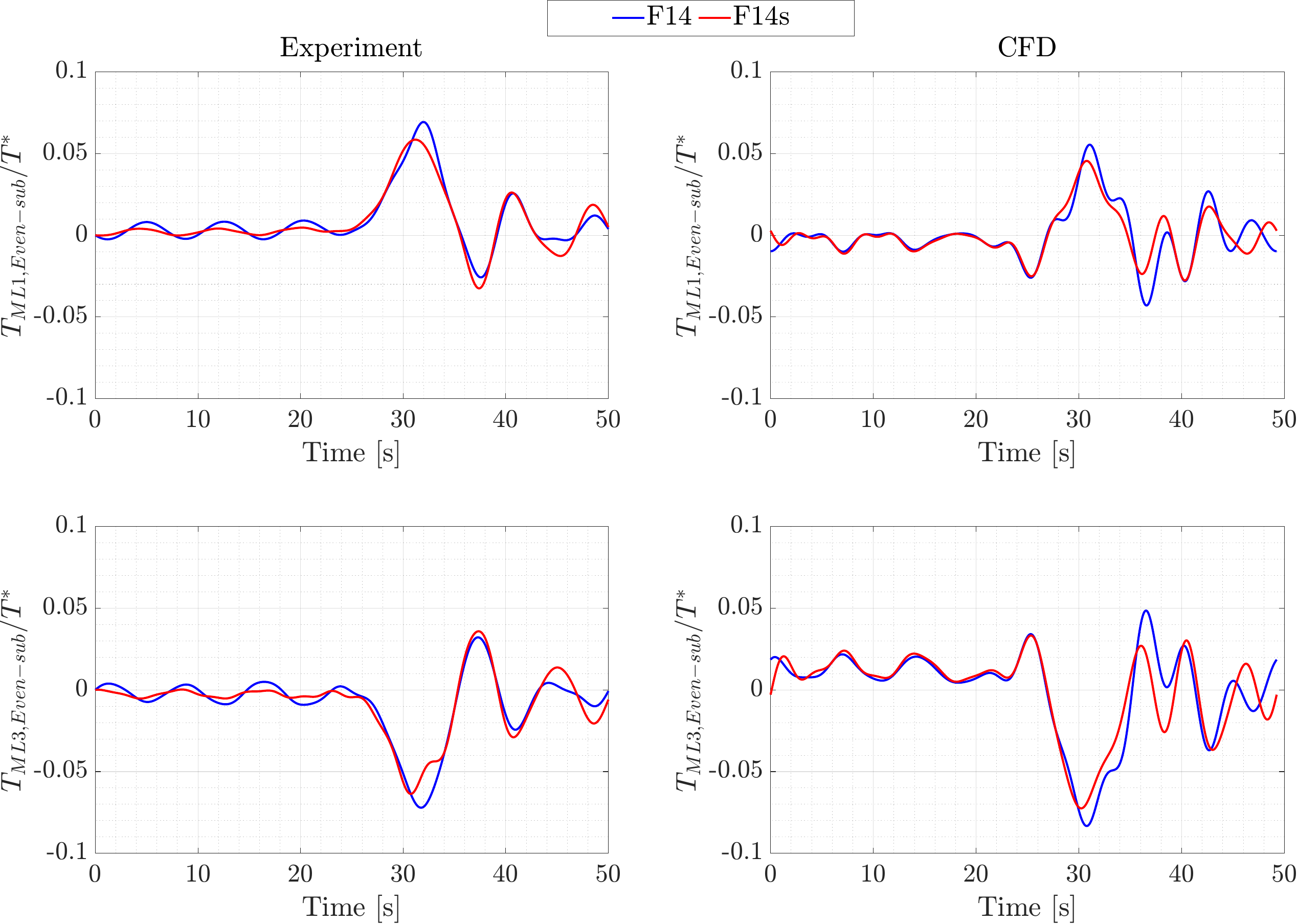}
    \caption{F14 and F14s: Row 1 shows experimental and CFD results for ML1 even subharmonic responses, followed by ML3 responses in Row 2, comparing cases with and without spreading\label{fig:ML13evensub0and20}}
\end{figure}

\begin{figure}[p]
    \centering
	    \includegraphics[width=0.825\textwidth]{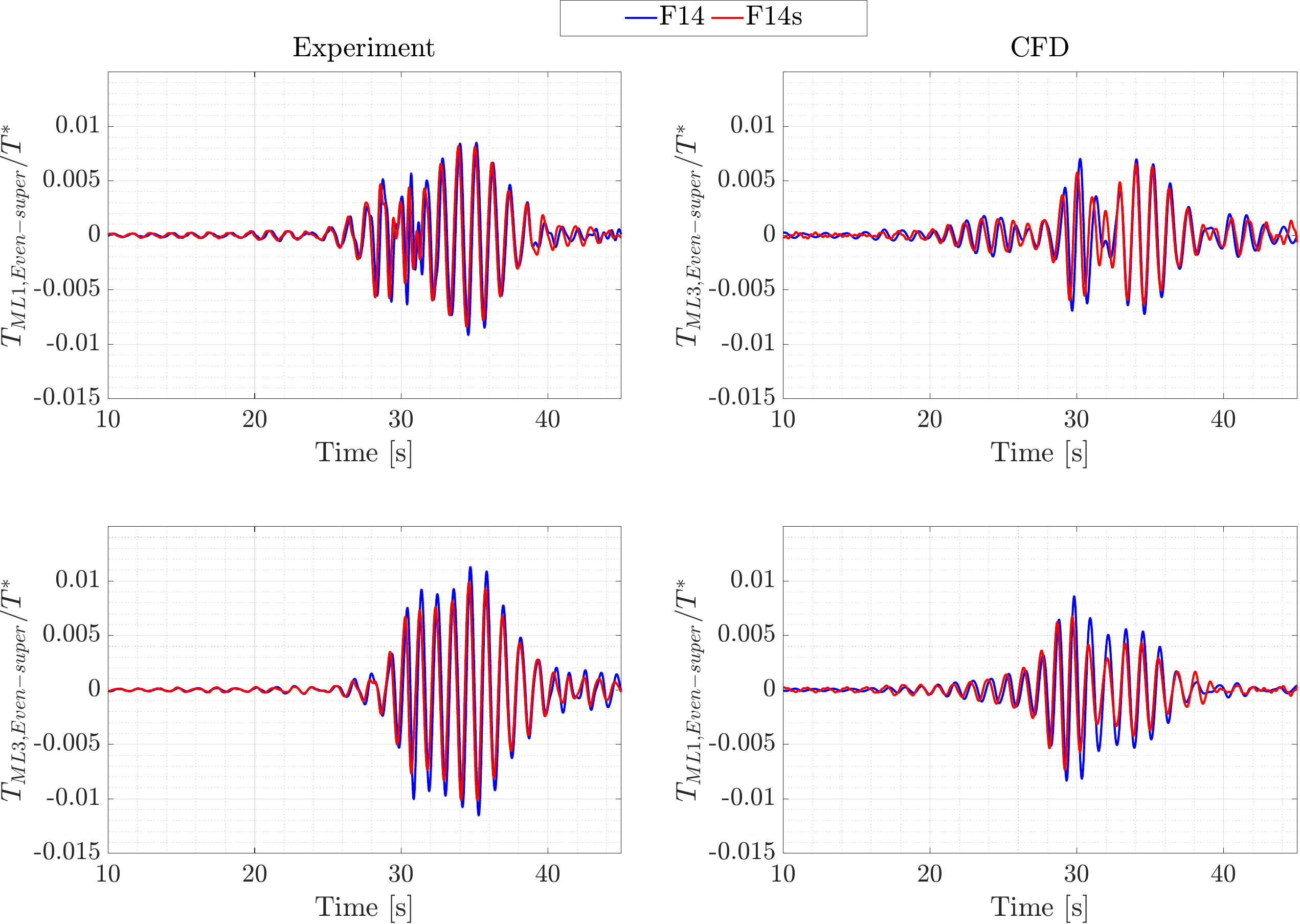}
     \caption{F14 and F14s: Row 1 shows experimental and CFD results for ML1 even super responses, followed by ML3 responses in Row 2, comparing cases with and without spreading\label{fig:ML13evensuper0and20}}
\end{figure}

\section{Effect of nonlinearity and wave spreading on mooring line tensions}\label{Section9}

Recalling the mooring system configuration described in Section \ref{Section2} is helpful to provide context for the following discussion. In this setup, the mooring system comprises four lines: ML1, ML2, ML3, and ML4. The mooring lines ML1 and ML2 are connected to a shared fairlead point at the front of the tank, while ML3 and ML4 are connected to separate fairlead points at the rear of the floating foundation. In the following analysis, the tension variations in ML1 and ML3 represent the line tensions in the front and rear mooring lines, respectively, and ML1 reflects the behavior of ML2 and ML3 that of ML4. From this point forward, ML1 and ML3 will be referred to as front and back mooring lines, respectively, to enhance clarity and understanding. Figure \ref{fig:ML13Odd0and20} presents the odd (linear) harmonics of the dynamic mooring line tensions subjected to the spread and non-spread {F14} focused wave group. Row 1 shows results for the front, and Row 2 shows results for the back mooring lines, including both the experiment and CFD results. Generally, the back mooring lines experience higher tension fluctuations than the front ones. Specifically, the maximum peak tension observed in the back mooring lines is approximately twice that of the front. The influence of wave spread remains minimal until the main wave passes, after which a minor reduction in tension is seen in both the front and back mooring lines for the spread case. 

Another observation is that the tension fluctuations of the front and back mooring lines are 180 degrees out of phase and the signal shape is different.  Although the back mooring lines exhibit a clear group structure, with the main peak being the highest, the front mooring lines have the first and third peaks higher than the main peak, possibly indicating a phase-related asymmetry between the front and back mooring lines. Typically, in a laterally symmetric mooring system with pretension in all lines, the front mooring lines tend to stretch when the back mooring lines slack. This can explain the phase difference. In addition, the inclined orientation of the mooring lines leads to a geometric coupling between surge and pitch. Hence, the front and back fairleads experience effects from the two distinct rigid modes of surge and pitch. The CFD simulations confirm the mentioned experimental characteristics, though a minor difference is observed in the front mooring lines after the focusing event.

Figure \ref{fig:ML13evensub0and20} presents the even subharmonics of the mooring line tensions for the same event. Also, for the subharmonic response, the front and back line tensions are of opposite phases. As the wave passes over the floating foundation, the front mooring lines display a positive tension, while the back mooring lines show a reduction. After the wave has passed, both the front and rear mooring lines display decay behavior as the floating foundation stabilizes. Similarly to the subharmonic motion, the associated mooring tensions are slightly less dominant than the odd harmonic line tension. The subharmonic mooring tensions in the front lines attain up to $80\%$ of the peak amplitude observed in the odd harmonic line tension. Furthermore, the lowest minimum tension observed in the back mooring lines is $50\%$ lower than the corresponding minimum with the odd harmonic line tension. The CFD predictions for mooring line tensions closely match the experimental results with respect to the trends and amplitudes. However, between $36$ and $38$ s, the CFD data reveal an additional peak that was not observed in the experiments, which slightly alters the subsequent decay characteristics of the floating foundation. The impact of wave spreading is visible through a reduction in peak amplitude in both front and back lines and differences in the motion history after the main peak. These characteristics are reproduced in the CFD results which indicate a slight overall reduction in tension due to wave spreading. 

We now turn to Figure \ref{fig:ML13evensuper0and20}, which presents even superharmonics in the mooring line tension. As expected, the superharmonic mooring line tensions are less dominant than the odd and subharmonic tensions. The maximum superharmonic amplitude in the front mooring lines is $10\%$ of the maximum amplitude of the odd tension. In the back mooring lines, the amplitude is $5\%$ of the tensions observed at the maximum trough amplitude of the odd harmonic tension. An interesting observation is that the front mooring line apparently experiences the tensions from the spurious free waves slightly earlier than the back mooring lines.  This can be related to the cancelation of the superharmonic tensions of the free and bound waves around $32$ s in the front mooring lines, and this cancelation is not evident in the back mooring line tensions. Similarly to the motions of the floating foundation, wave spreading leads to a reduction in the superharmonic tensions. Here, up to $10-15\%$. Although the reduction is less pronounced in the front mooring lines, it is more evident in the back mooring lines, mainly for the bound response at the time of the group peak. These behaviors are also reflected in the CFD results.

\section{Effect of wave steepness on the sub and superharmonic response}\label{Section10}
\begin{figure}[htb]
    \centering
	    \includegraphics[width=\textwidth]{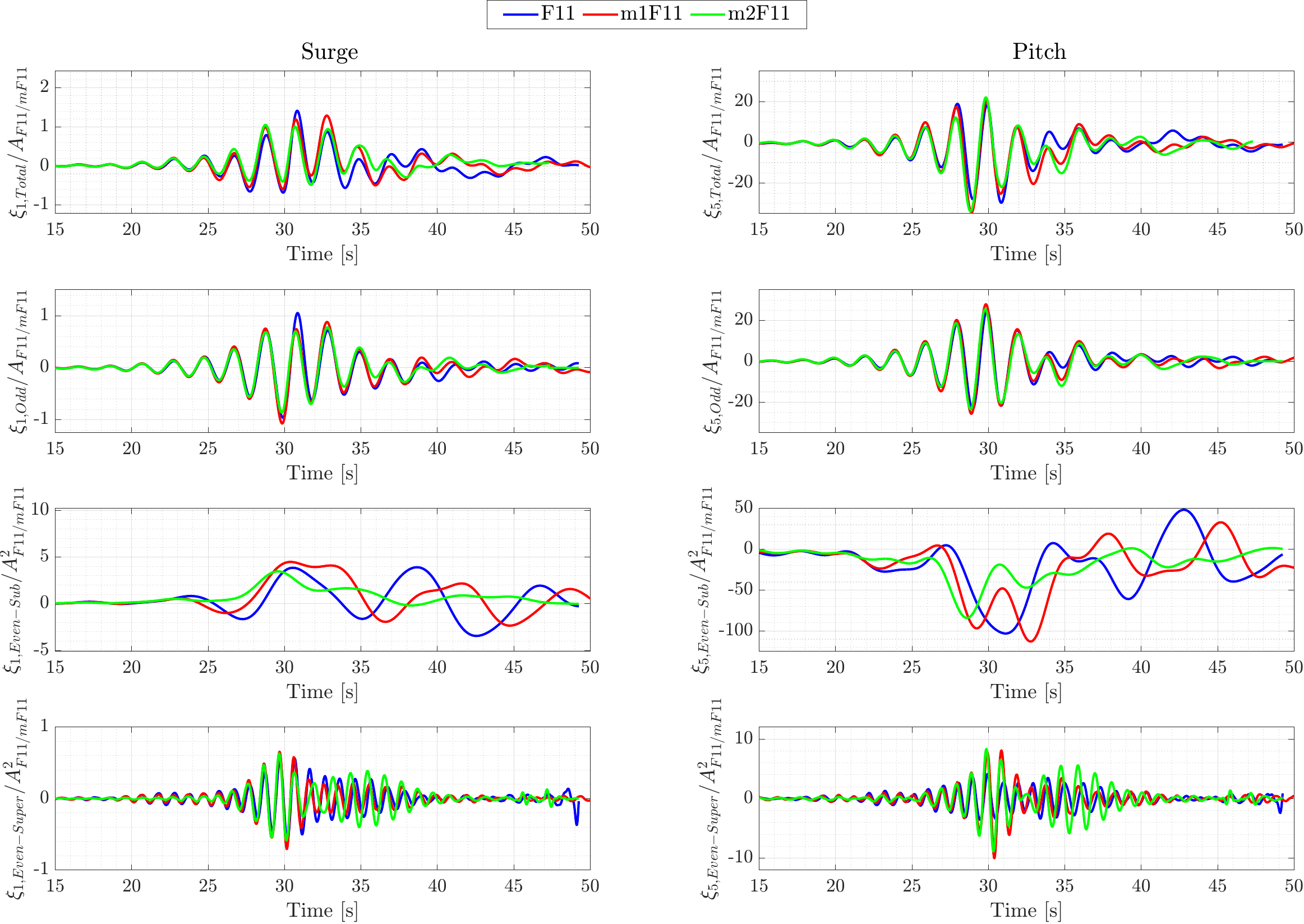}
    \caption{F11, m1F11 and m2F11: Column 1 shows Surge ($\xi_1$) responses for original and modified wave heights ($1.2H_s$), while Column 2 shows Pitch ($\xi_5$) responses normalised by amplitudes ($A_{F11}$). Rows represent total, odd, and even sub- and superharmonic responses.\label{fig:SurgePitch0and20H20}}
\end{figure}

\begin{figure}[htb]
    \centering
	    \includegraphics[width=\textwidth]{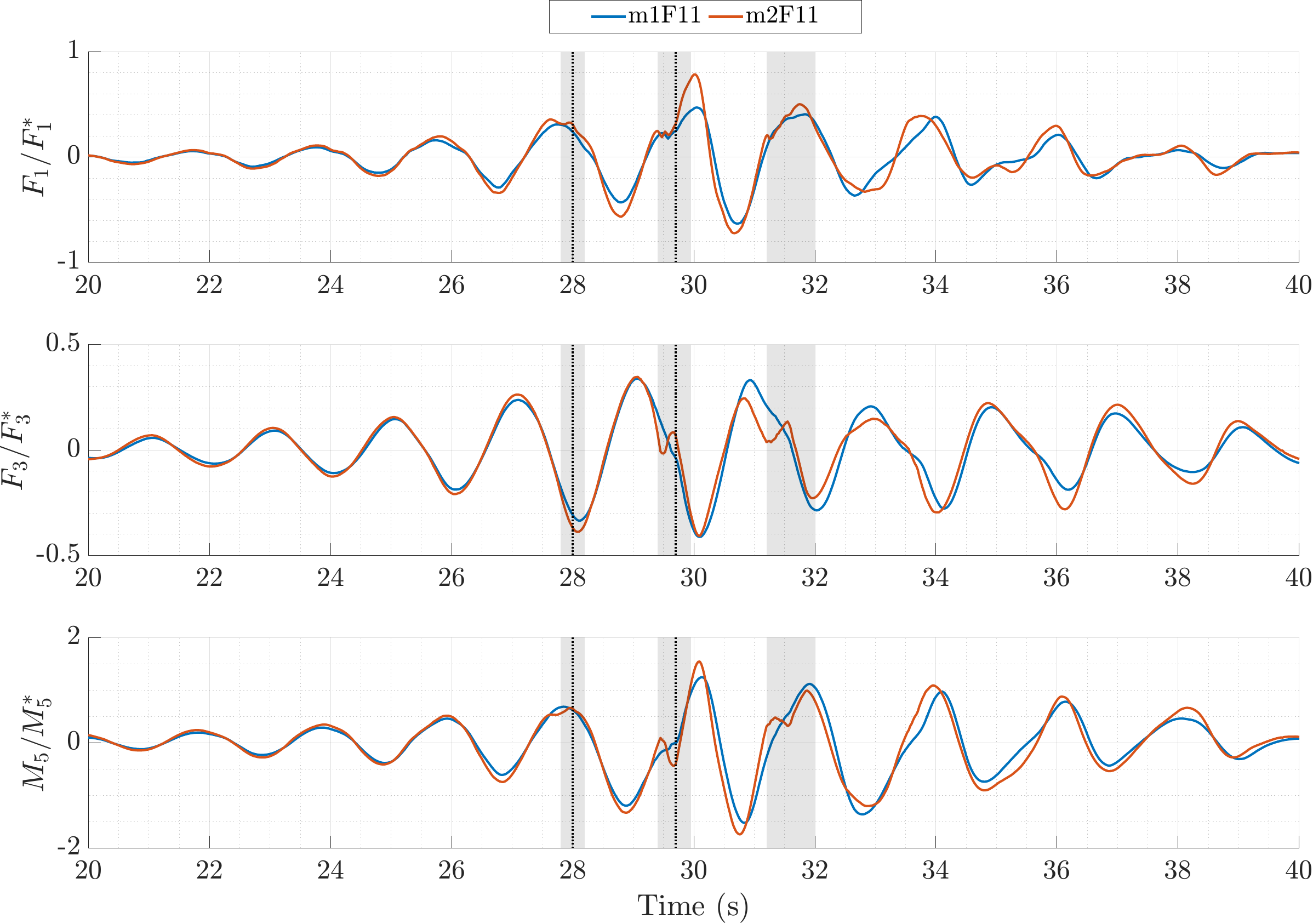}
    \caption{m1F11 and m2F11: Row 1 and row 2 shows Surge ($F_1$) and heave ($F_3$) forces for modified wave heights, while row 3 shows pitch moment ($M_5$) normalized by  constants ($F^*$ and $M^*$).\label{fig:SurgePitchForces}}
\end{figure}

\begin{figure}[ht]
    \begin{subfigure}[b]{0.47\textwidth}
        \centering
        \includegraphics[width=\textwidth,height=0.2\textheight]{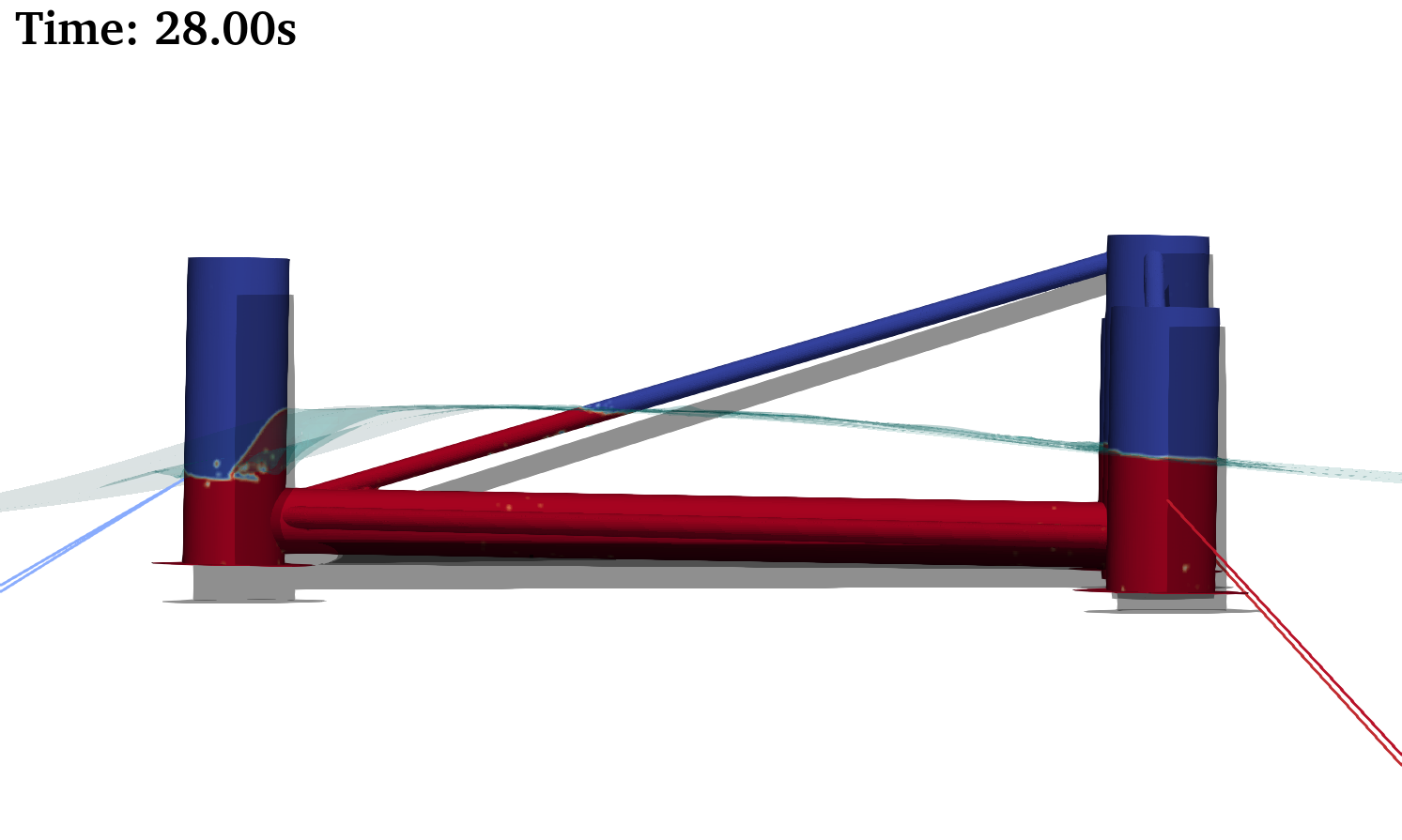}
    \end{subfigure}
    \begin{subfigure}[b]{0.49\textwidth}
        \centering
        \includegraphics[width=\textwidth,height=0.2\textheight]{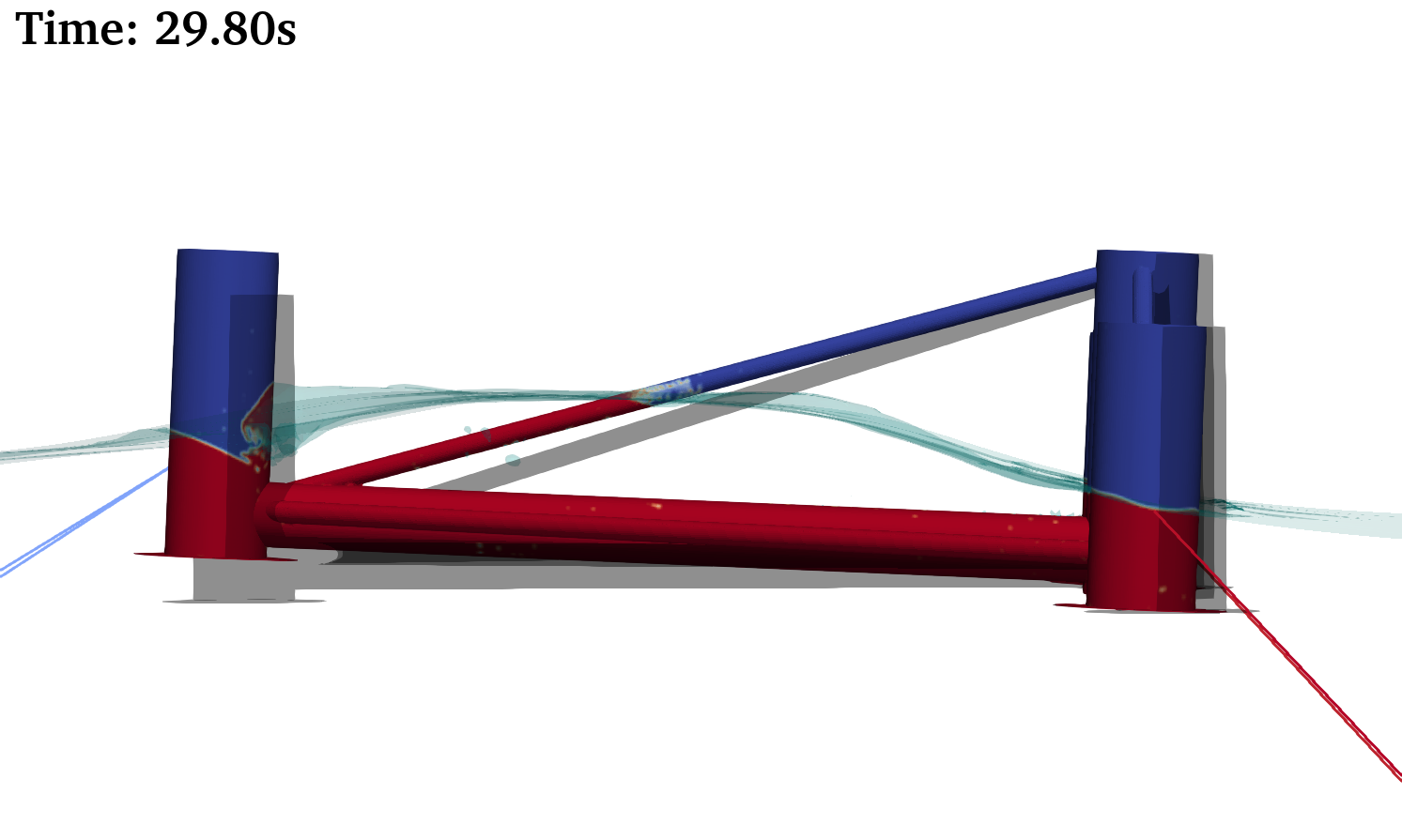}
    \end{subfigure}

    \caption{Visual time snaps showing the dynamic response of the floating foundation as the focused wave group progresses, at three distinct times, represented by the dotted black lines in Figure \ref{fig:SurgePitchForces}.}

    \label{fig:TimeSnapsWaveSeverity}
\end{figure}

In the following analysis, we examine the impact of amplitude on the harmonic response behavior by comparing three cases with different wave steepness. Using the previously validated F11 case as a baseline, we increased the wave steepness from $2.57\%$ to $3.1\%$ and $3.7\%$ by increasing the nominal crest amplitude by $20\%$ and $44\%$, respectively. This adjustment was applied to both the non-spread and spread groups to generate modified focused wave groups ({m1F11} and {m2F11} in Table \ref{tab:TestMatrix}). Due to similar observations in the spread cases and to avoid redundancy, only the comparisons for the non-spread groups are presented here. 

Figure \ref{fig:SurgePitch0and20H20} displays the surge and pitch responses in two columns. The top row shows the total response, the second row presents the odd harmonics, the third row displays the even subharmonics, and the bottom row the even superharmonics. Since data analysis allows a clean separation between the odd harmonic response (predominantly linear) and even (predominantly second order), we further test the scaling of these responses relative to amplitude(‘A’) for the two focused wave groups. Odd harmonic responses are normalized by the amplitude of the crest height of the focused wave group (A\(_{F11/mF11}\)). In contrast, even harmonics are scaled by the square of this amplitude (A\(^2_{F11/mF11}\)), consistent with its second harmonic nature. It is known that larger wave steepness can increase the wave propagation speed due to non-linear effects, causing steeper waves to travel slightly faster than their less steep counterpart of the same period. This aligns with our results, where the m2F11 case reaches the floating foundation slightly ahead of m1F11 and F11 (Figure \ref{fig:SurgePitch0and20H20}). Also, increasing the wave steepness leads to an amplified total response, particularly during the the main group focusing time ($28$ to $38$ s) for both surge and pitch responses. The curves for normalised odd harmonic responses largely overlap, confirming the linear response behavior of these signals. Although the amplitude of the odd harmonics in the surge response diminished at the main peak, possibly due to nonlinear third-harmonic content, it remains consistent in the preceding and following smaller waves, following linear behavior, as expected. Superhamonic surge follows $A^2$ scaling for the bound part that coincides with the main wave group, with some increase in the free harmonics that follow the group. However, for the superharmonic pitch motion, both the bound and the free part is seen to increase with the wave steepness. The largest effects of wave steepness are seen in the even subharmonic response, where normalized motion $A^2$ decreases in both the surge and pitch for increased steepness. This behavior was also found for the subharmonic response in random sea states, where a $A^1$ scaling aligned better with the results \citep{Aref2023}. In addition to the unexpected amplitude scaling, the response histories are seen to change markedly between the groups. Increasing the wave steepness enhances the pitch energy in the floating foundation responses, shifting the dominant frequency content from primarily low frequency (surge) to pitch during the focusing time. Furthermore, as a third effect of steepness, the damping effect becomes more pronounced in both surge and pitch, showing stronger damping at higher wave steepness (m2F11).

These observations led us to investigate the reasons behind the increased pitch energy and stronger damping, using the forces observed for cases m1F11 and m2F11, as shown in Figure \ref{fig:SurgePitchForces}. The surge and heave forces (F1, F3) and the pitch moment (M5) are normalized by the constants $F^*$ and $M^*$. To better understand the force variations, three shadowed rectangles are used to highlight the important key event as the wave steepness increases. A dotted line in the center of two rectangles marks a specific time instant, and the corresponding CFD rendering for the m2F11 case at that time is shown in Figure~\ref{fig:TimeSnapsWaveSeverity}. As the leading wave of the focused wave group passes over the front tank (at $28$ s in Figure \ref{fig:TimeSnapsWaveSeverity}), the wave crest positions itself just behind the tank and induces a local force peak in the surge inline force (F1), similar to the secondary load cycle for monopiles examined by \citet{ghadirian2020detailed}. The secondary load cycle is expected to be more prominent when the main wave group passes. However, when the main wave group reaches the front tank (at $29.8$ s in Figure \ref{fig:TimeSnapsWaveSeverity}), the trough preceding the main wave passes over the central tank, leading to a pronounced wave-structure interaction in the free surface zone. A clear fluctuation in the heave and pitch force is observed around this instant and can be linked to this interaction. Upon further analysis of the CFD-generated results, strong radiated wave patterns emanating from the forward tank following the wave passage were observed. Furthermore, these deep-trough passing and complex free-surface interactions in aft tanks appear to play an essential role in the increased damping observed in the even subharmonic response of m2F11 relative to F11. When the trailing wave of the focused wave group passes, a similar effect occurs, though with a reduced magnitude compared to the main wave.


\section{Conclusion}\label{Section11}

A detailed experimental and numerical analysis was conducted to investigate the effect of wave spreading on the linear and nonlinear motion response of a design variant of Stiesdal Offshore's  TetraSub semi-submersible floating foundation. The floating foundation used an inclined taut mooring system with lateral symmetry but no symmetry between the front and back. Focused wave groups of two distinct sea states, with and without wave spreading, were investigated using both in-phase and phase-shifted wave paddle signals. The measured responses were numerically reproduced using FloatStepper, a non-iterative OpenFOAM-based rigid body solver that integrates the incompressible Navier–Stokes equations with the geometric volume of fluid method. We applied a porous model-based mechanism in addition to the standard wave absorption boundary condition for wave absorption. This approach was validated for directional focused wave group cases and effectively mitigated wave reflections. The floating foundation numerical model showed reasonable accuracy in natural frequency and damping; however, a slight discrepancy in the natural frequency of the surge prompted additional validation against experiments of regular wave interaction. Odd and even harmonics were separated by applying harmonic separation to pairs of phase-manipulated responses. Furthermore, the even harmonics were divided into subharmonics and superharmonics, providing a complete assessment of the harmonics in both experimental and CFD results.

\textbf{Effect of wave severity}: The comparison between the milder focused wave group and the severe focused wave group showed a notable increase in odd harmonic amplitudes. Subharmonic surge responses reached about $75\%$ of the odd harmonic amplitudes, while pitch subharmonics were around $45\%$ for the test cases. The increase in wave severity activated pitch energy, showing a shift from primarily low-frequency surge responses to the inclusion of pitch frequencies, further altering the floating foundation's decay behavior. The superharmonic response showed a trend and amplitude similar to those of the milder focused wave groups, as there were minimal differences between the superharmonic waves of the milder and severe focused wave groups. Additionally, superharmonic spurious free wave responses reached up to $12\%$ of the maximum odd harmonic amplitudes in both surge and pitch, double the magnitude observed in milder focused wave groups. The CFD model demonstrated effective reproduction of both odd and even superharmonics; however, experiments indicated stronger damping for small amplitude motions than CFD across both subharmonic responses.

\textbf{Effect of wave spreading}: The total surge response showed larger peaks in the non-spread case than in the spread case, mainly due to odd and even superharmonics, with minimal contribution from even subharmonics. In both spread and non-spread cases, the total pitch response remained similar up to the focusing time. Beyond this point, the spread case showed a small decrease, again influenced by odd and even superharmonics, suggesting the predominance of wave contribution over natural frequencies. The CFD model effectively reproduced the similar spreading effects, with only a slight discrepancy in the even superharmonics.

\textbf{Effect of higher harmonics and wave spreading on mooring lines}: There is an asymmetry between the front and back mooring lines, resulting in different tension response patterns. The peak tension amplitude in the back mooring lines was roughly twice that of the front lines for odd harmonic mooring tensions. The front mooring lines experienced positive crests for even subharmonic mooring line tensions, while the back mooring lines showed negative troughs. The superharmonic tensions were minimal in comparison, contributing only a small fraction to the overall harmonic tension. For spurious free waves, the front mooring lines witnessed the tensions slightly earlier than the back mooring lines. Wave spreading slightly reduced tensions in both front and back mooring lines after the focusing event, with a more pronounced effect on even subharmonic tensions. Superharmonic tensions decreased by $10$-$15$\%, and the spreading effect was more noticeable in the back mooring lines.

\textbf{Effect of wave steepness}: This analysis revealed that increasing wave steepness amplified both surge and pitch responses, with notable changes observed during the focused wave group and floating foundation interaction. Specifically, while odd harmonic responses remained consistent with linear behavior, even subharmonics and superharmonics exhibited notable shifts, with pitch energy becoming more prominent and altering decay characteristics compared to smaller wave steepness cases. In addition, increased wave steepness introduces a clear fluctuation in the pitching moment, accompanied by stronger damping effects.

\section{Acknowledgement}
The research was carried out as part of the FloatLab project, funded by Innovation Fund Denmark (IFD) under grant no. 2079-00082B. This funding is gratefully acknowledged. The authors thank Aref Moalemi, Mathilde Howard Wagner, and Bjarne Jensen for their contributions to conducting the tests. Special thanks to Stiesdal Offshore for providing the test floater, supporting the experiments, and engaging in insightful discussions about the results.

\printcredits

\bibliographystyle{elsarticle-harv}

\clearpage

\bibliography{cas-refs.bib}

\appendix

\section{Superharmonic free waves} \label{AnnexA}
To analyze the relationship between the group speeds of primary waves and second-harmonic free waves, we start with the dispersion relation for primary waves. For a primary wave with angular frequency \(\omega_1\) and wavenumber \(k_1\) in deep water, the dispersion relation is:

\begin{equation}\label{eq:k1ref}
\omega_1^2 = gk_1 \rightarrow k_1=\frac{\omega_1^2}{g}
\end{equation}
The phase speed of these waves is thus $c_1=\omega_1/k_1 = g/\omega_1$ and the group speed at deep water is half of this:
\begin{equation}
c_{g1} = \frac{g}{2\omega_1}
\end{equation}
 This group velocity is also valid for the super harmonic bound waves, since they are bound to the primary waves. 

Next, we consider spurious superharmonic free waves, which have angular frequency \(\omega_2=2\omega_1\) and wave number \(k_2\). The dispersion relation for these waves is:
\begin{equation}\label{eq:k2ref}
\omega_2^2 =4\omega_1^2 =gk_2 \rightarrow k_2=\frac{4\omega_1^2}{g}
\end{equation}
The phase speed of these free waves is thus $c_{2}=2\omega_1/(4\omega_1^2 g)=g/(2\omega_1)$ and the group speed is then
\begin{equation}
c_{g2} = \frac{g}{4\omega_1}
\end{equation}
Thus, the group speed \(c_{g2}\) of the second-harmonic free waves is half the group speed \(c_{g1}\) of the primary waves:

\begin{equation}
c_{g2} =  \frac{c_{g1}}{2}
\end{equation}

For the sea states F11 and F14, the primary waves travel at group speeds of $3.1$ {m/s} and $3.9$ {m/s}, respectively, and are expected to reach the focusing location $7.73$ m from the wavemaker at $30^{th}$ s. In contrast, spurious superharmonic free waves, with half the group speeds of $1.58$ {m/s} and $1.75$ {m/s}, take approximately $4.89$ s and $4.41$ s more to cover the same distance. As a result, the peaks of these spurious waves are expected to reach the focusing location between $34$ s and $35$ s.

\end{document}